\documentclass[aps,prd,showpacs,twocolumn,floatfix,nofootinbib,superscriptaddress,
amsmath,amssymb,amsfonts,]{revtex4-1} 

\usepackage{bm}
\usepackage{slashed}
\usepackage{graphicx}
\usepackage{subfigure}
\usepackage[svgnames]{xcolor}
\usepackage{placeins}
\usepackage{xspace}
\def\LEPT{\texttt{LEPTO}\xspace} 
\def\MLEPT{\texttt{\lowercase{m}LEPTO}\xspace} 
\def\PYTH{\texttt{PYTHIA6}\xspace} 
\def\MPYTH{\texttt{\lowercase{m}PYTHIA}\xspace}

\def\fR{\varphi_R}
\def\fS{\varphi_S}

\def\fT{\varphi_T}
\def\kT{\vect{k}_T}

\def\PT#1{\vect{P}_{#1T}}
\newcommand{\vT}{\vect{P}_T}
\newcommand{\vect}[1]{\boldsymbol{#1}}
\newcommand{\bfk}{\vect{k}}
\newcommand{\bfq}{\vect{q}}

\newcommand{\bfS}{\vect{S}_{_T}}

\newcommand{\al}[1]{\begin{align} #1 \end{align}}
\newcommand{\non}{\nonumber}
\newcommand{\vf}{\varphi}

\newcommand{\Ge}{\mathrm{GeV}}
\newcommand{\Gs}{\mathrm{GeV}^2}

\newcommand{\ImM}{0.27\paperwidth}
\newcommand{\ImMM}{0.28\paperwidth}

\newcommand{\Eq}[1]{Eq.~(\ref{#1})}

\begin{document}

\title{Predictions for Sivers single spin asymmetries in one- and two-hadron electroproduction at CLAS12 and EIC}

\preprint{ADP-15-5/T907}

\author{Hrayr~H.~Matevosyan}
\affiliation{ARC Centre of Excellence for Particle Physics at the Tera-scale,\\ 
and CSSM, Department of Physics,\\
The University of Adelaide, Adelaide SA 5005, Australia
\\ http://www.physics.adelaide.edu.au/cssm
}

\author{Aram~Kotzinian}
\affiliation{Yerevan Physics Institute,
2 Alikhanyan Brothers Street,
375036 Yerevan, Armenia
}
\affiliation{INFN, Sezione di Torino, 10125 Torino, Italy
}

\author{Elke-Caroline~Aschenauer}
\affiliation{
Physics Department,
Brookhaven National Laboratory,
Upton, NY 11973,
USA
}

\author{Harut~Avakian}
\affiliation{
Thomas Jefferson National Accelerator Facility, 
12000 Jefferson Ave. Suite 5, 
Newport News, VA 23606,
USA
}

\author{Anthony~W.~Thomas}
\affiliation{ARC Centre of Excellence for Particle Physics at the Tera-scale,\\     
and CSSM, Department of Physics,\\
The University of Adelaide, Adelaide SA 5005, Australia
\\ http://www.physics.adelaide.edu.au/cssm
}

\begin{abstract}
The study of the Sivers effect, describing correlations between the transverse polarization of the nucleon and its constituent (unpolarized) parton's transverse momentum, has been the topic of a great deal of experimental, phenomenological and theoretical effort in recent years. Semi-inclusive deep inelastic scattering measurements of the corresponding single spin asymmetries (SSA) at the upcoming CLAS12 experiment at JLab and the proposed Electron-Ion Collider will help to pinpoint the flavor structure and the momentum dependence of the Sivers parton distribution function describing this effect. Here we describe a modified version of the \PYTH Monte Carlo event generator that includes the Sivers effect. Then we use it to estimate the size of these SSAs, in the kinematics of these experiments, for both one and two hadron final states of pions and kaons. For this purpose we utilize the existing Sivers parton distribution function (PDF) parametrization extracted from  HERMES and COMPASS experiments. Using this modified version of \PYTH, we also show that the the leading order approximation commonly used in such extractions may provide significantly underestimated values of Sivers PDFs, as in our Monte Carlo simulations the omitted parton showers and non-DIS processes  play an important role in these SSAs, for example in the COMPASS kinematics.
\end{abstract}


\date{\today}                                           

\maketitle

\section{Introduction}
\label{SEC_INTRO}
%
 The study of the structure of the nucleon has long been a focus of both theory and experiment within the medium and the high energy hadronic physics community, mostly concentrating on the distributions of the longitudinal momentum fraction of partons within nucleon described by the transverse momentum integrated ("collinear") parton distribution functions (PDF). Lately, exploring the transverse structure of the nucleons in momentum space via transverse momentum dependent (TMD) PDFs and the generalized parton distribution functions in impact parameter space has attracted great interest. The spin of the nucleon and its constituent partons plays an important role when describing the transverse structure of the nucleon. Here we will focus on the so-called Sivers TMD PDF that describes the modification of the fully unpolarized TMD PDF when considering unpolarized partons inside of a transversely polarized nucleon~\cite{Sivers:1989cc}. The knowledge of the Sivers PDF is crucial in our understanding of the spin structure of the nucleon. It can  be accessed in a variety of experiments in the so-called deep inelastic scattering (DIS) regime, where the probes intact directly with a parton inside of the target, such as the Drell-Yan process and semi-inclusive hadron production in polarized hadron-hadron collisions, or semi-inclusive DIS (SIDIS) of a lepton off a transversely polarized nucleon. 
 
 In a SIDIS process with a single detected final state hadron, the Sivers PDF can be accessed through a term in the single spin asymmetry (SSA) modulated with the sine of the so-called Sivers angle~\cite{Anselmino:2005nn}, defined as the azimuthal angle between the transverse momentum of the detected hadron and the transverse spin of the nucleon. The amplitude of this modulation is given as a convolution of the Sivers PDF with the TMD unpolarized fragmentation function (FF) divided by a convolution of the unpolarized TMD PDF and TMD FF. Experimental measurements  of these asymmetries have been performed by several collaborations: HERMES~\cite{Airapetian:2009ae,Airapetian:2004tw}, COMPASS~\cite{Adolph:2012sp,Adolph:2014zba} and JLab HALL A~\cite{Qian:2011py}, extracting sizable effects for both pion and kaon final states. The Sivers PDF has then been extracted in Ref.~\cite{Anselmino:2008sga} using phenomenological parametrizations of the involved TMD PDFs and TMD FFs to fit the experimentally measured SSAs, and more recently with the inclusion of the TMD QCD evolution~\cite{Collins:2014jpa} in Refs.~\cite{Anselmino:2012aa,Echevarria:2014xaa}. Unfortunately these parametrizations suffer from  relatively large uncertainties arising from both theory and experiment.  These uncertainties come from the experimental measurements themselves, the lack of knowledge of unpolarized TMD FFs (especially those for the unfavored channels~\cite{Matevosyan:2011zza}, as well as those involving the strange quark~\cite{Matevosyan:2010hh}), and assumptions about the shape and the flavor dependences of both unpolarized and Sivers PDFs and unpolarized FFs. This is relevant, as in calculating SSAs one integrates over the transverse momentum and then sums over the type of struck parton. Moreover, in such extractions there are unquantified uncertainties associated with using the leading order expressions for the SSAs that ignore any parton showering (note that TMD evolution partially accounts for this effect) or non-DIS process contributions (such as vector-meson-dominance) to the cross section of the inclusive hadron production in lepton scattering off a polarized nucleon target.
 
 Recently, it has been suggested that the Sivers PDF can be also measured in SIDIS production of two hadrons in the fragmentation of the same struck parton~\cite{Kotzinian:2014lsa,Kotzinian:2014gza,Kotzinian:2014hoa}. Here the SSA has two terms involving the Sivers PDF, modulated with respect to the sine of the Sivers angles of the total and the relative transverse momenta of the pair. In each of these terms, the Sivers PDF is convoluted with a corresponding unpolarized dihadron fragmentation function (DiFF), that is not  yet known. \LEPT Monte Carlo (MC) generator was modified to study these SSAs by using the Sivers PDF parametrizations of~\cite{Anselmino:2008sga,Anselmino:2012aa} and the Lund string fragmentation model employed in \LEPT to describe the semi-inclusive production of two hadrons. Using this MC generator, it was shown that the corresponding SSAs are of a similar size to those for the single hadron production in the kinematics of the COMPASS experiment. In addition, the two hadron final state provides a larger basis than the single hadron case for disentangling the flavor dependence of the Sivers PDF, since for $N$ types of detected hadrons we can measure $N$ single hadron SSAs compared with $N\cdot (N+1)/2$ dihadron SSAs (including those with the same type of hadrons in the pair). Typically, due to the design constraints of a particular detector, $N$ is not large enough to determine the flavor dependences. On the other hand, as shown in Sec.~\ref{SEC_MC_ONE_HADRON} and Sec.~\ref{SEC_MC_TWO_HADRON}, a particular choice of hadrons in a pair may allow one to gain more sensitivity to the contributions of a quark with given flavor. 
 
 Thus the measured dihadron SSAs will help to put additional constraints on the flavor dependence of the Sivers PDFs, and ideally extract them from their convoluted sums in SSAs without any assumptions on their flavor dependences. Moreover, SSA measurements in two-hadron SIDIS will be actively pursued in the upcoming CLAS12  and proposed SoLID experiments for accessing the transversity PDF through the interference fragmentation channel~\cite{JLAB12:CLAS12, JLAB12:SOLID}, making Sivers two-hadron studies a relatively easy addition to the analysis program. Together with the one hadron SIDIS and other measurements this will help to map out the full momentum and flavor dependence of Sivers PDFs. This is, of course, critically hinged on our detailed knowledge of unpolarized TMD PDFs, TMD FFs and DiFFs, that also need to be mapped out in future experiments. Finally, dihadron measurements will help to investigate the correlations in the hadronization process that have been discussed recently~\cite{Kotzinian:2014uya}.
  
 The exploration of the transverse structure of the nucleon, and the Sivers function in particular, is one of the main focuses of the CLAS12 experiment at the soon to be operational JLab at $12~\Ge$ and the proposed Electron Ion Collider (EIC)~\cite{Accardi:2012qut}. Thus it is important to make projections for the sizes of the expected SSAs using the current information on the Sivers PDF extracted from experiment, as well as to analyze what are the important physics questions that can be posed in different centre-of-mass energy regimes. Here, we use a modified version of the \PYTH (\MPYTH) MC event generator that incorporates the Sivers effect (as described in Sec.~\ref{SEC_MC_ONE_HADRON}) to make such projections for the CLAS12 experiment at JLab and for two sets of beam energies for EIC.
 
 This paper is organized in the following way. In Sec.~\ref{SEC_THEORY} we will give a brief overview of the SIDIS production of one and two hadrons including the Sivers effect. In Sec.~\ref{SEC_MC_ONE_HADRON}, we present our predictions for one hadron Sivers SSAs for both CLAS12 and EIC experiments, while in Sec.~\ref{SEC_MC_TWO_HADRON} we present the similar results for the two hadron production. Finally, in Sec.~\ref{SEC_CONC} we present our conclusions and outlook.

\vspace{-0.5cm}
\section{The Sivers effect in SIDIS}
\label{SEC_THEORY}

\begin{figure}[b]
\begin{center}
\includegraphics[width=1\columnwidth]{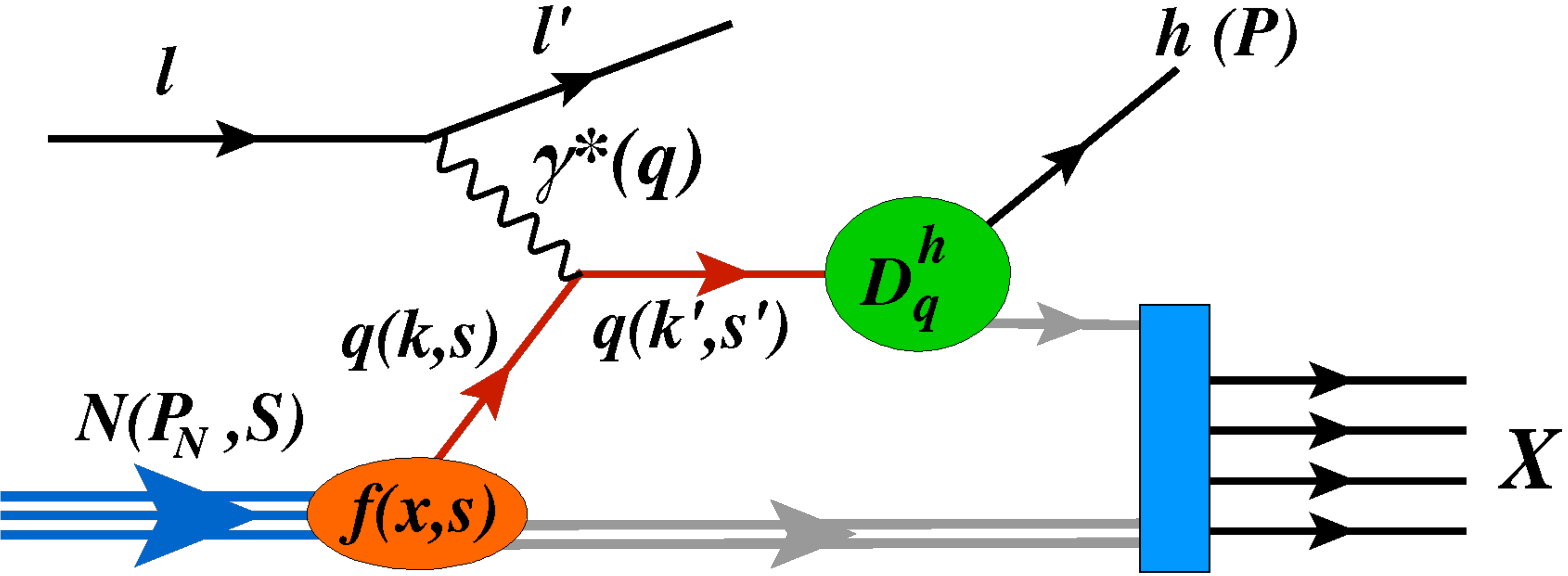}
\caption{The leading order diagram for one hadron production in the current fragmentation region of SIDIS.}
\label{FIG_ONEHADRON_CFR}
\end{center}
\end{figure}

In the SIDIS process a high energy lepton with momentum $l$, scatters off a nucleon target, $N$, producing several final state particles along with the scattered lepton with momentum $l'$
\al{
\label{EQ_DIS}
\ell (l) + N({P_N},S) \to \ell (l')   + X\,,
}
where $P_N$ and $S$ denote the momentum and the spin four-vectors of the nucleon. The leading-order DIS process in the factorized framework, where the lepton hard scatters of a single parton inside of a nucleon via single virtual photon exchange $\gamma^*$, is schematically depicted in Fig.~\ref{FIG_ONEHADRON_CFR}. Here the momenta and the spin four-vectors of the initial and the scattered partons are denoted by $k,s$ and $k',s'$. 

\begin{figure}[h]
\begin{center}
\includegraphics[width=1\columnwidth]{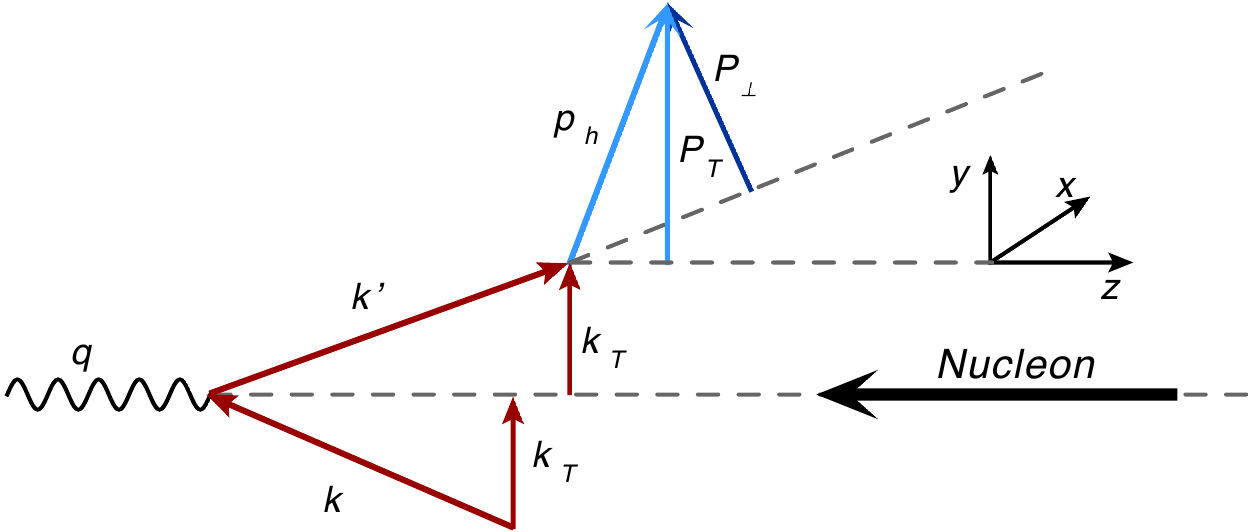}
\caption{$\gamma^*-N$ center of mass frame.}
\label{FIG_GAMMA-N_FRAME}
\end{center}
\end{figure}

In describing the relevant kinematic variables and cross-section, we adopt the $\gamma^*-N$ center of mass frame depicted in Fig~\ref{FIG_GAMMA-N_FRAME}. The $z$ axis is taken along the direction of the virtual photon three-momentum $\bfq$, and the $x$ axis is along the component of the lepton's momentum transverse to $\vect{q}$. The transverse components of the momenta in this frame are defined with respect to the $z$ axis with subscript $_T$ and the transverse momenta with respect to the fragmenting quark's direction with subscript $_\perp$.

 The usual DIS variables are defined as:
\al{
\label{EQ_KIN}
&
q=l-l',\  Q^2 = -q^2, \ x = \frac {Q^2}{2P_N \cdot q},
\\&\non
\, y = \frac {P_N \cdot q}{P_N \cdot l}, \ W^2=  (P_N+q)^2,
}
where $P_N$ is the momentum of the nucleon and $x$ is the Bjorken variable. It is also straightforward to see that in the  $\gamma^*-N$ system the transverse momenta of the initial and  fragmenting quarks are the same, $\vect{k}'_T=\vect{k}_T$, and $\vect{q}_T\equiv 0$, by definition.

The Sivers effect describes the modification of the TMD PDF of unpolarized partons inside a transversely polarized nucleon, $f_\uparrow^q(x,k_T)$,  with respect to that for the unpolarized nucleon $f_1^q(x,k_T)$ (we omit the $Q^2$ dependence of these PDFs for notational convenience). This modification is quantified by the Sivers PDF,  $f_{1T}^{\perp q}(x,k_T)$, which arises from the correlation between the active quark transverse momentum, $\vect{k}_T$, and the transverse polarization of the nucleon, $\mathbf{S}_T$. The polarized PDF is then expressed as (see e.g.,~\cite{Anselmino:2005nn,Kotzinian:2014lsa})
\al{
\label{EQ_POL_PDF}
f_\uparrow^q(x,\vect{k}_T)=f_1^q(x,k_T)+\frac{[\bfS \times \bfk_T]_3}{M}f_{1T}^{\perp q}(x,k_T),
}
where $M$ denotes the nucleon mass and the subscript $3$ denotes the $z$ component of the vector.

\vspace{-0.5cm}
\subsection{The Sivers effect in single hadron production}
\label{SUB_SEC_XSEC_1H}

 In the single hadron production in SIDIS,
\al{
\label{EQ_1H_SIDIS}
\ell (l) + N({P_N},S) \to \ell (l') + h(P)  + X\,,
}
one of the final state hadrons, $h$, with momentum, $P$, is detected along with the scattered lepton, $l$, carrying momentum $l'$, as schematically depicted in Fig.~\ref{FIG_ONEHADRON_CFR}. Here two of the frequently used scaling variables are
\al
{
z= \frac{P_N \cdot P}{P_N \cdot q},
}
describing the fraction of the virtual photon's energy carried by the produced hadron in the target rest frame, and
\al
{
x_F \equiv \frac{2P_z}{W},
}
the fraction of the maximum possible longitudinal momentum of the produced hadron in the $\gamma^*-N$ center-of-mass CM system (Feynman $x$). In QCD factorized approach in the so-called current fragmentation region (CFR) defined as $x_F>0$, this hadron is primarily produced by the fragmentation of the scattered parton. Although, within a dynamic string fragmentation picture, it is hard to make a strict distinction between the hadrons produced by the scattered parton's fragmentation and those by the nucleon remnant's fragmentation, especially for hadrons with relatively small momentum components along the lepton-nucleon axis in the $\gamma^*-N$ CM system. Nevertheless, a formal distinction is assigned by considering hadrons with $x_F>0$ in the CFR and those with $x_F<0$ in the target fragmentation region (TFR).

 Then, in the leading order approximation of QCD factorized approach for hadrons produced in the CFR, the relevant cross section can be expressed as~\cite{Bacchetta:2006tn}
\al
{
& \frac{d\sigma^{h}}{dx\, d{Q^2}\, d{\fS}\, dz\, d^2\PT{}} =C(x,Q^2)  \big({\sigma_U^h} + {\sigma_{S}^h}\big),
\\ 
\label{EQ_1H_SIG_UNP}
& \sigma_U^h = \sum_q e_q^2 \int d^2 \kT \ f_1^q\ D_q^{h},
\\
\label{EQ_1H_SIG_SIV}
&\sigma_{S}^h = \sum_q e_q^2 {\int {{d^2}{\kT}} 
\frac{S_T k_T }{M} \sin(\vf_k-\vf_S) f_{1T}^{ \perp q}\ D_{1q}^{h}},
\\
&C(x,Q^2) = \frac{\alpha^2(1+(1-y)^2)}{Q^4},
}
where $\vf_k$ and $\vf_S$ are the azimuthal angles of the quark's  transverse momentum, $\vect{k}_T$, $S$ the nucleon spin and $\alpha$ the fine-structure constant. $D_{1q}^{h}(z,P_\perp^2)$ denote the TMD fragmentation functions (FF), where $\vect{P}_\perp$ is the transverse momentum of the hadron with respect to the fragmenting quark, acquired in the hadronization process.

  The Sivers SSA for this process can be expressed as
\al
{
\label{EQ_SSA_1H_SIV}
A_{Siv}^h \equiv 2 \frac{\int d\vf_S d\vf_h\ \sigma^{h}_S \sin(\vf_h-\vf_S)} {\int d\vf_S d\vf_h\ \sigma^{h}_U }.
}

 Note, that the Sivers SSA of \Eq{EQ_SSA_1H_SIV} is calculated using the cross-section formulae (\ref{EQ_1H_SIG_UNP},\ref{EQ_1H_SIG_SIV}) entail that we only use the leading order approximations (in twist expansion and $\alpha_s$ expansion for the hard scattering, etc.). This approach considers corrections from gluon radiation by the initial or scattered quark, or when the virtual photon interacts with the nucleon via vector-meson-dominance (VMD) type processes, to be small. These assumptions might not strictly hold for real world experiments, and we will explore this issue using MC generators for COMPASS and EIC energies in the next section.

\vspace{-0.5cm}
\subsection{The Sivers effect in two hadron production}
\label{SUB_SEC_SIV_2H}
%
\begin{figure}[tbh]
\begin{center}
\includegraphics[width=1\columnwidth]{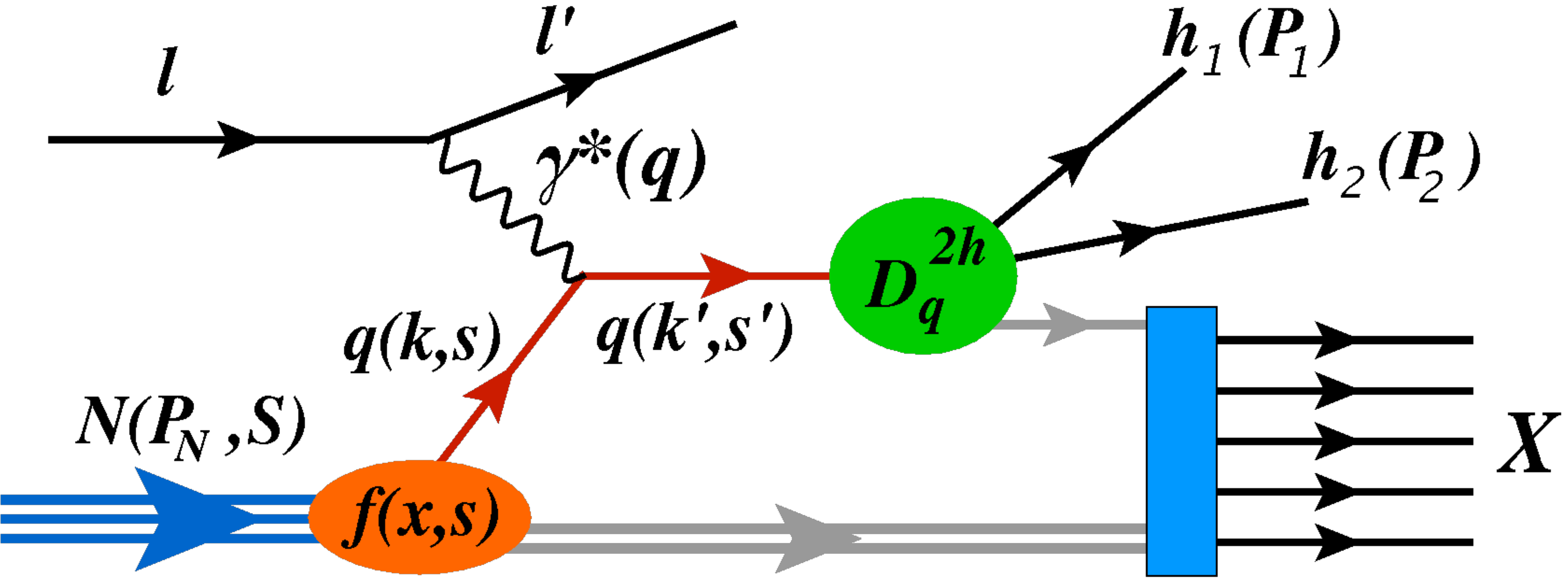}
\caption{The leading order diagram for two hadron production in the current fragmentation region of SIDIS.}
\label{FIG_DIHADRON_CFR}
\end{center}
\end{figure}
%

 Here we consider the SIDIS process, where along with the scattered lepton we also detect two of the produced hadrons $h_1$ and $h_2$, with momenta $P_1$ and $P_2$
\al{
\label{EQ_2H_SIDIS}
\ell (l ) + N({P_N},S) \to \ell (l') + h_1(P_1) + h_2(P_2) + X\,.
}
The leading order factorized schematic diagram for this process is depicted in Fig.~\ref{FIG_DIHADRON_CFR}, where we assume both hadrons are produced in CFR. 

  In the experimental and phenomenological analysis of two hadron SIDIS, the transverse components of the two hadron momenta, $\PT{1}$ and $\PT{2}$, are often replaced by their linear combinations~\cite{Bianconi:1999cd, Airapetian:2008sk,Adolph:2012nw}
\al
{
\label{EQ_T_R}
&
\vT=\vect{P}_{1T}+\vect{P}_{2T},
\\&
\vect{R}=\frac{1}{2}\left(\vect{P}_{1T}-\vect{P}_{2T}\right),
}
where the corresponding azimuthal angles are denoted as $\vf_T$ and $\vf_R$.  The cross section terms relevant to the Sivers effect in this process can then be expressed as~\cite{Kotzinian:2014lsa,Kotzinian:2014gza}
\al{
\label{EQ_2H_X_SEC_RT}
&\frac{d\sigma^{h_1 h_2}}{dx d{Q^2} d{\fS} dz_1 dz_2 d^2\vT d^2\vect{R}\,} = C(x,Q^2) 
(\sigma_U^{h_1 h_2} + \sigma_{S}^{h_1 h_2}),
\\ \label{EQ_2H_SIG_UNP}
& \sigma_U^{h_1 h_2} = \sum_q e_q^2 \int d^2 \kT \ f_1^q\ D_q^{h_1h_2},
\\
\label{EQ_2H_SIG_SIV}
&\sigma_{S}^{h_1 h_2} = \sum_q e_q^2 {\int {{d^2}{\kT}} 
\frac{[\bfS \times \bfk_T]_3}{M} f_{1T}^{ \perp q}\ D_{1q}^{h_1 h_2}},
}
where the virtual photon energy fractions $z_1$ and $z_2$ carried by two hadrons are defined similarly to the one hadron case. $D_{1q}^{h_1h_2}$ is the fully unintegrated unpolarized dihadron fragmentation function. The terms $\sigma_U$ and $\sigma_S$ depend on $x,\ Q^2, \ z_1, \ z_2, \ P_T, \ R$ and  $\vT\cdot\vect{R}=P_T R\cos(\vf_T-\vf_R)$. The Sivers term can be expressed as
\al
{
\sigma_{S} = S_T \Big[\sigma_T\frac{P_T}{M}\sin(\vf_T-\fS) + \sigma_R\frac{R}{M}\sin(\vf_R-\fS)\Big]
\label{EQ_2H_SIG_S_SIV}
}

 The cross section in Eq.~(\ref{EQ_2H_X_SEC_RT}) is usually measured after integrating over the azimuthal angle (and often magnitude) of the relative or total transverse momentum, $\fR$ or $\fT$ respectively:
\al{
\label{EQ_2H_X_SEC_INT_R}
\frac{d\sigma^{h_1h_2}}{d^2\vT RdR} &= C(x, Q^2)
\left[
\sigma_{U,0}  \vphantom{\frac{1}{1}} \right.
\\ \non
&\left. +
S_T \left(\frac{P_T}{M}\sigma_{T,0}+\frac{R}{2M}\sigma_{R,1} \right)\sin(\vf_T-\fS)
\right],
\label{EQ_2H_X_SEC_INT_T}
\\ 
\frac{d\sigma^{h_1h_2} }{P_T dP_T d^2\vect{R}} &= C(x, Q^2)
\left[
 \sigma_{U,0} 
\vphantom{\frac{1}{1}} \right.
\\ \non&
\left. +
S_T \left(\frac{P_T}{2M}\sigma_{T,1}+\frac{R}{M}\sigma_{R,0} \right)\sin(\vf_R-\fS)
\right],
}
where $\sigma_{U,i}, \sigma_{T,i}$ and $\sigma_{R,i}$ are the zeroth ($i=0$) and the first ($i=1$) harmonics of the $\cos(n(\vf_T-\vf_R))$ Fourier expansions of the corresponding structure functions. The superscripts ${h_1 h_2}$ are omitted.

 Then the relevant Sivers SSA can be obtained for the sine modulation of the cross section with respect to $\vf_{T,R}-\vf_S$
\al
{
A_{Siv}^{h_1 h_2}  \equiv 2 \frac{\int d\vf_S d\vf\ \sigma^{h_1 h_2}_S \sin(\vf-\vf_S)} {\int d\vf_S d\vf\ \sigma_U^{h_1 h_2} }
}
%

\section{Monte Carlo predictions for one hadron SSA}
\label{SEC_MC_ONE_HADRON}
 
  The Monte Carlo (MC) event generators are immensely valuable tools for both phenomenology and experiment in hadronic and high energy physics. They allow one to understand various features in the complicated DIS processes that are induced by the strong nonperturbative as well as perturbative effects, kinematical limitations coming from both underlying process and detector performance used in a particular experiment. Another advantage provided by MC event generators, versus simply employing the analytical expressions for the relevant cross sections, is the full coverage of the available phase space for produced particles. Furthermore, the dynamical full hadronization models employed in MC generators provide a unified description of the process in all regions of $x_F$, where the formalism for SIDIS in CFR for $x_F>0$ involves different functions (PDFs and FFs) to those in TFR for $x_F<0$ (Fracture Functions~\cite{Trentadue:1993ka, Anselmino:2011ss,Kotzinian:2011av}).
  
   Here we start with the well-established MC generator \PYTH and modify it (\MPYTH) to include the Sivers effect, which yields a modulation of the azimuthal angle of the struck quark's transverse momentum with respect to the nucleon's spin. For this we use the phenomenological parametrization of the Sivers PDF of Refs.~\cite{Anselmino:2008sga,Anselmino:2012aa} extracted from fits to SIDIS SSAs measured in the HERMES~\cite{Diefenthaler:2005gx,Airapetian:2009ae} and COMPASS~\cite{Alexakhin:2005iw,Bradamante:2011xu} experiments.  The modified version of the \LEPT MC generator, \MLEPT, has already been used in previous work~\cite{Kotzinian:2005zs,Kotzinian:2005zg,Kotzinian:2014lsa,Kotzinian:2014gza,Kotzinian:2014hoa} to estimate the SSAs in both single and dihadron SIDIS production, mostly in the COMPASS kinematics. For the present study \PYTH has been chosen as a more modern generator that among other new features uses double precision to operate with the particle momenta which is better suited for EIC studies. 
   
  The Sivers effect has been implemented in \MPYTH by modifying the part of the \PYTH code that generates the azimuthal angle of the struck quark in the $\gamma^*-N$ CM system, where the $x$ and $k_T$ have already been selected according to the cross-section formula for the lepton-nucleon DIS scattering and the selected intrinsic transverse momentum parametrization (we chose Gaussian), respectively. Here, instead of the uniform distribution for the case of an unpolarized nucleon, we sample the azimuthal angle according to the \Eq{EQ_POL_PDF}. We use one of the standard parametrizations for the unpolarized PDF accessible in \PYTH. We take the Sivers function parametrization from Refs.~\cite{Anselmino:2008sga,Anselmino:2012aa} ("DGLAP evolution" set) and slightly modify it to best reproduce COMPASS measurements of the corresponding asymmetries. The latter is expressed as a product of the unpolarized PDF and two functions, where one depends only on $x$ and the other only on $k_T$. The parametrization actually involves only these two multiplicative functions. Further, we use a Lorentz transform to determine $\vect{S}_T$ in the  $\gamma^*-N$ CM system in order to calculate the Sivers term. These modulations of the struck quark azimuthal angle are then transferred to the final state hadrons by the hadronization process. Moreover, these azimuthal modulations are also transferred to the nucleon remnant via momentum conservation.  Note that here we only study the Sivers effect for the light quarks, as currently there are no available parametrizations for the Sivers PDFs of a gluon or for heavy flavor quarks.
    
\begin{figure}[tb]
\centering 
\subfigure[] {
\includegraphics[width=\ImM]{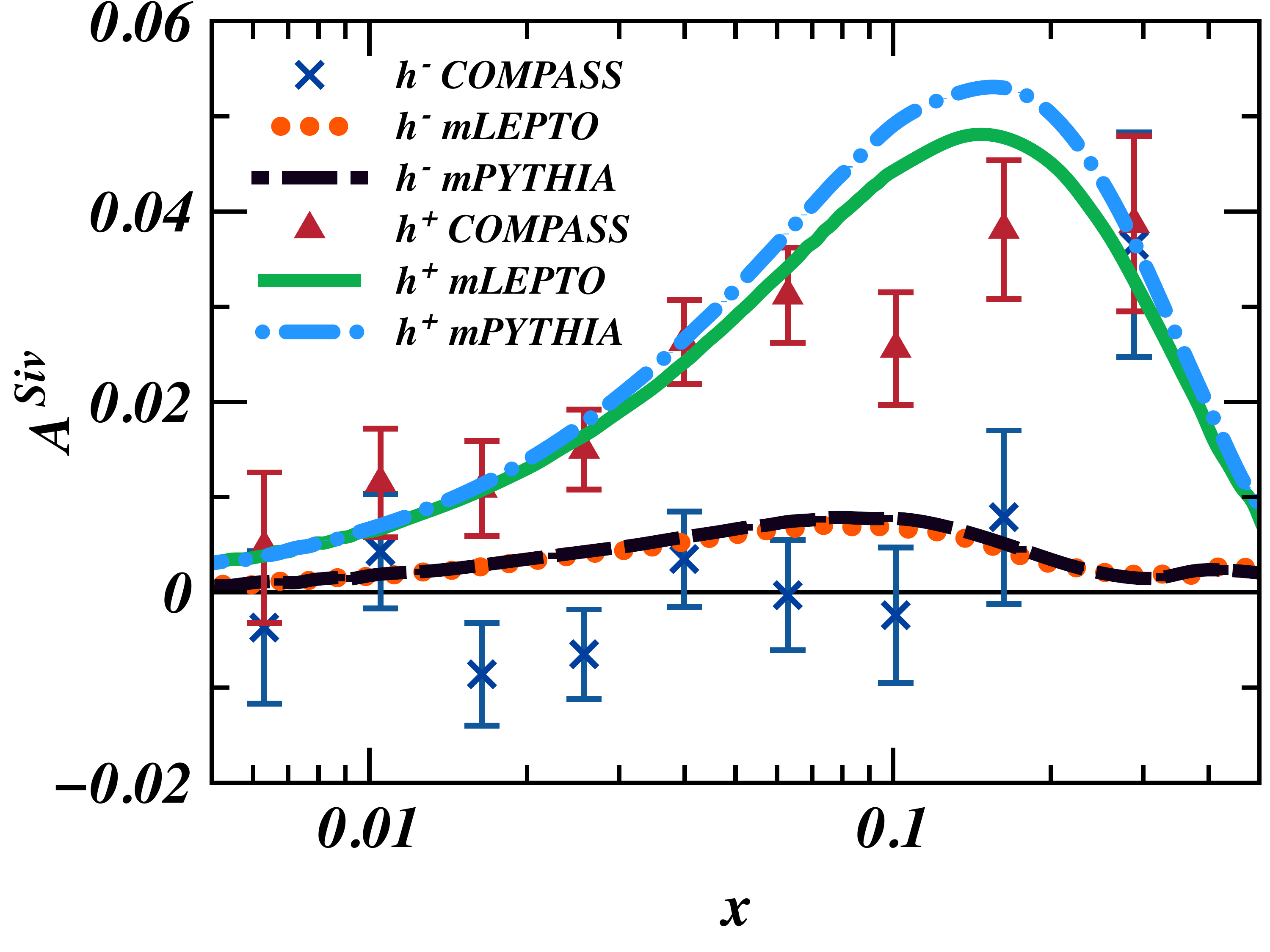}}
\\\vspace{-0.2cm}
\subfigure[] {
\includegraphics[width=\ImM]{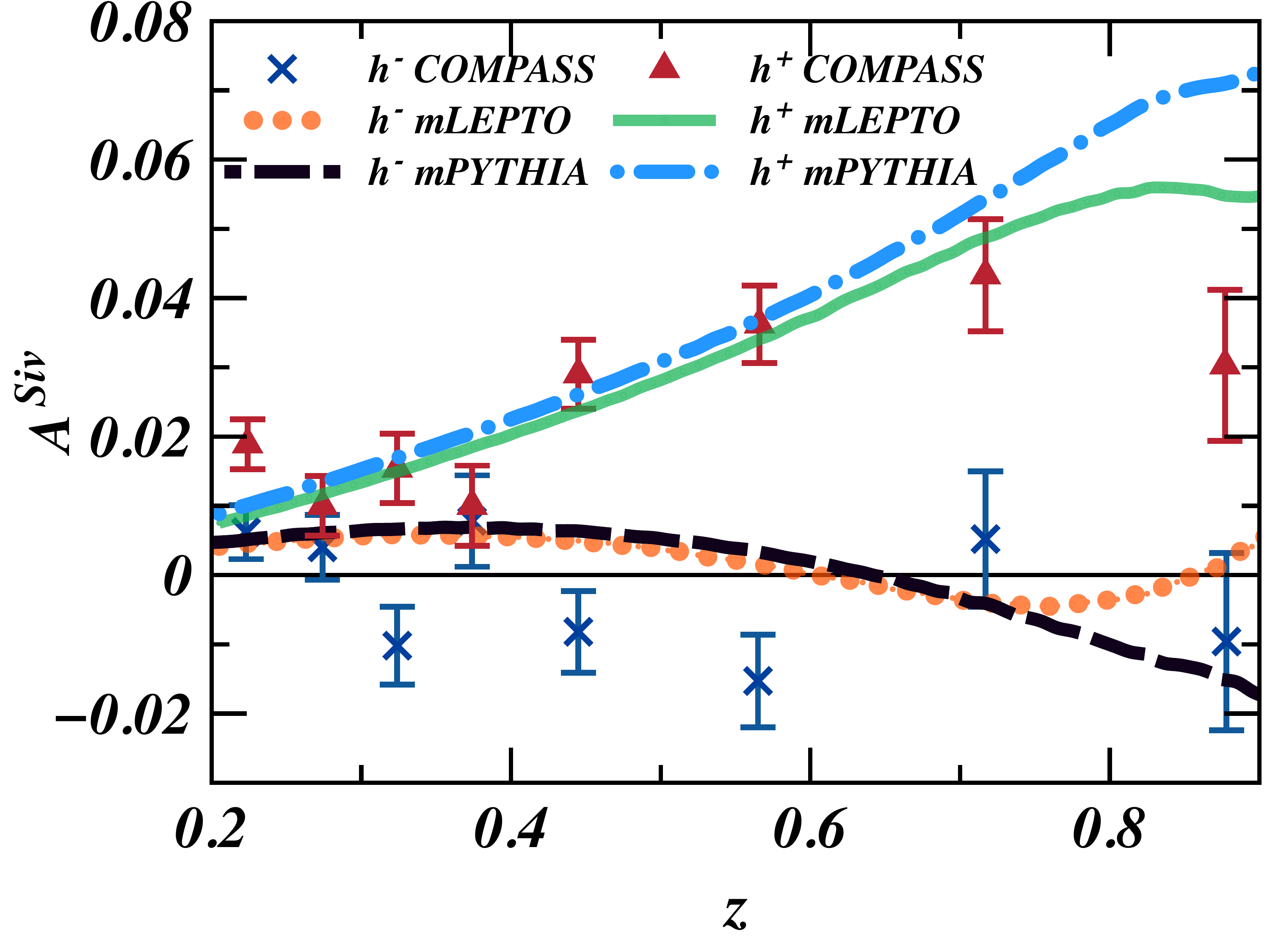}
}
\\\vspace{-0.2cm}
\subfigure[] {
\includegraphics[width=\ImM]{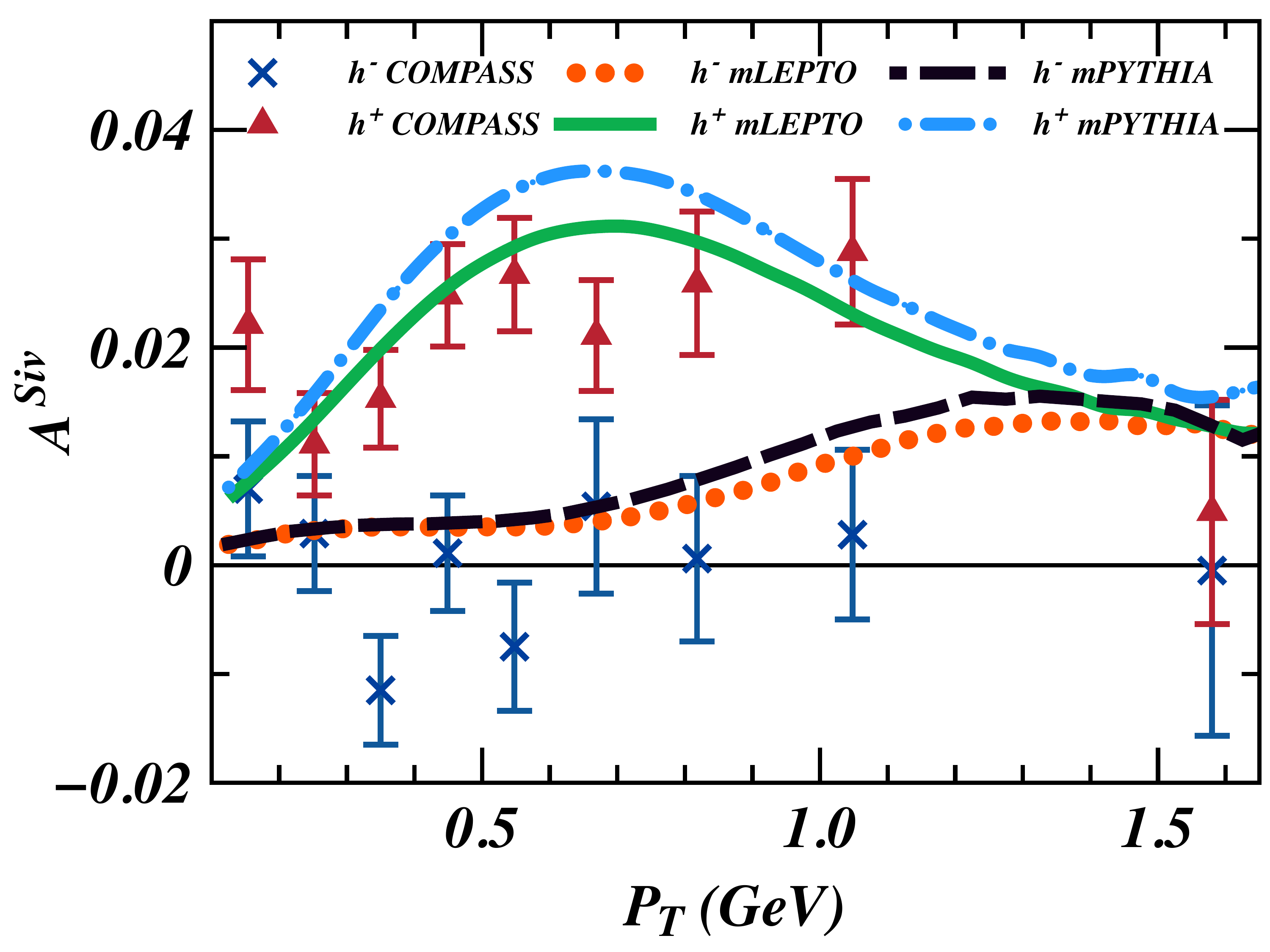}
}
\\\vspace{-0.2cm}
\caption{Sivers SSAs vs (a) $x$,  (b) $y$ and (c) $P_T$ from COMPASS data, and calculated with \MLEPT and \MPYTH.}
\label{PLOT_COMPASS_MPYTHIA}
\end{figure}

 Here we use \MPYTH to study the Sivers SSAs for single hadron production in SIDIS at the soon to be operational CLAS12 and planned EIC experiments. In this initial study, we do not include the TMD evolution of the Sivers PDF, as it would require major rewrites of these MC codes to also include the TMD evolution in the unpolarized PDFs and the hadronization model. Moreover, here we only use some basic kinematics limits coming from the angular coverage of the detectors in each experiment, without taking into account the realistic acceptance effects in these experiments. Nonetheless, one should keep in mind that our results are aimed at  establishing order-of-magnitude estimates for the observables.

 First we performed a cross-check between \MPYTH and \MLEPT for the COMPASS kinematics, where a transversely polarized proton asymmetry has been measured using $E_\mu=160~\Ge$ muon beam~\cite{Adolph:2012sp}. In MC simulations we allowed only the leading order DIS process and excluded any parton showering effects. The results of this simulation, along with the COMPASS data~\cite{Adolph:2012sp} and \MLEPT results are depicted in Fig.~\ref{PLOT_COMPASS_MPYTHIA}. We see that there is overall good agreement between \MLEPT and \MPYTH, with both describing the experimental data reasonably well. The small differences arising in $z$-dependence between the two MC generators are due to the differences in the hadronization modeling and parameters.

\FloatBarrier
\subsection{The experimental kinematics}
\label{SUBSEC_EXPERIMENTS}

 We use the kinematic limits and cuts for CLAS12 described in Ref.~\cite{CLAS12:2012DH}, where an $11~\Ge$  electron beam scatters from transversely polarized proton target. The relevant ranges for the kinematical variables are  $1~\Gs < Q^2 < 6.3~\Gs$ and $0.075\leq x\leq 0.532$. The polar angle (in the lab frame) of the scattered electron  $7^\circ \le \theta_{e'} \le 35^\circ$ and the polar angles of the hadrons $7^\circ \le \theta_{h} \le 100^\circ$ are imposed by the detector coverage. We additionally apply a cut $W^2 > 4~\Gs$ to select mostly DIS events.  For single hadron SSAs, we impose $0.2< z <0.8$ . For dihadron pairs we impose $0.4< z <0.8$ for the total energy fraction of the pair.  Additionally, to exclude the exclusive baryon resonance production region, we require the missing mass for both one- and two-hadron production, $M_{(e+p) - (e' +h)}$ and $M_{(e+p) - (e' +h_1 +h_2)}$, to be larger than $1.5~\Ge$.
  
 The EIC~\cite{Accardi:2012qut} proposal has several possible implementations at different existing and planned facilities. Here we choose the kinematics of the experiment based on the RHIC facility~\cite{EIC:BNL}, where electron and polarized proton beams collide.  The predictions for EIC have been carried out for two sets of electron $l$ and proton $P_N$ beam momenta, thus we will denote the corresponding energies in the text below as $l\times P_N$ and skipping the units always taken to be in~$\Ge$ (for example  $5\times50$).
The following kinematic cuts are used:
\al
{
0.01<y<0.95,\ 
Q^2>1\Ge^2,\ 
W>5\Ge,
}
 as well as those described in Ref.~\cite{EIC:BNL} for SIDIS. Additionally, we identify the hadrons in the CFR as those with $x_F>0$ in the $\gamma^*-N$ CM system. 

\FloatBarrier
\subsection{EIC toy model studies}
\label{SUBSEC_EIC_TOY_MPYTHIA}

\begin{figure}[htb]
\centering 
\subfigure[] {
\includegraphics[width=\ImM]{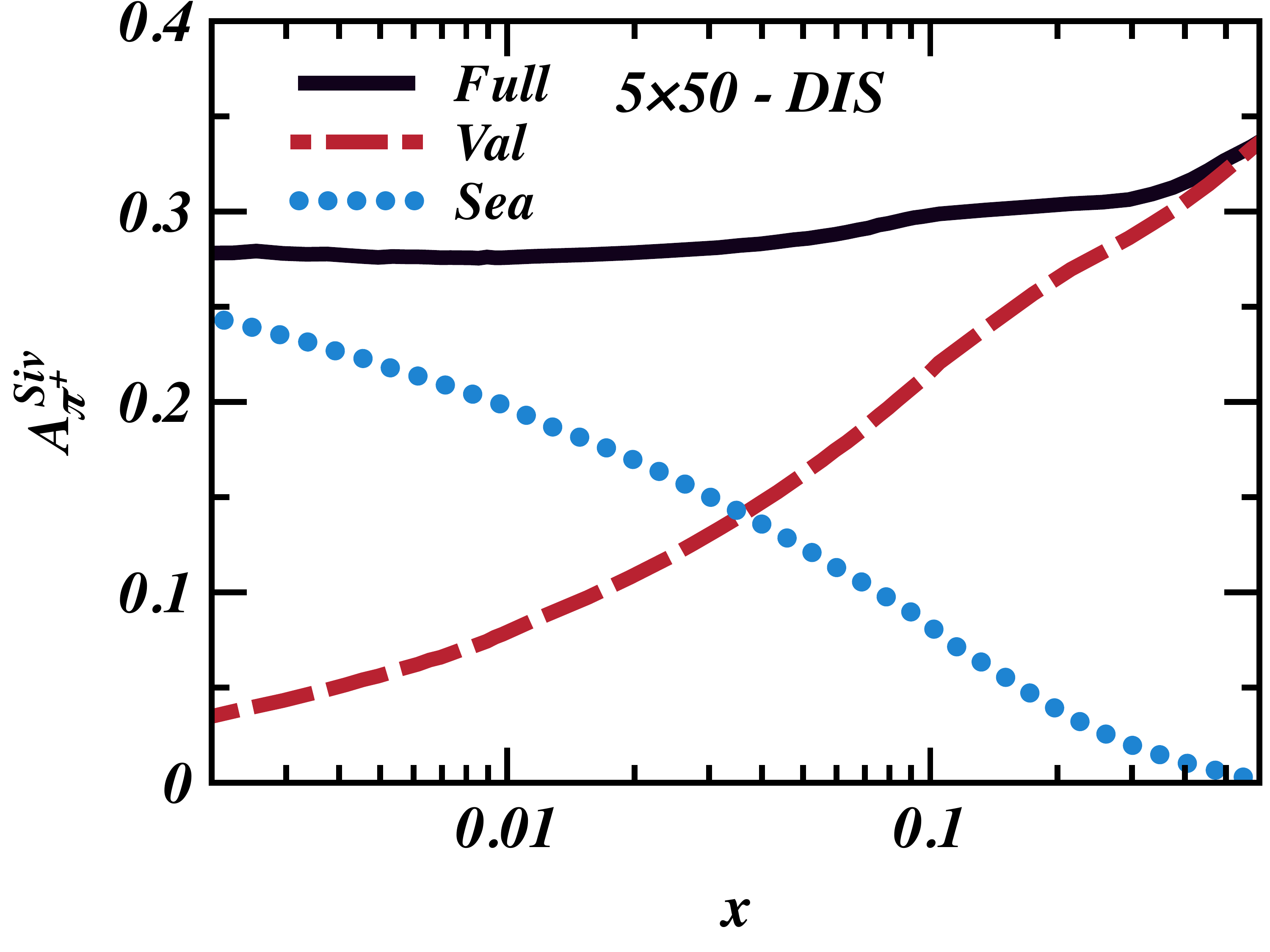}}
\\\vspace{-0.2cm}
\subfigure[] {
\includegraphics[width=\ImM]{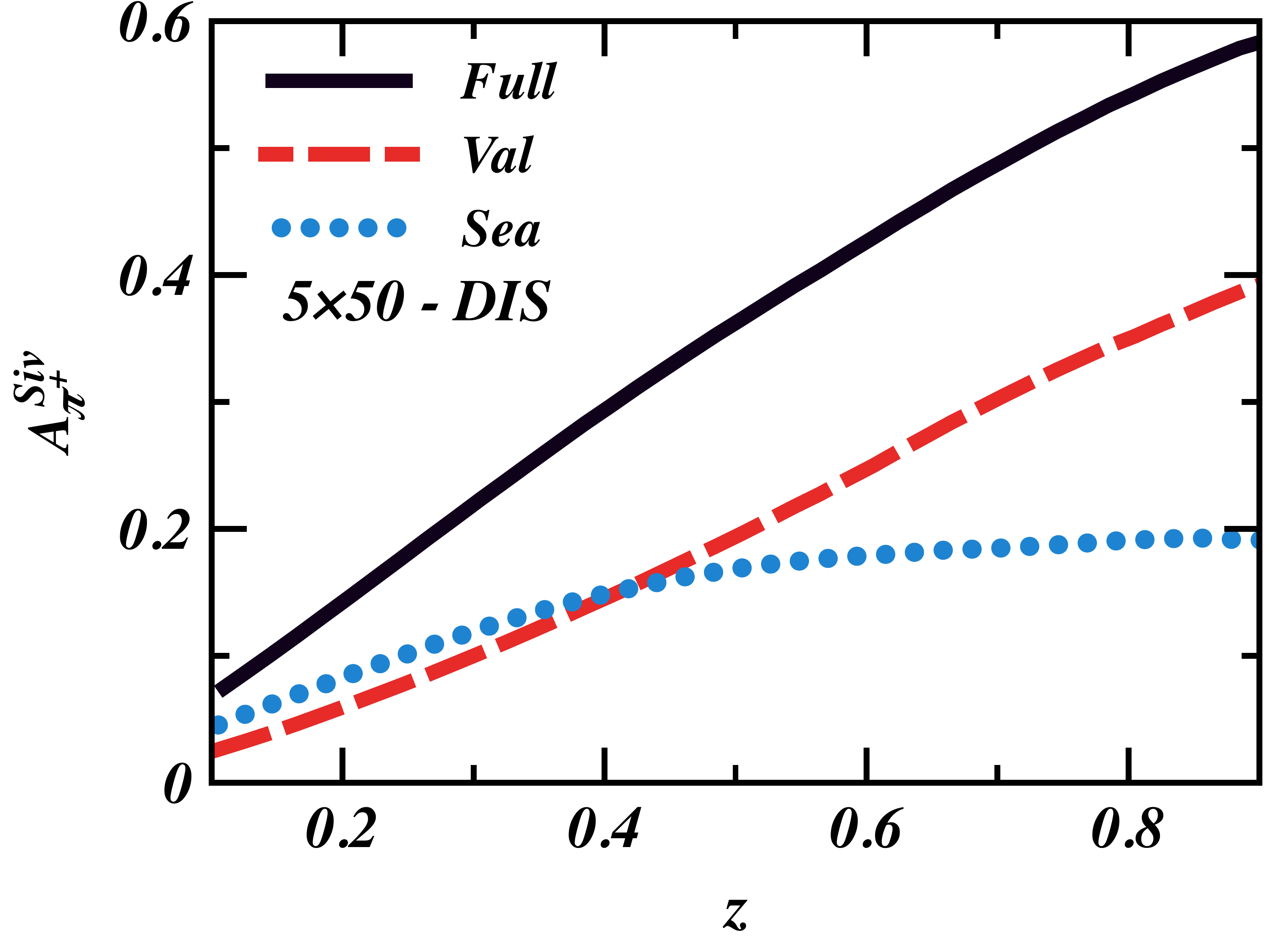}
}
\caption{EIC toy model SSAs vs (a) $x$ and (b) $z$ for $5\times50$ SIDIS kinematics for $\pi^+$. Comparison of sea and valence quark contributions in pure DIS events.}
\label{PLOT_EIC_TOY_SSA_X_Z}
\end{figure}

 We employ a toy model for the Sivers PDF to study the general features of the generated SSA. In particular, it is interesting to study the relative contributions of the sea and valence quarks to Sivers SSAs in various regions of $x$ and $z$. This will help to identify the most relevant experimental regimes for studying sea or valence Sivers PDF. For a toy model Sivers PDF, we choose it proportional to the unpolarized PDF of the corresponding quark flavor, with a large coefficient to maximize the generated  SSA while satisfying the positivity of the polarized PDF. The plots in Fig.~\ref{PLOT_EIC_TOY_SSA_X_Z} depict the results for $5\times50$ SSAs' dependence on (a) $x$  and (b) $z$. Here the label "Full" denotes the calculations where we set the Sivers terms using the toy model expression
$({S_T k_T}/{M}) f_{1T}^{\perp q}(x,k_T)=0.9\ f_1(x,k_T)$
for all the light flavored quarks, while for calculations with the label "Val" we only use the toy model Sivers terms for the valence $u$ and $d$ quarks and set them to zero for the sea quarks.  Finally, the label "Sea" denotes the scenario where we use the toy model Sivers terms for the sea quarks and set them to zero for the  valence quarks. It is apparent that the valence quarks produce the predominant part of the signal in the large-$x$ region, while the sea does in the small-$x$ region, as one would naturally expect. Also, for all the cases, the SSAs increase as a function of $z$, as  one naturally expects that the hadrons with a large fraction of the fragmenting quark's energy also carry a large fraction of its transverse momentum. Thus their azimuthal modulations become larger. Finally, the plots labeled "Val+Sea", indicating the simple sum of the results when we have sea-only and valence-only contributions. These coincide with the "Full" simulations, serving as a simple cross-check.

\begin{figure}[htb]
\centering 
\vspace{-0.2cm}
\subfigure[] {
\includegraphics[width=\ImMM]{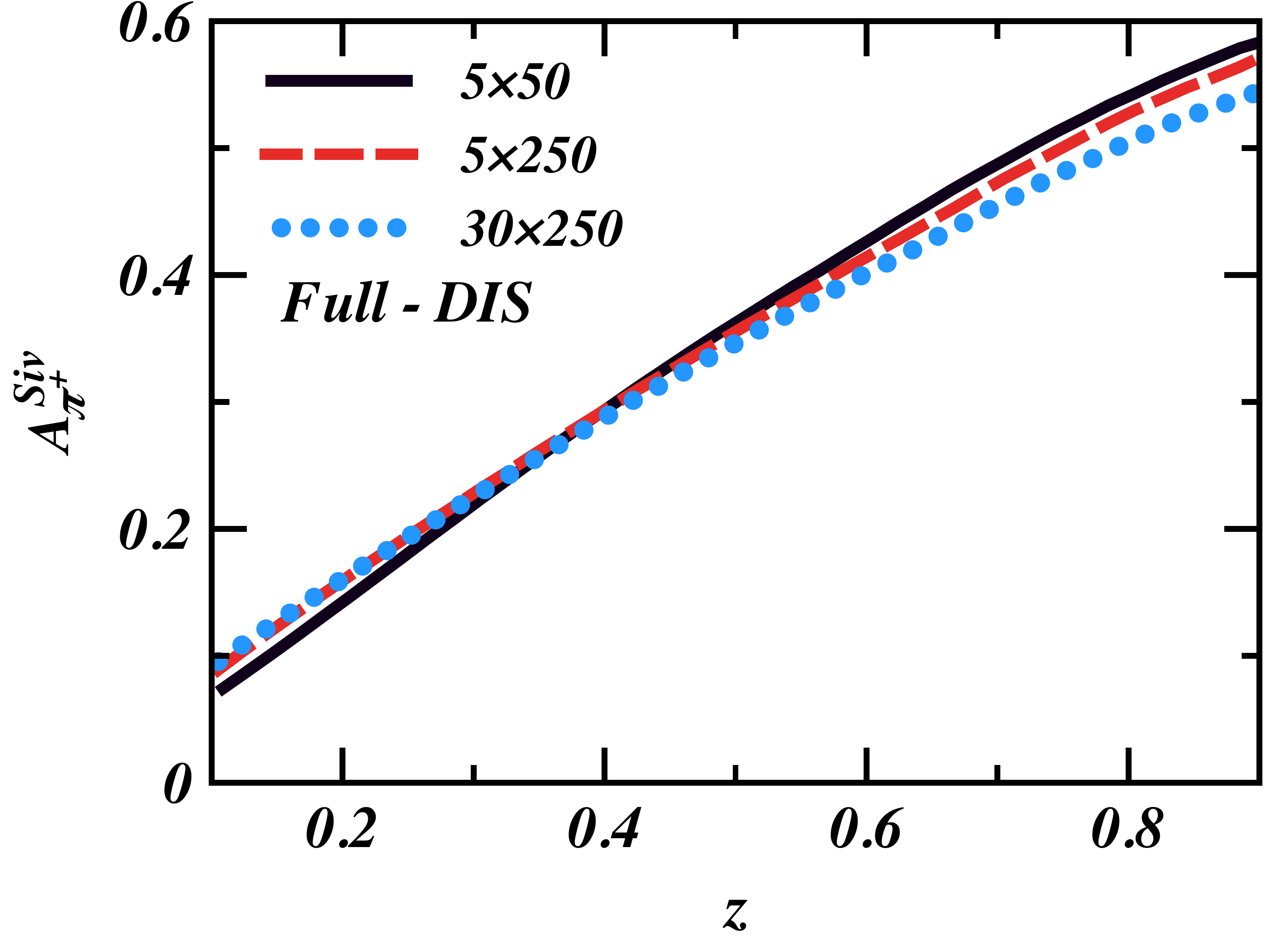}
}
\hspace{-0.2cm}
\subfigure[] {
\includegraphics[width=\ImMM]{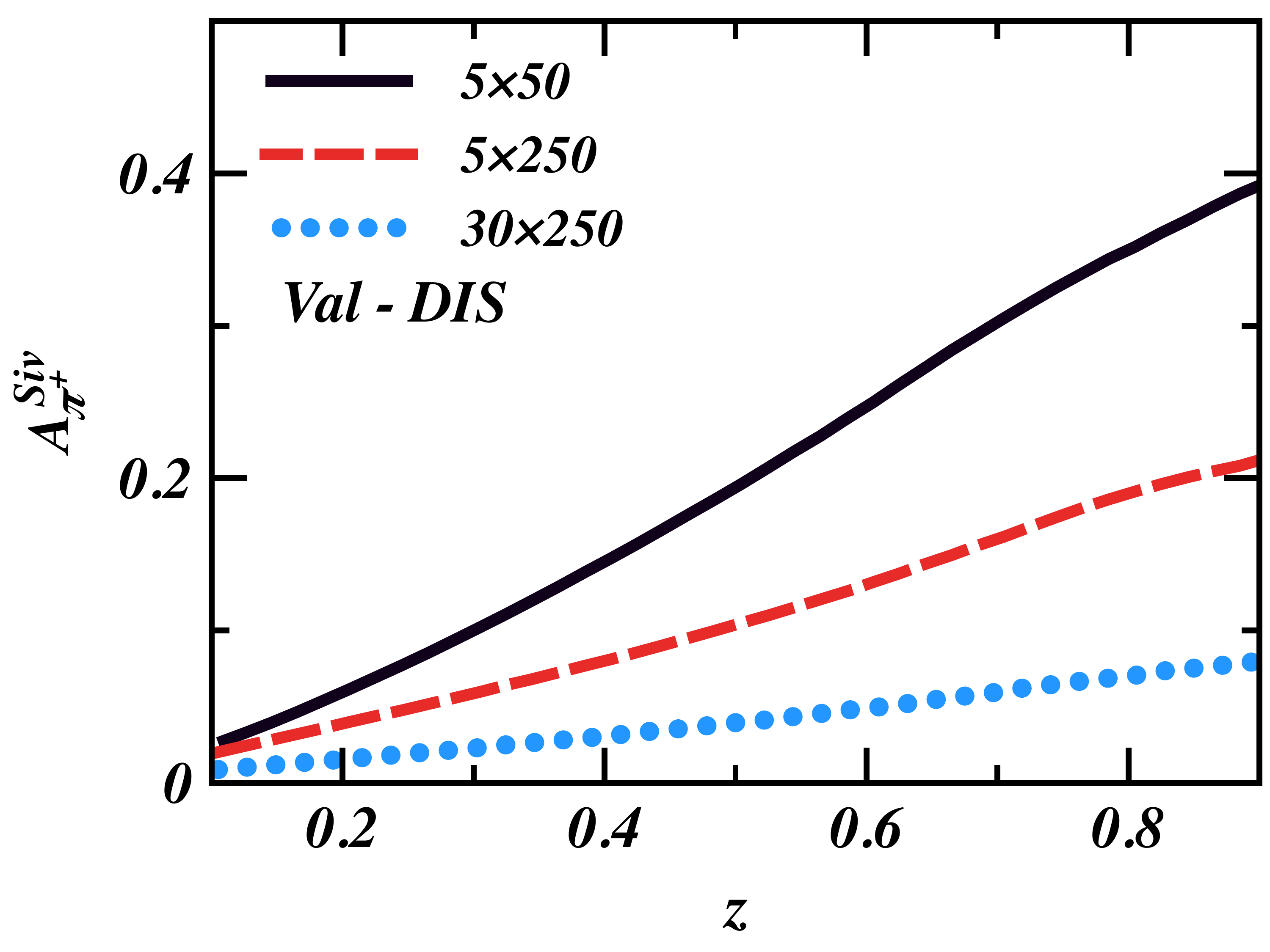}
}
\hspace{-0.2cm}
\subfigure[] {
\includegraphics[width=\ImMM]{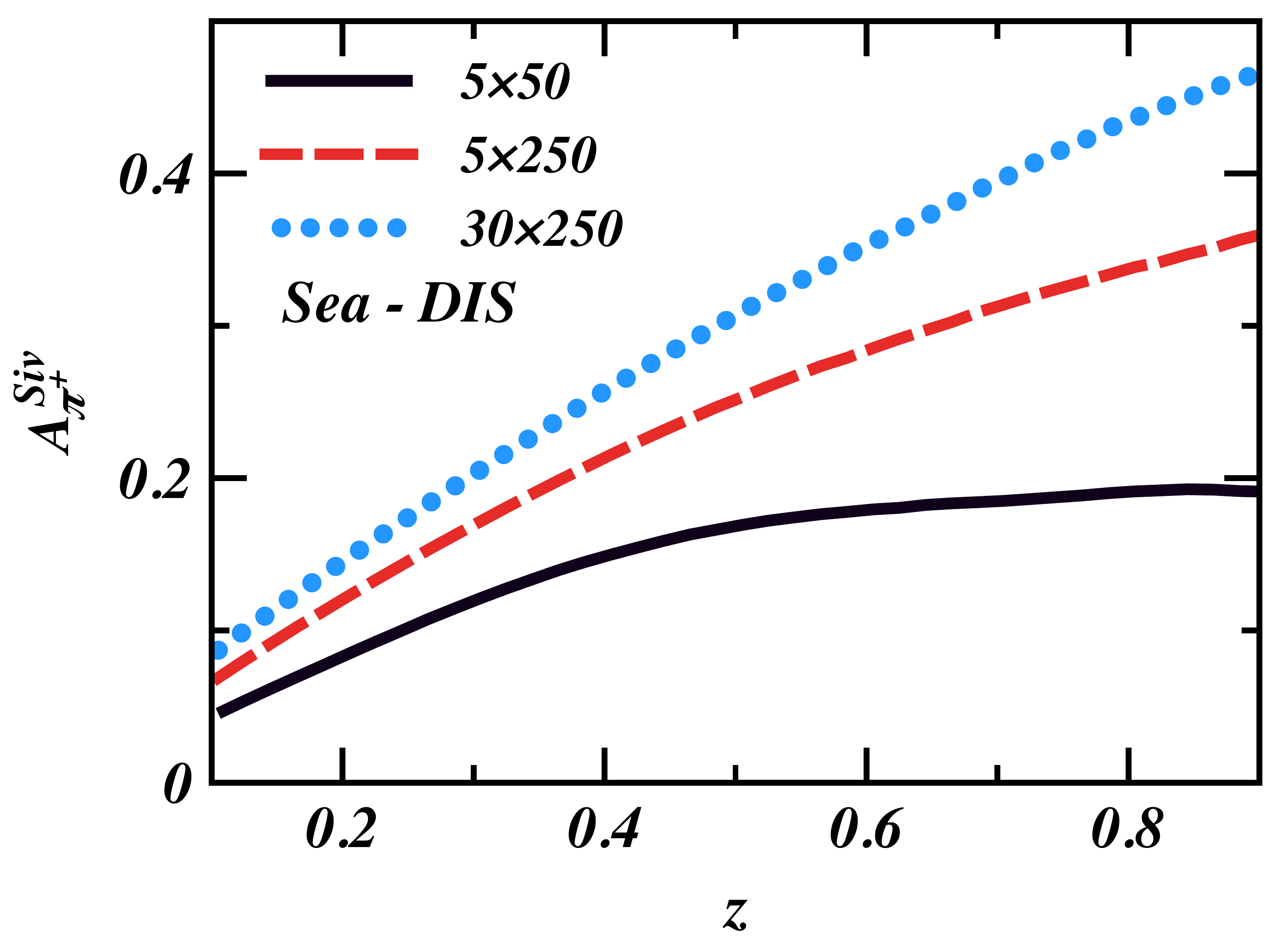}
}
\\\vspace{-0.2cm}
\caption{EIC toy model SSAs for $\pi^+$ for a range of beam energies. Comparison of (a) the full, (b) valence-only and (c) sea-only Sivers effect induced SSAs are shown for various energies vs $z$, where we only allow for the DIS process.}
\label{PLOT_EIC_TOY_SSA_EDEP}
\end{figure}

 Next, we study the dependence of the valence and sea quark contributions to SSAs for several different energies, as depicted in Fig.~\ref{PLOT_EIC_TOY_SSA_EDEP}. The results where we set all the light quarks' Sivers PDF to be the model value show little energy dependence, as only the increase in charm sea quark PDFs at higher CM energies slightly reduces the corresponding SSAs (here we do not allow a Sivers effect for the charm quarks). The valence and sea quarks, as expected, trade strength as the CM energy increases, and naturally the two results $5\times 250$ and $30\times 50$ with very similar CM energies also produce very close results for the corresponding SSAs. Thus one can conclude, that the high energy collision runs at EIC would be  much more sensitive to sea quark (and by the same to the logic gluon) Sivers PDF, while the relatively low energy ones will most effectively probe the valence quark region.

\begin{figure}[htb]
\centering 
\includegraphics[width=\ImMM]{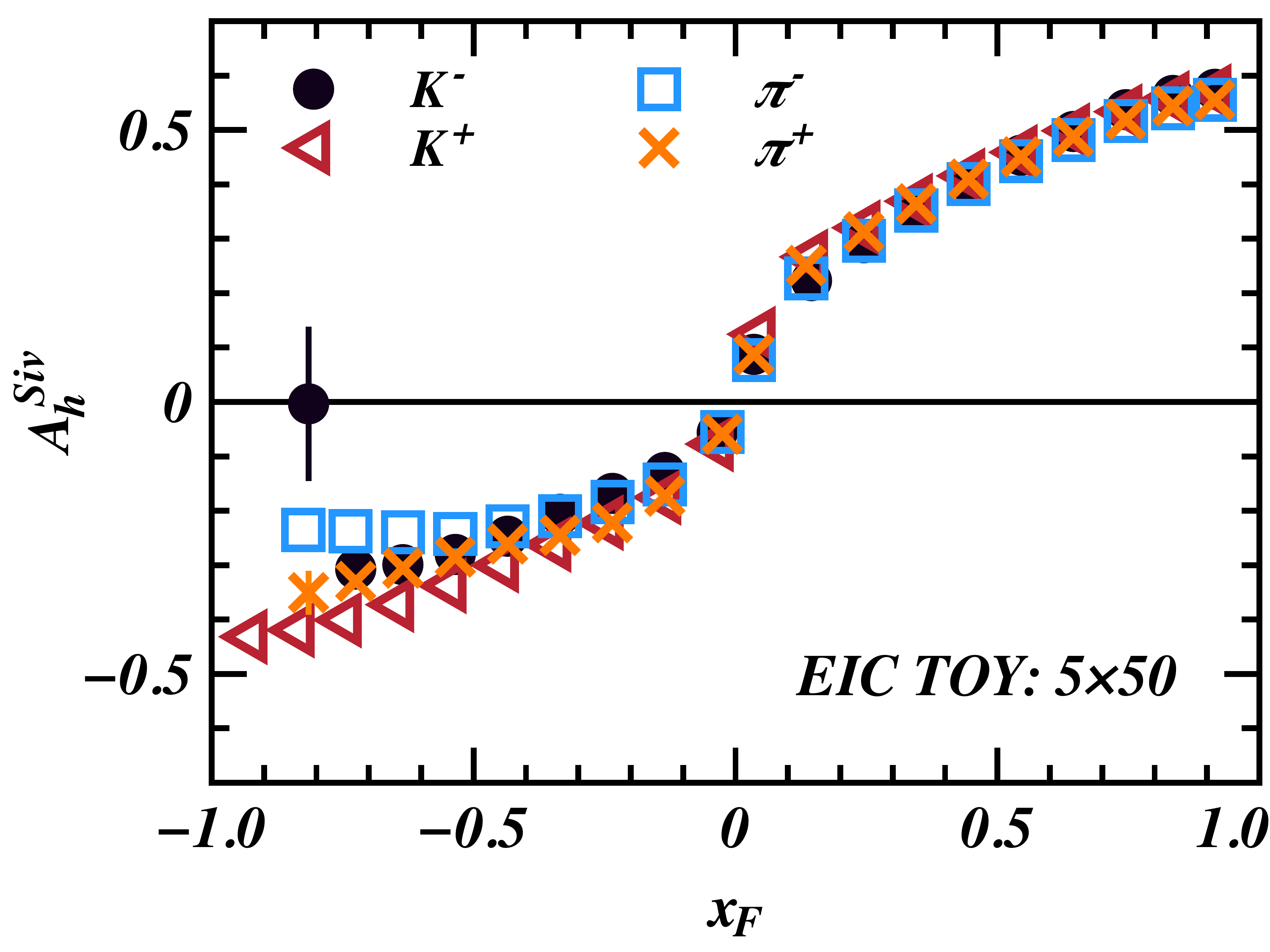}
\\\vspace{-0.2cm}
\caption{EIC toy model SSAs for $5\times50$ SIDIS kinematics for charged pions and kaons vs $x_F$. The Sivers asymmetry is present both in the current and target fragmentation regions.}
\label{PLOT_EIC_TOY_SSA_XF}
\end{figure}

 Studying the Sivers asymmetry in SIDIS in the TFR, $x_F<0$,  provides yet another avenue for accessing the spin-orbit correlations in the nucleon via so-called fracture functions~\cite{Anselmino:2011ss,Kotzinian:2011av}. Note, that in the leading order QCD factorized approach, based on  the fracture functions  formalism, the only single spin asymmetry allowed in the TFR is the Sivers-like one. Thus, by measuring both this and the Collins-like [$\sin(\vf_h+\vf_S)]$ asymmetries (the latter predicted to be vanishing) one can probe factorization for SIDIS in TFR.  Within the string model for hadronization adopted in the MC event generators, the inclusively measured hadron can be considered as originating from the target remnant side of the string. If the struck quark's transverse momentum is modulated by the Sivers effect, then so should be the transverse momentum of the remnant with a relative phase $\pi$ via momentum conservation. These modulations will be then transferred to the transverse momenta of the hadrons produced by the remnant, manifesting in the corresponding SSAs. Moreover, measuring the average transverse momentum square of the hadrons in the negative region, $x_F\lesssim -0.2$, allows greater sensitivity to accessing the quark average transverse momentum than in region $x_F\gtrsim 0.2$ ~\cite{Aschenauer:2014cki}.  The plot in Fig.~\ref{PLOT_EIC_TOY_SSA_XF} depicts the Sivers SSA for charged pions and kaons as a function of $x_F$. We notice that the Sivers SSA is also nonvanishing in the target fragmentation region. Moreover, it changes sign when $x_F$ crosses zero, as expected from the arguments about the relative signs of the target remnant and struck quark's transverse momenta. Thus we have  demonstrated that the Sivers SSA measurements in TFR are as viable as in CFR, modulo any specific experimental setup limiting the hadron detection at negative $x_F$.

\subsection{The effects of non-DIS  processes and parton showering on $A^{Siv}$ }
\label{SUB_SEC_NONDIS}

 In this article we adopt the leading order approximations for calculating the Sivers SSAs predictions for CLAS12 and EIC, where we ignore any possible non-DIS processes or gluon radiation corrections either by struck or fragmenting quarks (the initial- and final-state parton showering). Moreover, the same approach has been adopted when fitting the experimentally measured SSAs to extract the Sivers PDF~\cite{Anselmino:2005ea,Anselmino:2008sga,Anselmino:2012aa}. Thus it is very important to at very least make a qualitative study of the accuracy of this leading order treatment.  A rigorous treatment of this problem would require that one include all the terms in the cross section for all the relevant processes, which would require the calculation of both unpolarized and polarized QCD and QED radiative corrections to DIS process to several orders in the corresponding couplings and take into account possible interference. Such an approach will additionally require extensive modifications of the existing MC generators to incorporate all of the new physics. Here we choose to use the less strict but more practical approach of using the \PYTH generator with its built-in capabilities for treatment of the above effects to test the applicability of the leading order approximation.  In \PYTH, the polarized parton showering is not included, while the unpolarized showering is assumed to occur either before or after the hard photon-parton scattering, with no interference effects. Nevertheless, we think that a first step needs to be taken and the current machinery would suffice for an initial estimate.
 
 We start with the COMPASS kinematics and study four different cases. First we allow only the DIS single photon exchange process. Then, we also allow either parton showering (labeled "+Show") or non-DIS processes (labeled "+NonDIS") such as VMD and resolved photon contributions. Finally, we allow both parton showering and non-DIS processes alongside the DIS process. The details of the particular methods used in \PYTH to model all these processes can be found in Ref.~\cite{Sjostrand:2006za}.  
\begin{figure}[htb]
\centering 
\subfigure[] {
\includegraphics[width=\ImM]{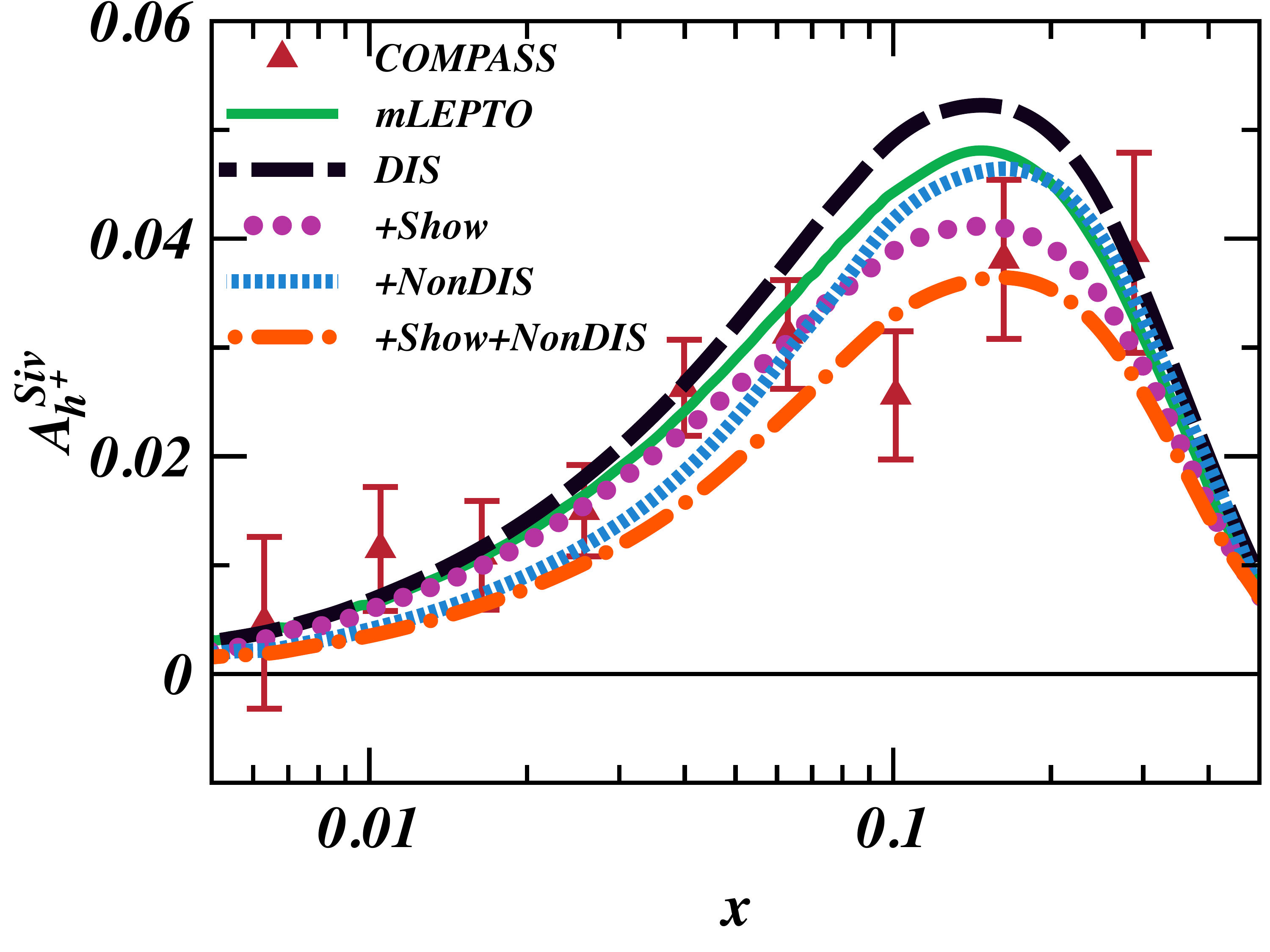}
}
\hspace{-0.2cm}
\subfigure[] {
\includegraphics[width=\ImM]{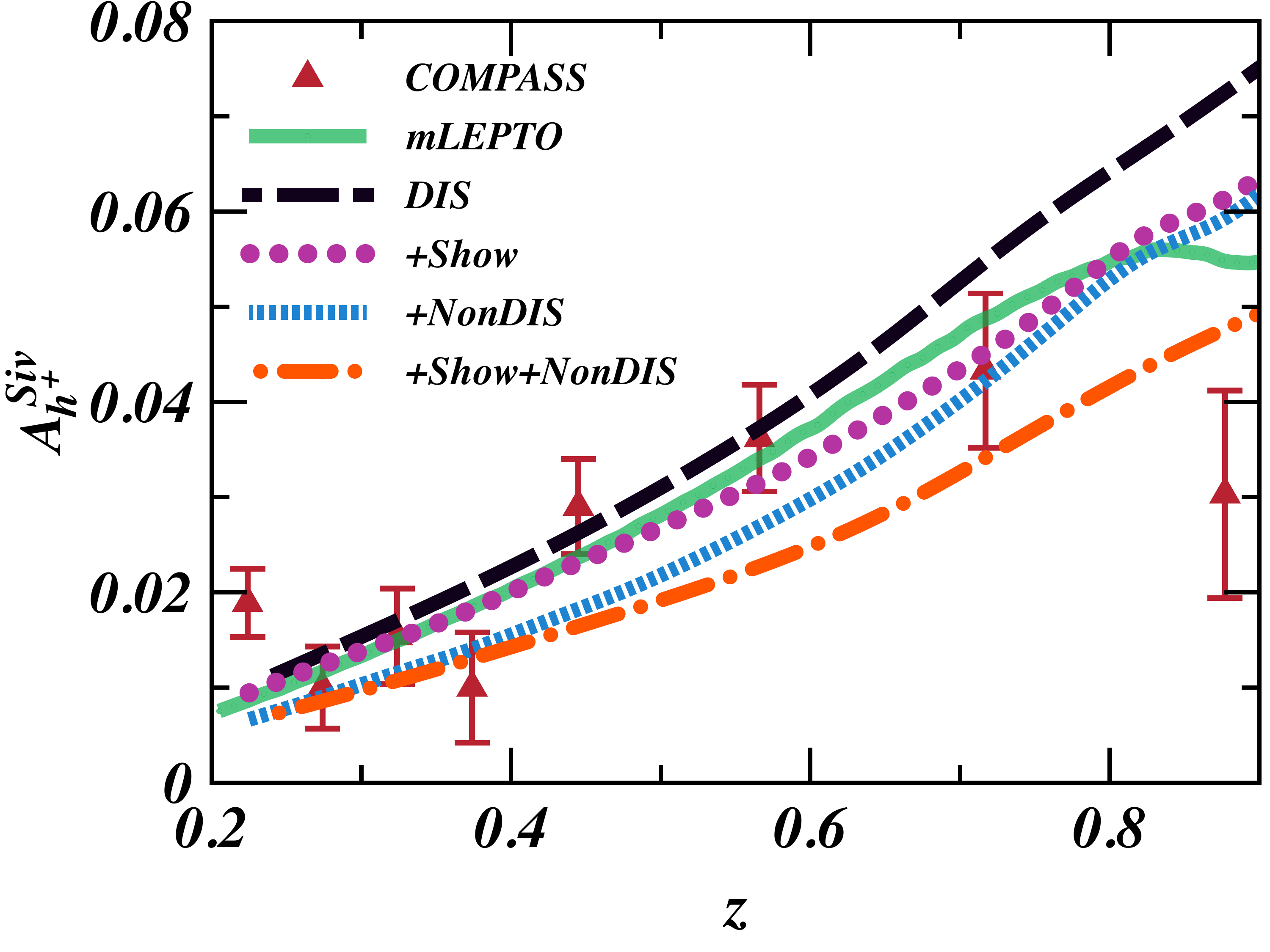}
}
\hspace{-0.2cm}
\subfigure[] {
\includegraphics[width=\ImM]{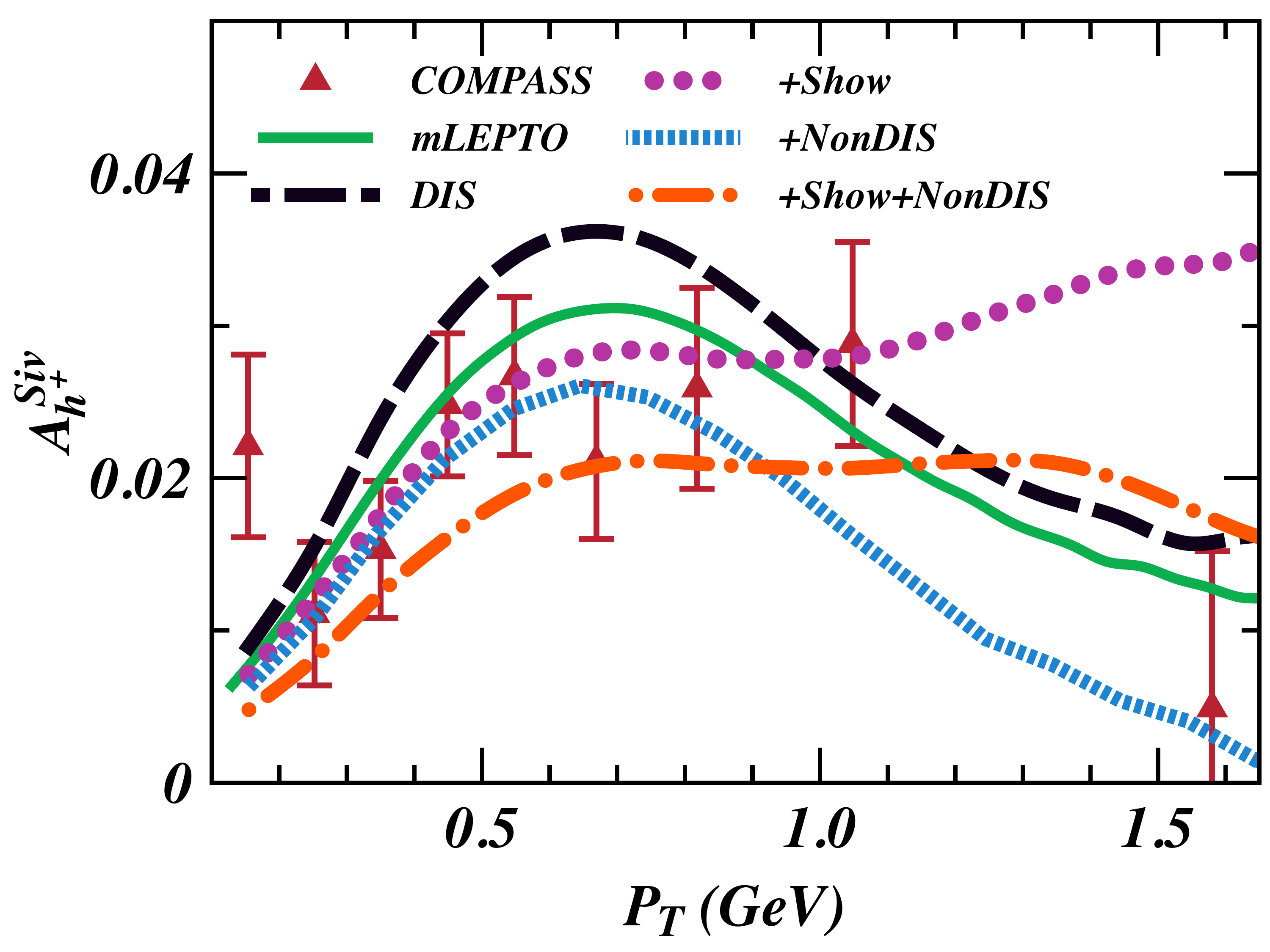}
}
\\\vspace{-0.2cm}
\caption{COMPASS SSAs for $h^+$ vs (a) $x$, (b) $z$ and (c) $P_T$  when various contributions beyond the DIS process are included.}
\label{PLOT_COMPASS_X_Z_PT_NON_DIS}
\end{figure}

The plots in Fig.~\ref{PLOT_COMPASS_X_Z_PT_NON_DIS}  depict the $x$, $z$ and $P_T$ dependence of the Sivers asymmetry as we include various contributions. Also depicted are the COMPASS measurements and the results obtained from the \MLEPT generator. Here we see that the non-DIS and parton showering contributions significantly decrease both the $x$ and $z$ dependence of the SSAs, as the denominators of the SSAs, \Eq{EQ_SSA_1H_SIV}, with the "unpolarized" terms increase. It is also curious, that the parton showering effects shift the strength of the SSAs from the low to high $P_T$ region. Intuitively, this can be understood as the fragmenting quark being given an additional kick, uniform in $\vf$, in the transverse direction by emitting a gluon or a photon, thus increasing the average $P_T$ of the produced hadrons. Nevertheless, all of the results describe the current data reasonably well since it has significant uncertainties. On the other hand, there might be a potential impact from these findings on the phenomenological extractions of the Sivers PDF~\cite{Anselmino:2005ea,Anselmino:2008sga,Anselmino:2012aa} from experimentally measured SSAs by HERMES and COMPASS. One can readily see from the Eq.~(35) in~\cite{Anselmino:2012aa}, that the LO expression for the Sivers SSA is used in the fits would underestimate the relative contribution of the Sivers term in the cross section. Thus we deduce that the current extractions of the Sivers PDFs are most likely to significantly underestimate the true values because of the effects described. The reevaluation of these parametrizations might pose a significant challenge, similar to those described earlier in this subsection. On the other hand, here our goal is to predict the SSAs to be measured at the future experiments. Thus we need to first understand what is the relative contribution of these processes at the kinematics of interest.

\begin{figure}[tb]
\centering 
\subfigure[] {
\includegraphics[width=\ImM]{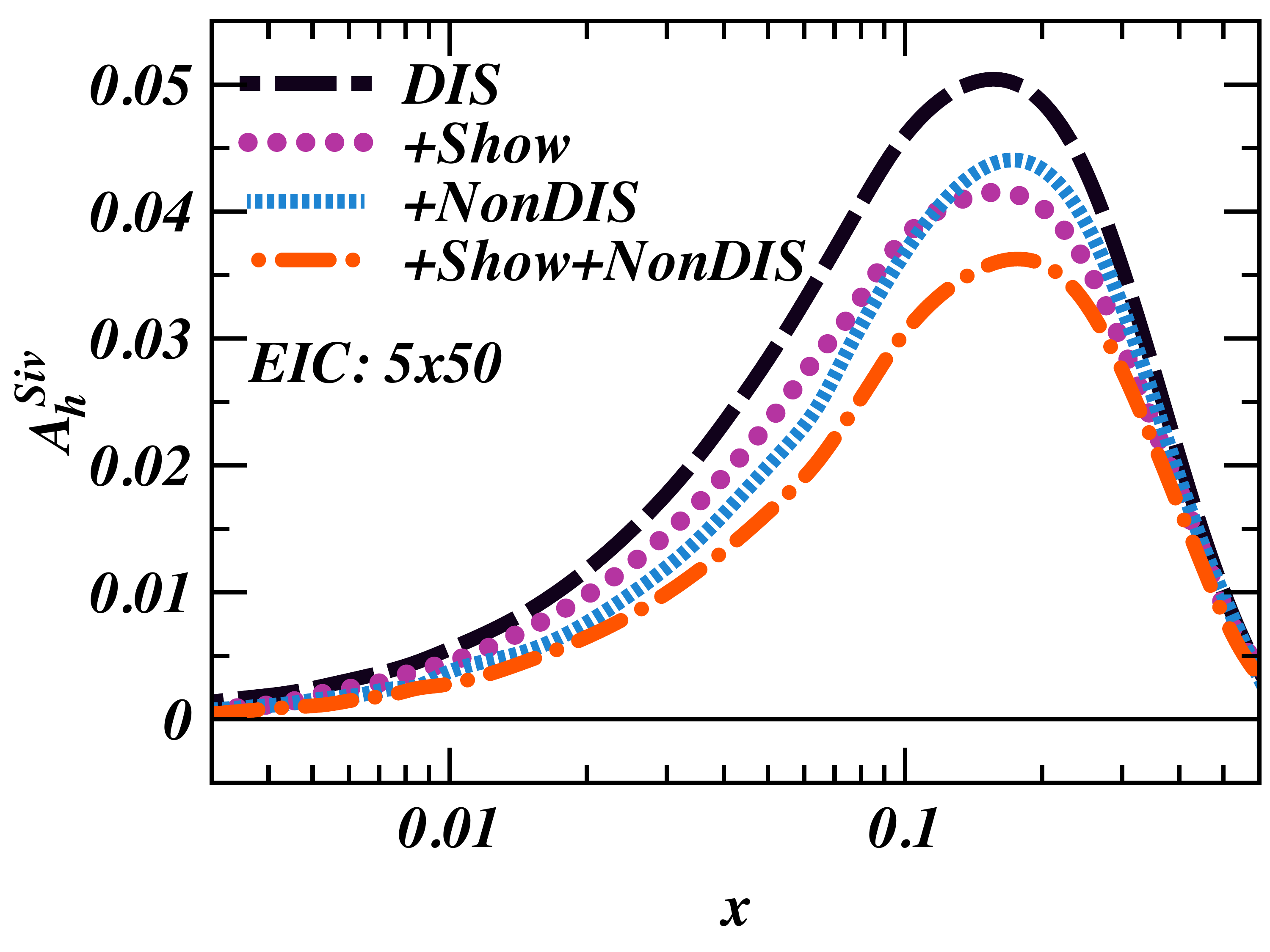}
}
\hspace{-0.2cm}
\subfigure[] {
\includegraphics[width=\ImM]{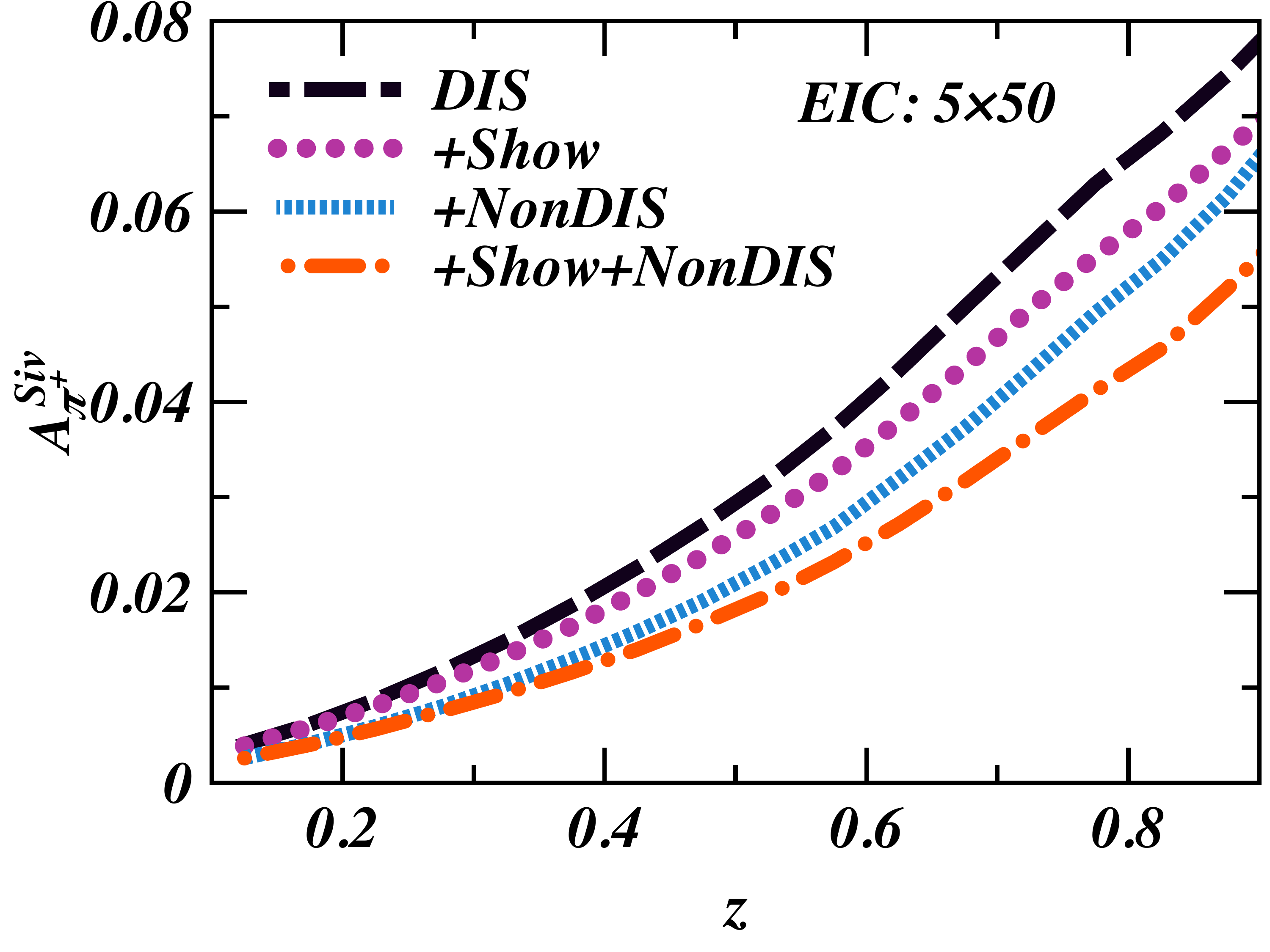}
}
\hspace{-0.2cm}
\subfigure[] {
\includegraphics[width=\ImM]{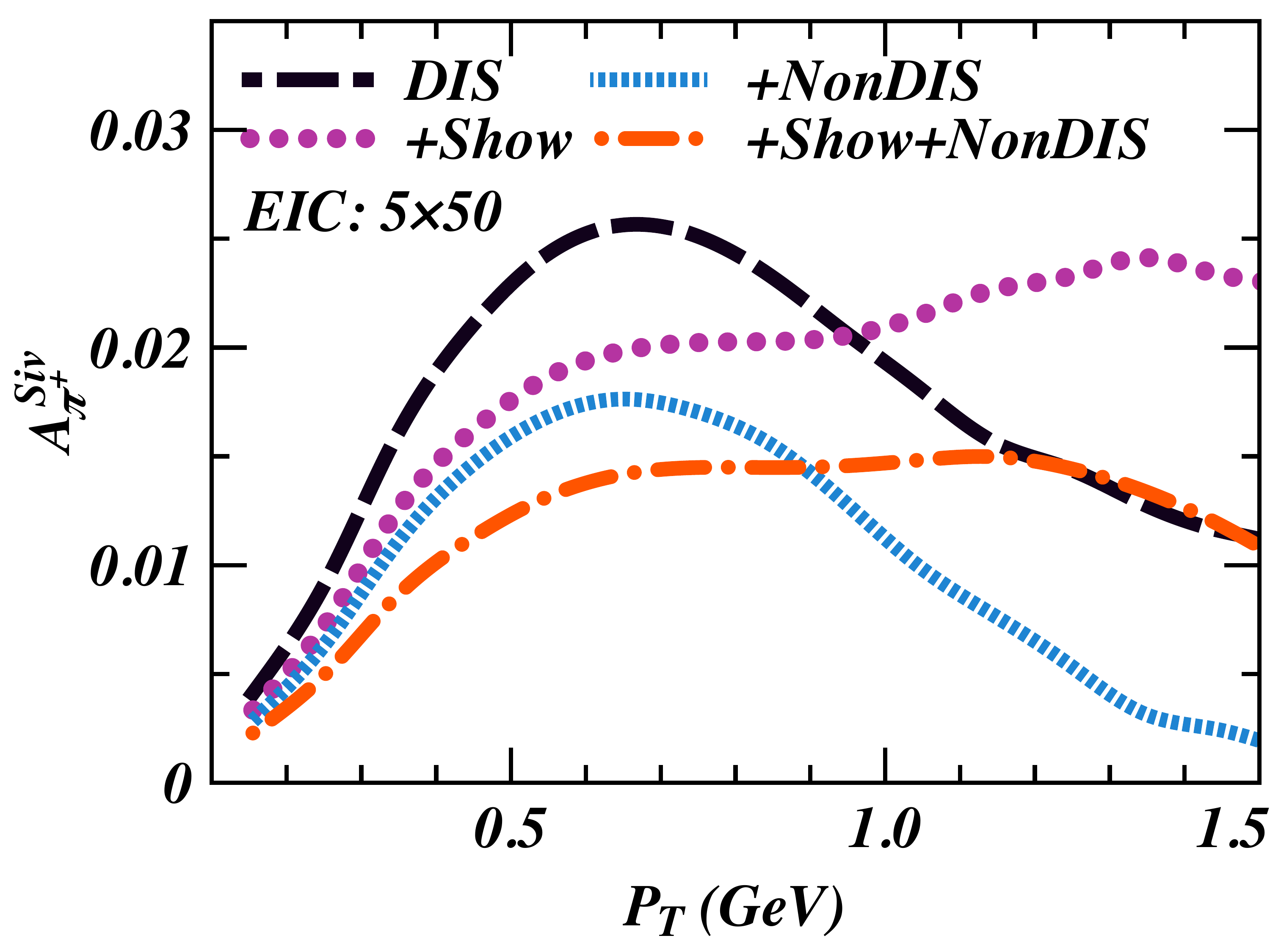}
}
\\\vspace{-0.2cm}
\caption{EIC($5\times50$) SSAs for $\pi^+$ vs (a)$ x$, (b) $z$ and (c) $P_T$ when various contributions beyond the DIS process are included.}
\label{PLOT_EIC_X_Z_PT_NON_DIS}
\end{figure}

 The plots in Fig.~\ref{PLOT_EIC_X_Z_PT_NON_DIS} are the analogues of those in Fig.~\ref{PLOT_COMPASS_X_Z_PT_NON_DIS}, calculated for EIC $5\times 50$ kinematics. Here we see trends in the behavior of the SSAs similar to those in COMPASS kinematics. This hints that the relative changes of the SSAs might be very similar at both COMPASS and EIC kinematics. To test this hypothesis we have calculated the ratios of both hadron count rates and SSAs calculated including all the non-DIS and parton showering processes to those calculated only including the DIS process.
 
\begin{figure}[tb]
\centering 
\subfigure[] {
\includegraphics[width=\ImM]{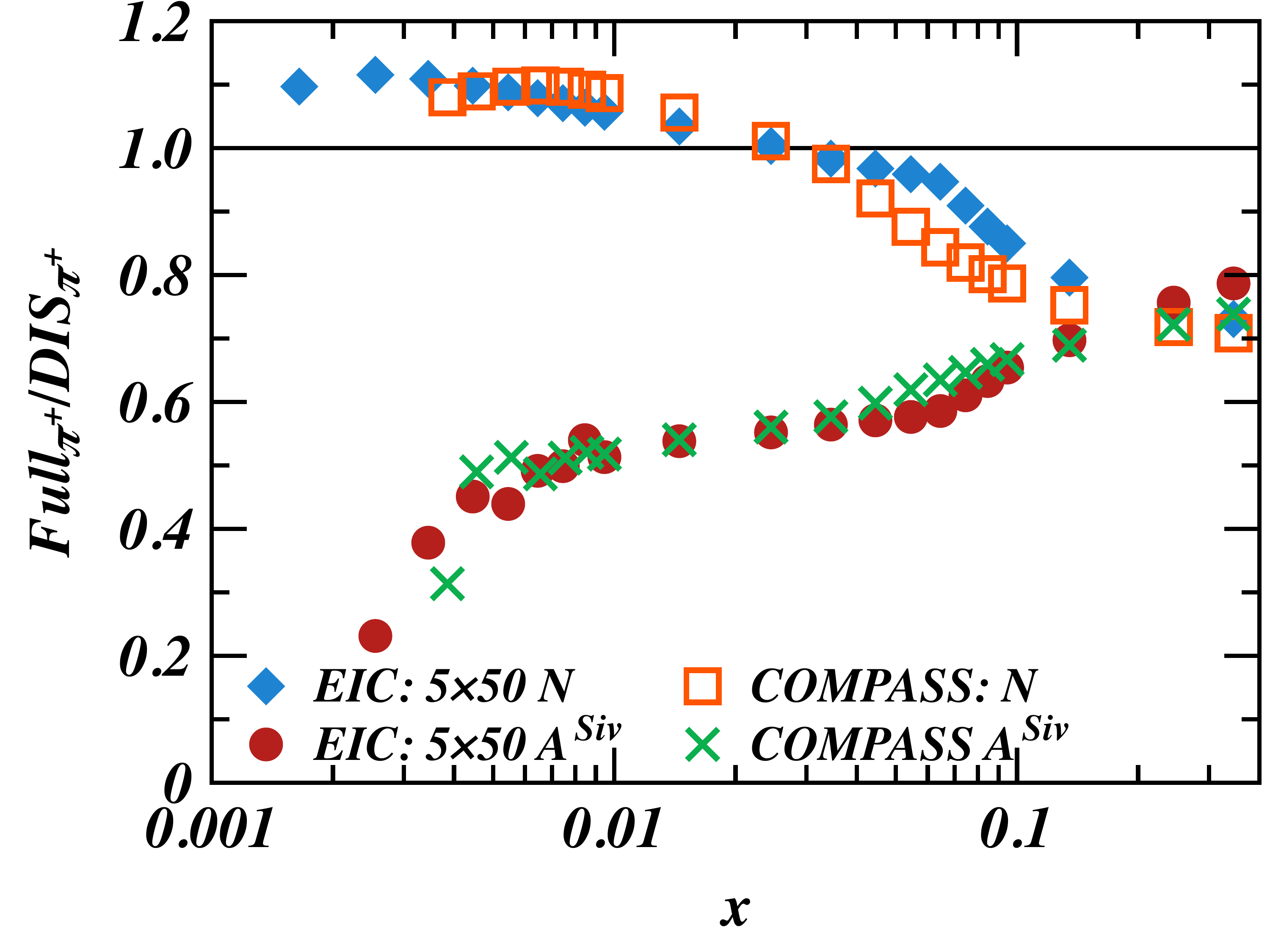}
}
\hspace{-0.2cm}
\subfigure[] {
\includegraphics[width=\ImM]{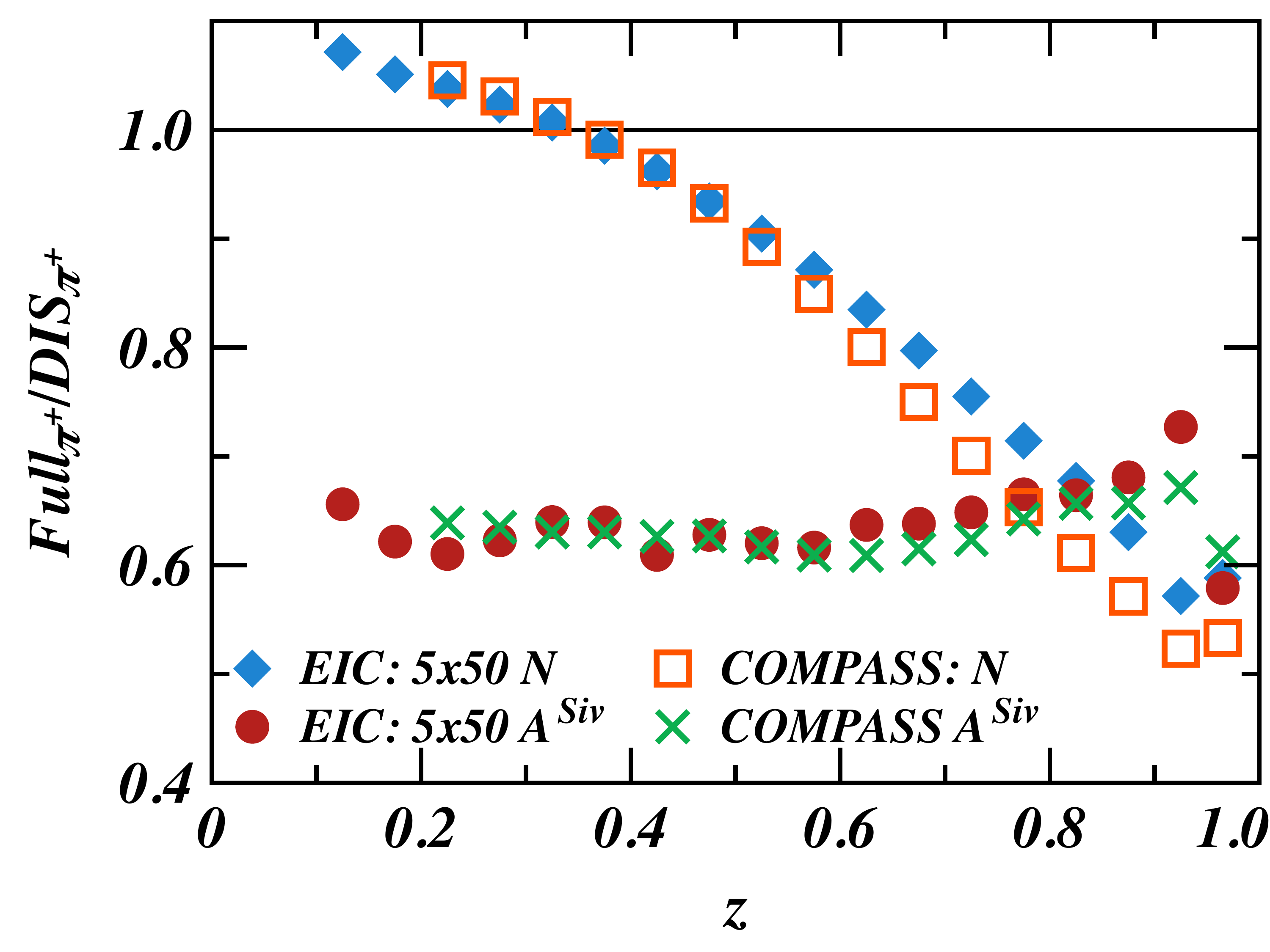}
}
\hspace{-0.2cm}
\subfigure[] {
\includegraphics[width=\ImM]{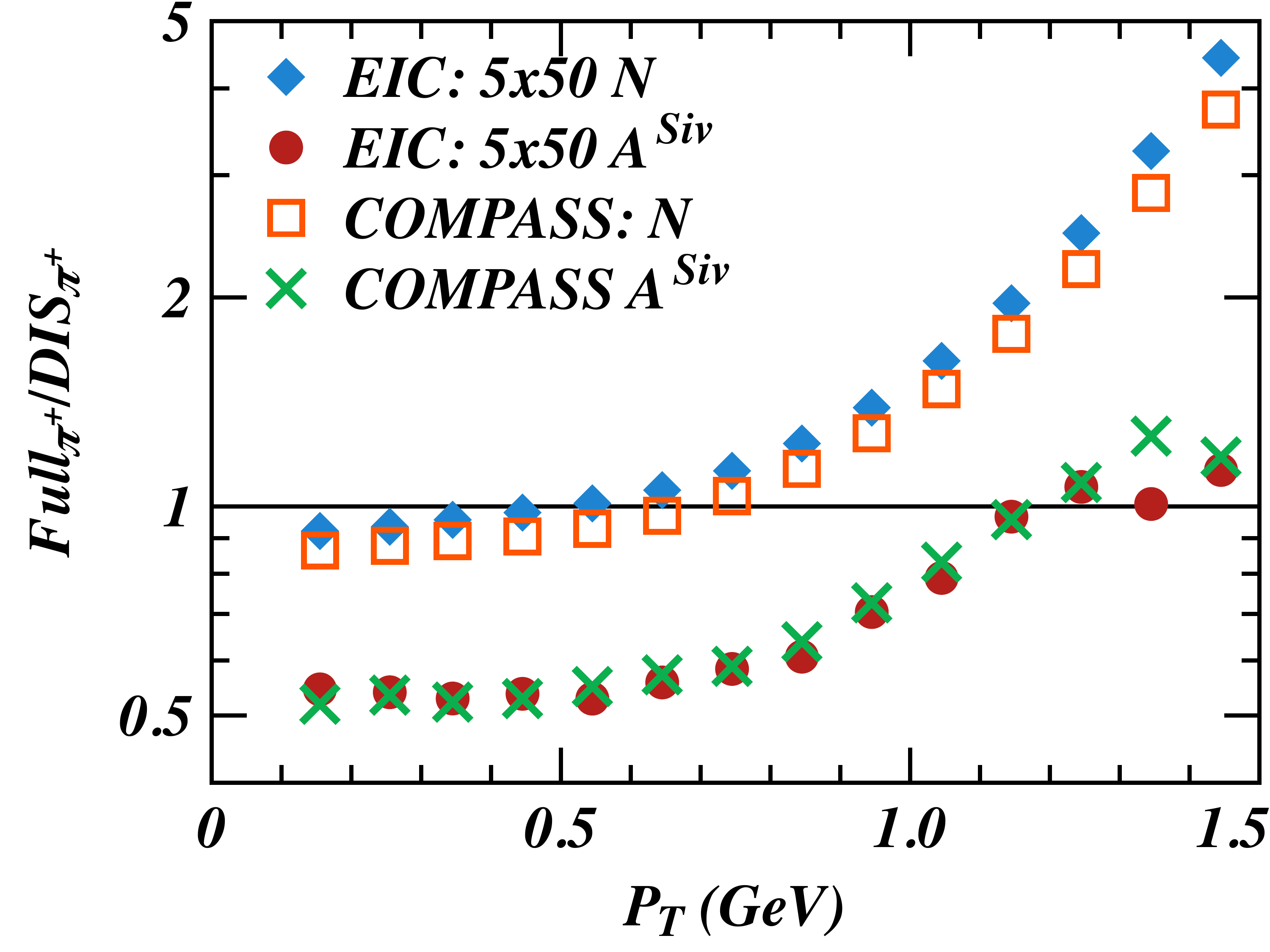}
}
\\\vspace{-0.2cm}
\caption{Ratios of full to DIS-only results for rates ($N$) and SSAs ($A^{Siv}$) for EIC($5\times50$) and COMPASS vs (a) $x$, (b) $z$ and (c) $P_T$ .}
\label{PLOT_RATIO_X_Z_PT_NON_DIS}
\end{figure}

The plots in Fig.~\ref{PLOT_RATIO_X_Z_PT_NON_DIS} depict the ratios of full and DIS-only results vs (a) $x$, (b) $z$ and (c) $P_T$ for EIC ( $\pi^+$ at $5\times50$) and COMPASS ($h^+$). We can see striking similarities for both COMPASS and EIC results, if we do not take into account the small-$x$ region for the SSAs, where in the ratios of two very small numbers the small statistical uncertainties create a large spread of results. These results are encouraging, as we can still use the Sivers PDF parametrization~\cite{Anselmino:2012aa} to obtain good estimates of the SSAs using only the DIS process. The reason is that, to obtain the COMPASS SSA induced by the DIS-only contributions to both unpolarised and Sivers terms, the measured SSA should be divided by the ratio of the full to DIS contributions for COMPASS depicted in Fig.~\ref{PLOT_RATIO_X_Z_PT_NON_DIS}.  On the other hand, to predict the size of the SSAs at EIC, the estimates that include only the Sivers PDF contributions should be multiplied by the ratio of the full to DIS contributions for EIC, also depicted in Fig.~\ref{PLOT_RATIO_X_Z_PT_NON_DIS}. We can readily see from Fig.~\ref{PLOT_RATIO_X_Z_PT_NON_DIS} that these two factors cancel each other to a very approximation. Moreover, our initial calculations~\cite{Kotzinian:2014gza} that only included the leading order DIS process were reproducing the measured COMPASS data well. A rigorous study of the effect of polarized and unpolarized parton showering, as well as TMD evolution of the Sivers PDF in MC simulations for calculating the observable SSAs is left for future work. 

\subsection{EIC projections}
\label{SUBSEC_EIC_DIHAD_MPYTHIA}

 Here we present the results for the EIC simulations using the kinematical  cuts given in the beginning of the section. In simulations we only allow pure DIS scattering. We present results for two sets of energies of the electron and proton beams, $5\times50$ and $5\times250$, that provide a reasonable variation in the CM energies.
 
\begin{figure}[tb]
\centering 
\subfigure[] {
\includegraphics[width=\ImM]{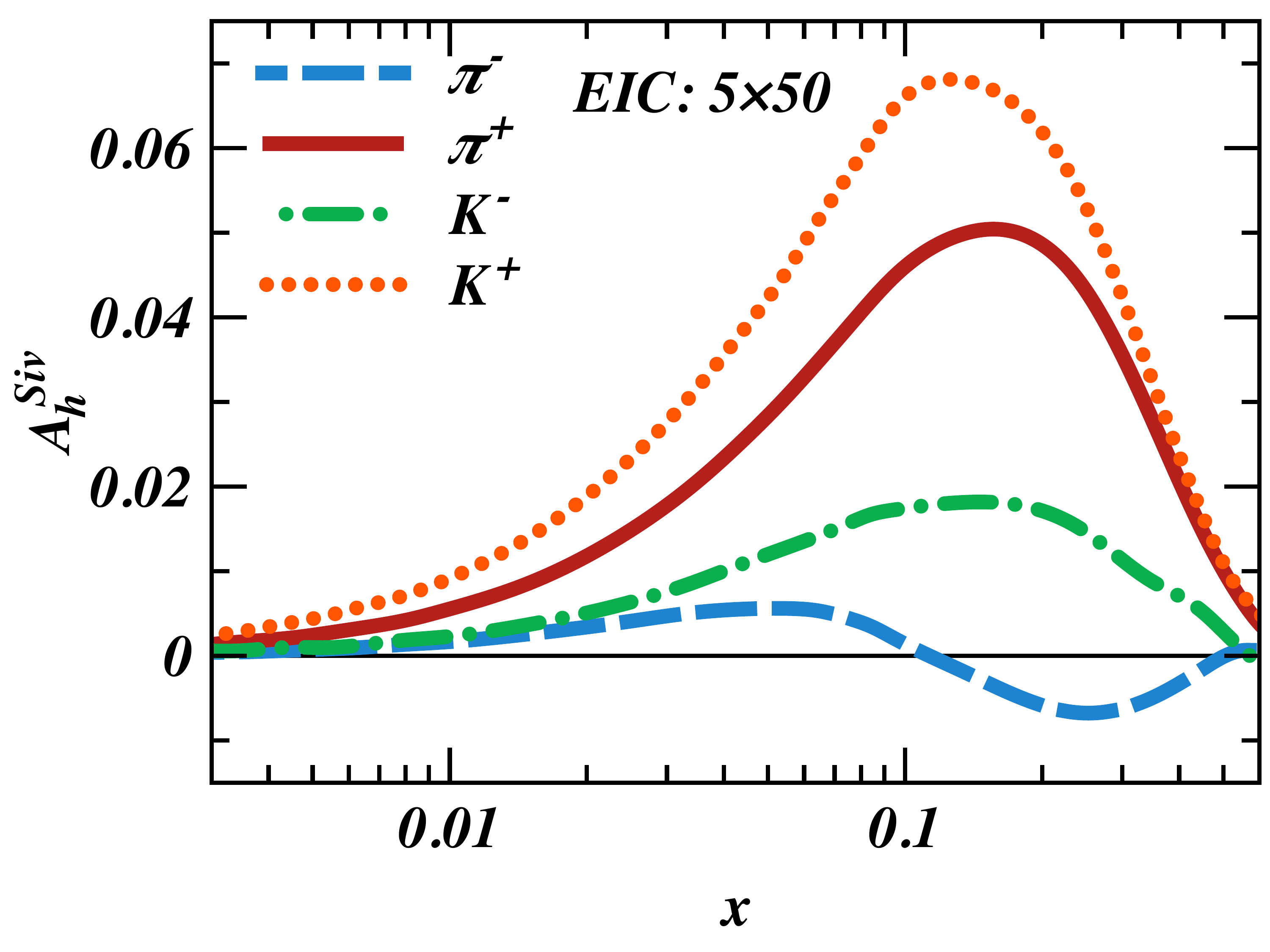}}
\hspace{-0.2cm}
\subfigure[] {
\includegraphics[width=\ImM]{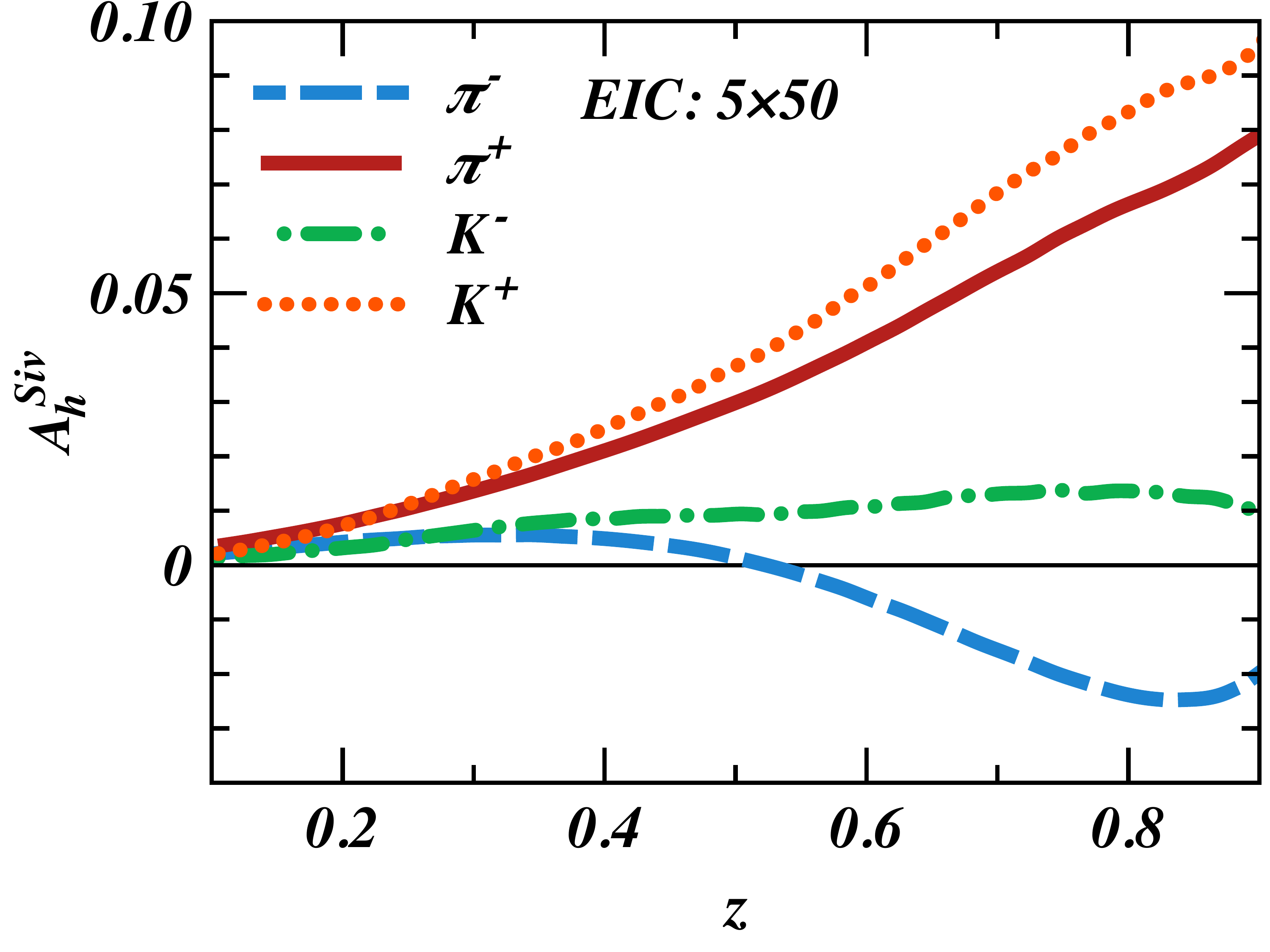}
}
\hspace{-0.2cm}
\subfigure[] {
\includegraphics[width=\ImM]{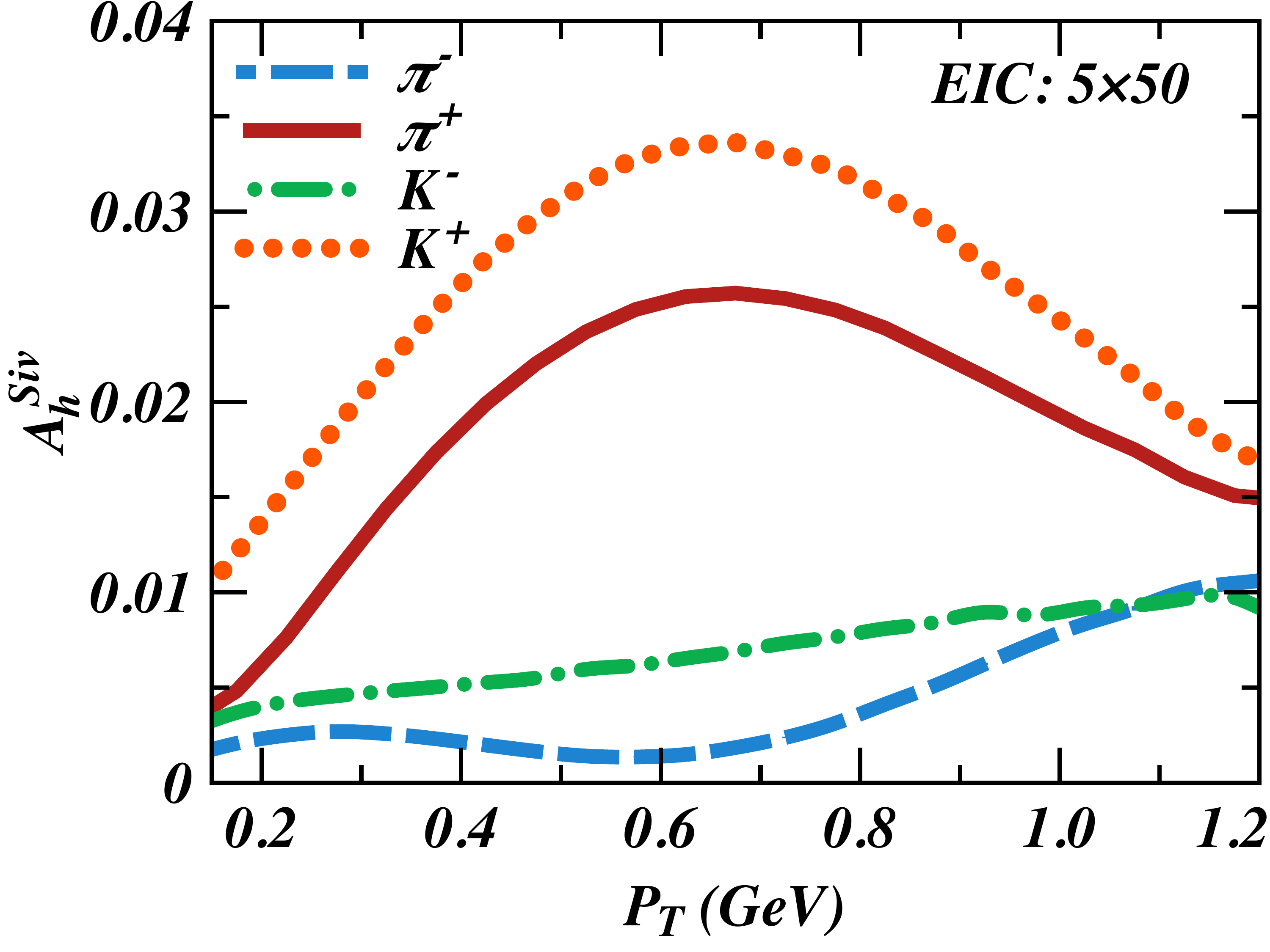}
}
\\\vspace{-0.2cm}
\caption{EIC SSAs vs (a) $x$, (b) $z$ and (c) $P_T$ for $5\times50$ SIDIS production of  charged pseudoscalars.}
\label{PLOT_EIC_SSA_X_Z_PT_5x50}
\end{figure}
  
 The plots in Fig.~\ref{PLOT_EIC_SSA_X_Z_PT_5x50} depict the SSAs for charged pseudoscalars as functions of (a) $x$, (b) $z$ and (c) $P_T$ for the $5\times50$ configuration. We see that the SSAs for $K^+$ are large, while those for $\pi^-$ and $K^-$ are relatively small for all cases. This can be understood in terms of a significant $u\to K^+$ FF that has a bump at relatively large $z$ ($z\approx0.7$), while $u\to \pi^+$ drops sharply with increased $z$ (though it stays larger in magnitude than that for the $K^+$). Thus, on average a larger portion of the fragmenting quark's transverse momentum is transferred to the $K^+$, leading to the large SSA.  Additionally, the contribution of the $\bar{s}$ quark to the $K^+$ asymmetry is quite large due to the parametrization of the Sivers function for $s$ and $\bar{s}$, which was driven by HERMES and COMPASS data. The contribution of the strange quarks to $\pi^+$ is negligible, as the corresponding fragmentation functions are very small compared to those for the up quark. We have checked explicitly, that when the strange Sivers PDF is set to zero, only the $K^+$ SSA is affected, it is reduced and becomes almost equal to that for the $\pi^+$.
 
\begin{figure}[tb]
\centering 
\subfigure[] {
\includegraphics[width=\ImM]{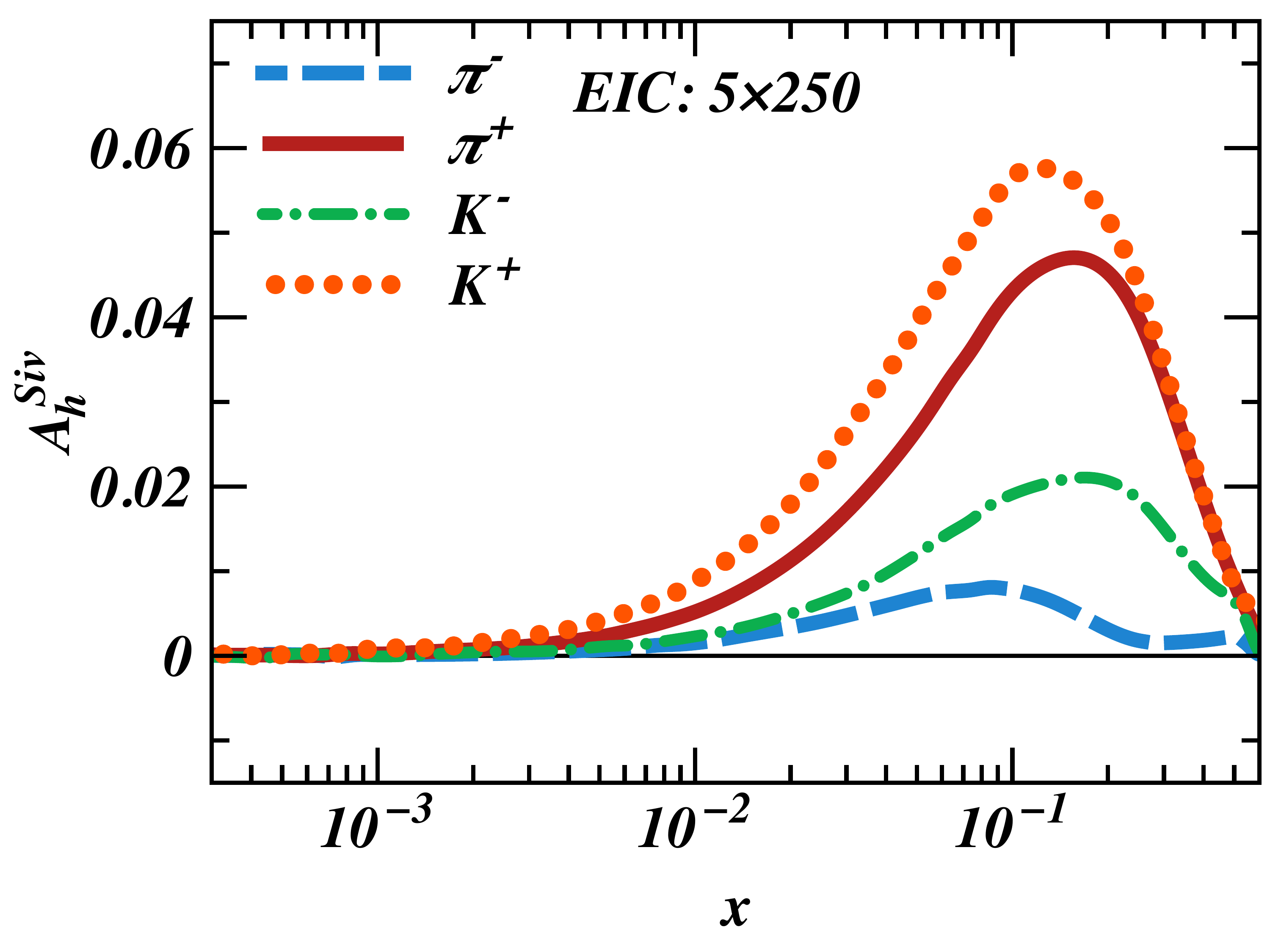}}
\hspace{-0.2cm}
\subfigure[] {
\includegraphics[width=\ImM]{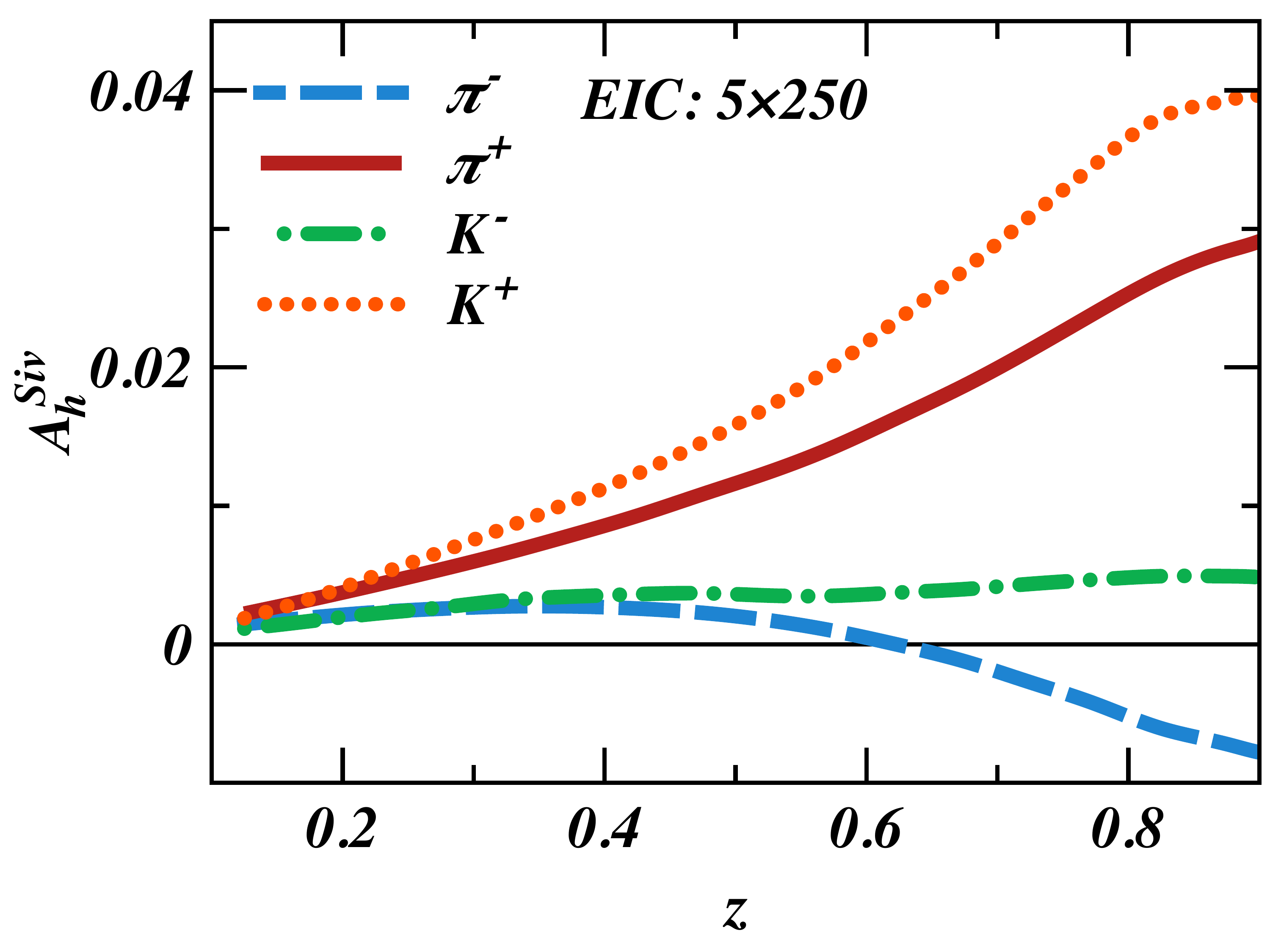}
}
\hspace{-0.2cm}
\subfigure[] {
\includegraphics[width=\ImM]{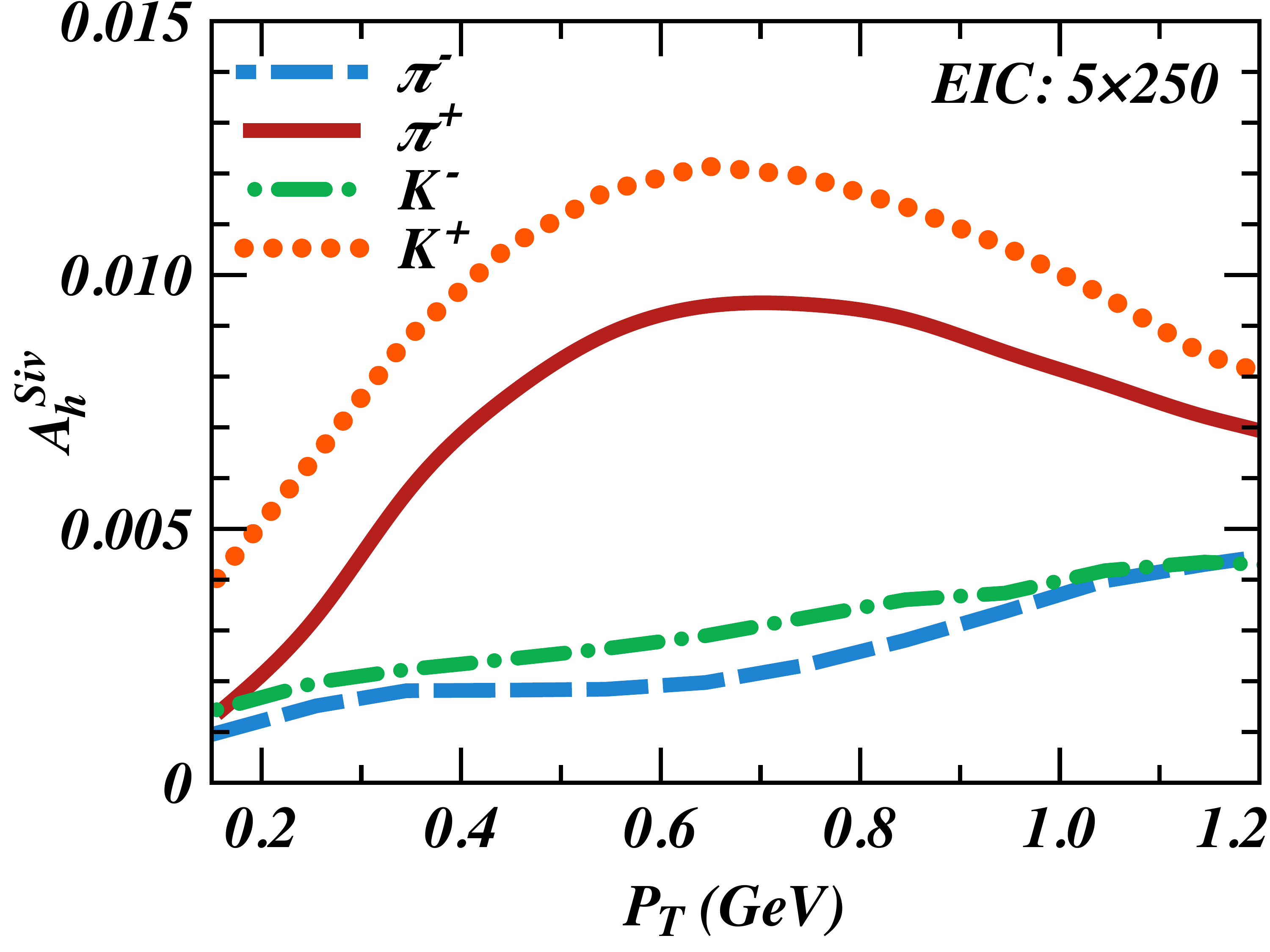}
}
\\\vspace{-0.2cm}
\caption{EIC SSAs vs (a) $x$, (b) $z$ and (c) $P_T$ for $5\times250$ SIDIS production of  charged pseudoscalars.}
\label{PLOT_EIC_SSA_X_Z_PT_5x250}
\end{figure}

\begin{figure}[t]
\centering 
\vspace{0.2cm}
\includegraphics[width=\ImMM]{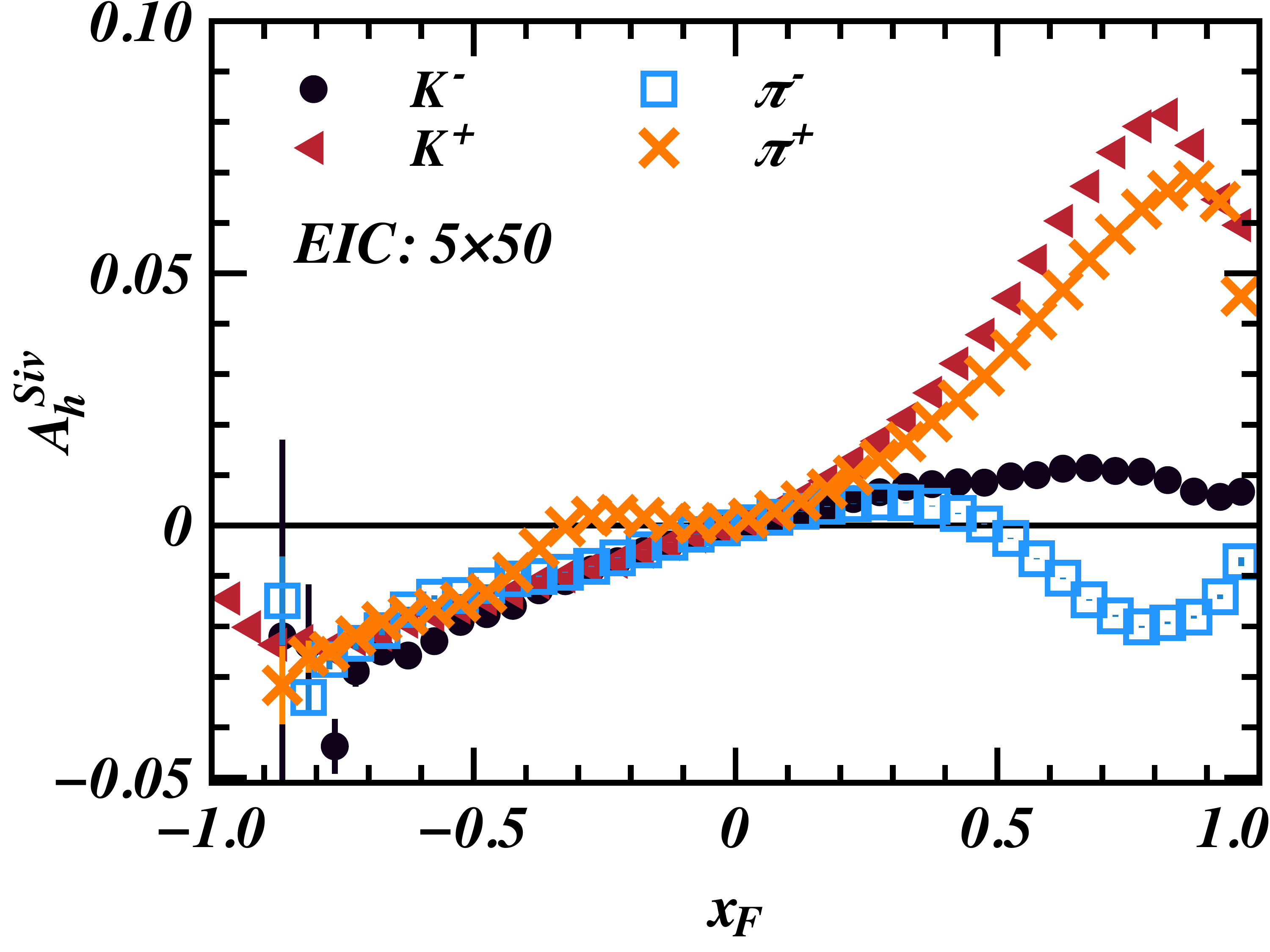}
\caption{EIC model SSAs for $5\times50$ SIDIS kinematics for charged pions and kaons vs $x_F$. The Sivers asymmetry is present both in the current and target fragmentation regions.}
\label{PLOT_EIC_SSA_XF}
\end{figure}

\begin{figure}[bh]
\centering 
\subfigure[] {
\includegraphics[width=\ImMM]{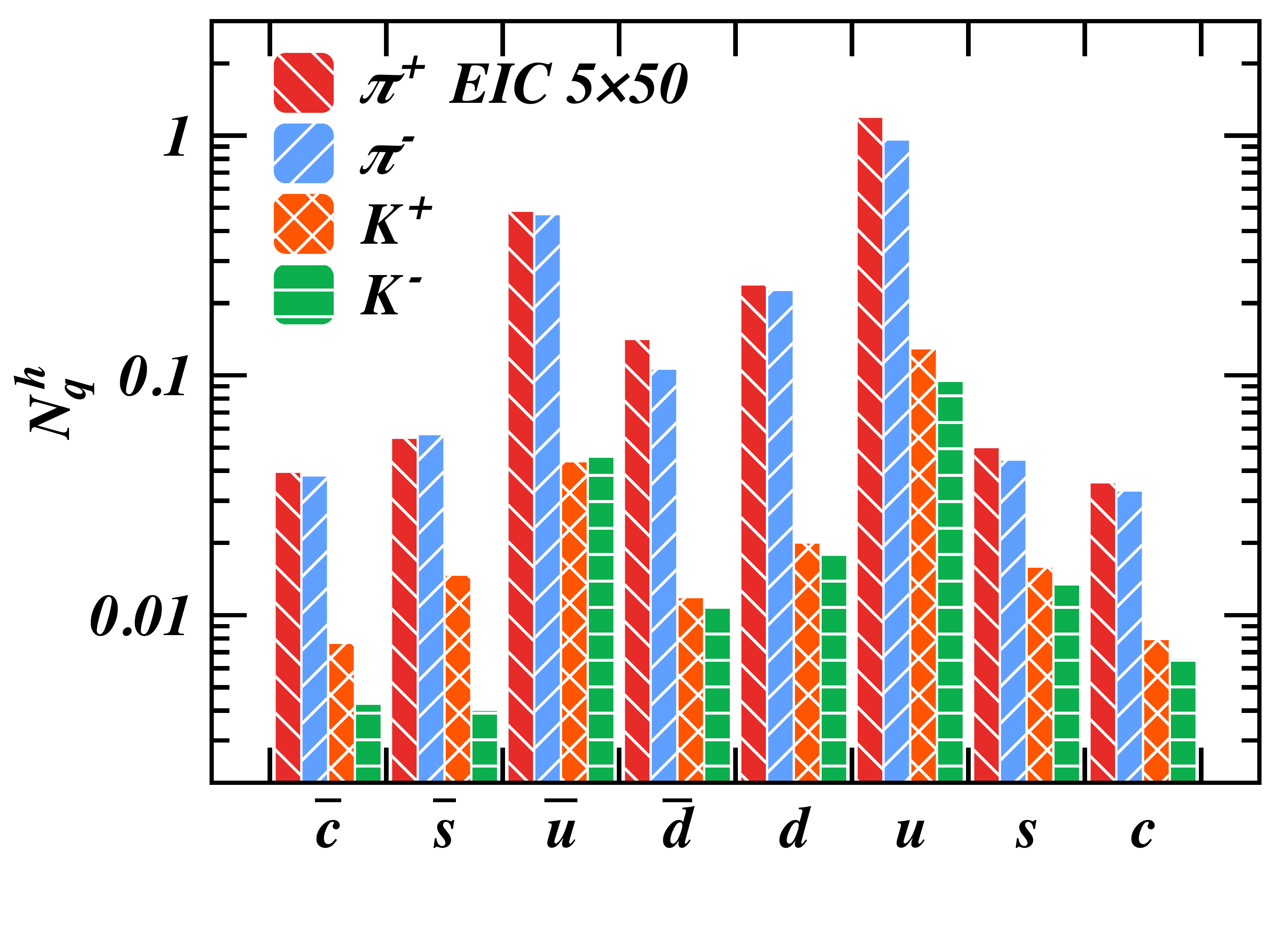}
}
%
\subfigure[] {
\includegraphics[width=\ImMM]{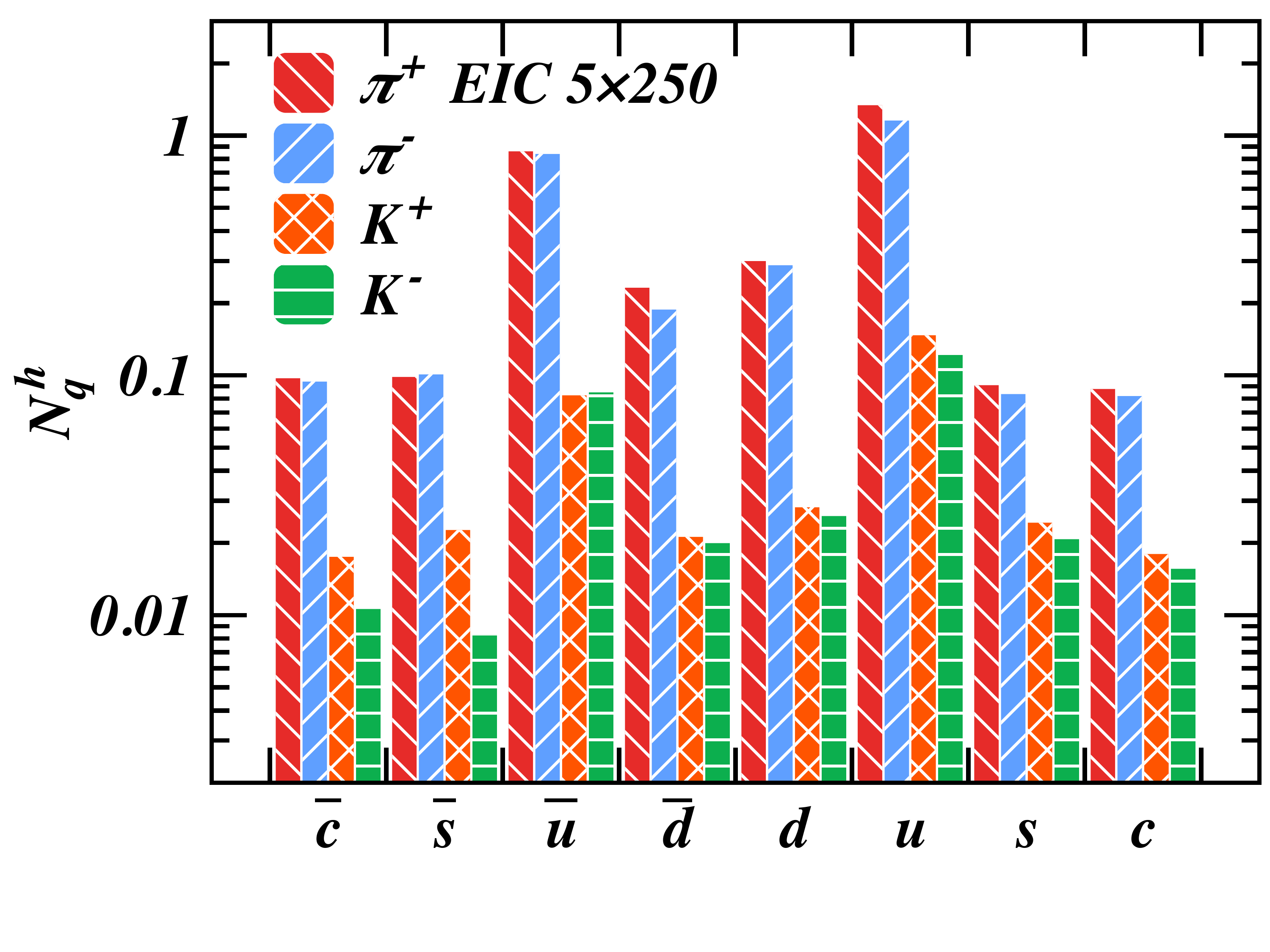}
}
%
\caption{EIC rates for the struck quark flavor for the selected events in SIDIS productions of charged pions and kaons for (a) $5\times 50$ and (b) $5\times 250$ energies.}
\label{PLOT_EIC_QFLAV}
\end{figure}

Next we want to test how these predictions would vary for increased CM energy, corresponding to the $5\times250$ configuration of beam energies. The plots in Fig.~\ref{PLOT_EIC_SSA_X_Z_PT_5x250} depict the results for the SSAs for $\pi^+$ as functions of (a) $x$, (b) $z$ and (c) $P_T$ for this energy. We see that the SSAs in general decrease as we increase the CM energy. This is because at larger energies we sample lower $x$ values, where the sea quarks give the largest contributions to the SSAs. Note that we have used the parametrizations extracted from COMPASS and HERMES data, where the Sivers PDF for the sea is not very well constrained and comes out small when compared to valence. Thus we can only conclude from these plots that if the sea Sivers PDFs are small compared to the valence ones, then the Sivers SSAs should become smaller and smaller as we increase the energy. On the other hand, measurements at higher energies would help to probe the small $x$ region and better measure the sea Sivers PDFs.
%
\begin{figure}[tb]
\centering 
\subfigure[] {
\includegraphics[width=\ImM]{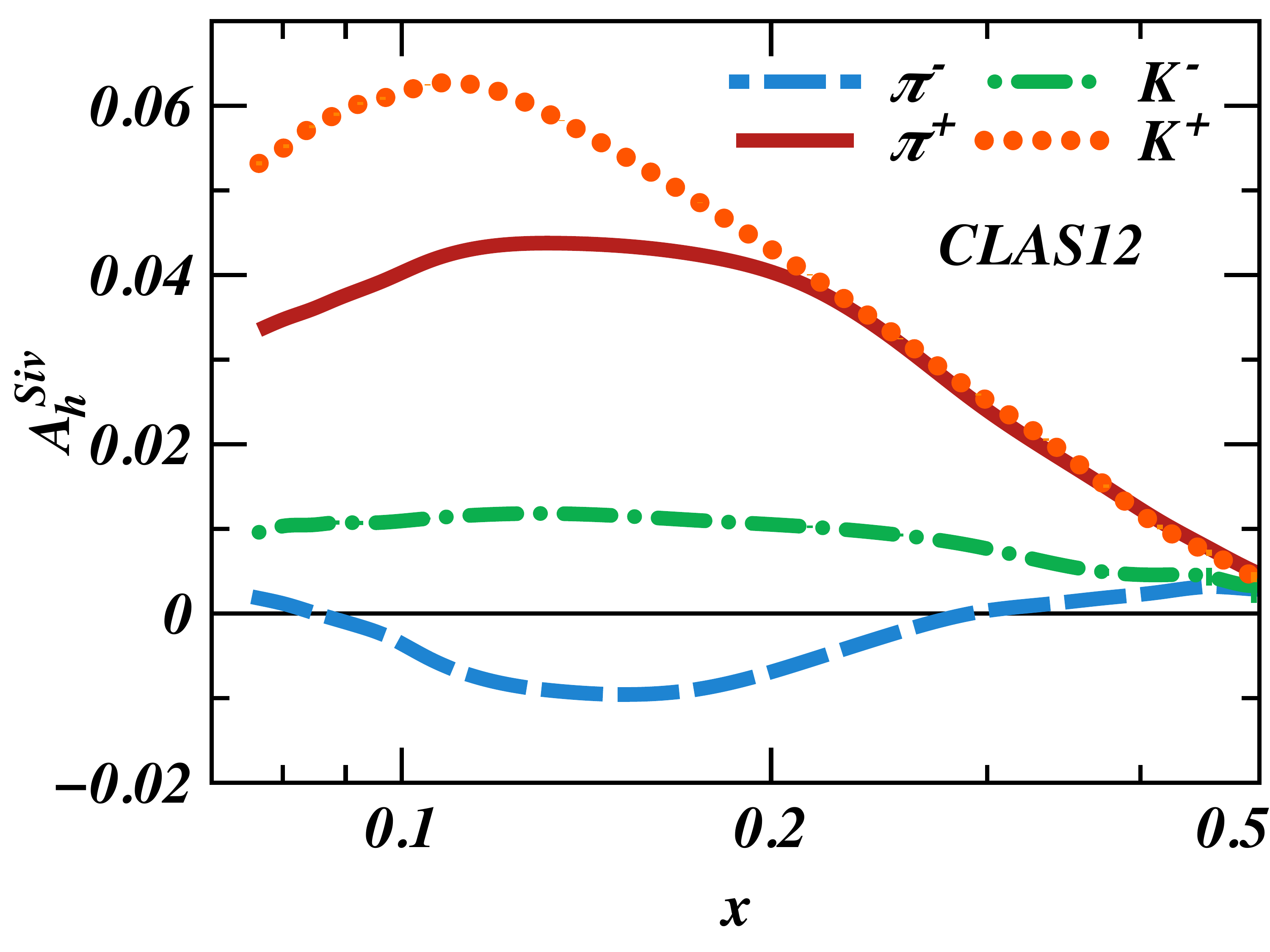}}
\\\vspace{-0.2cm}
\subfigure[] {
\includegraphics[width=\ImM]{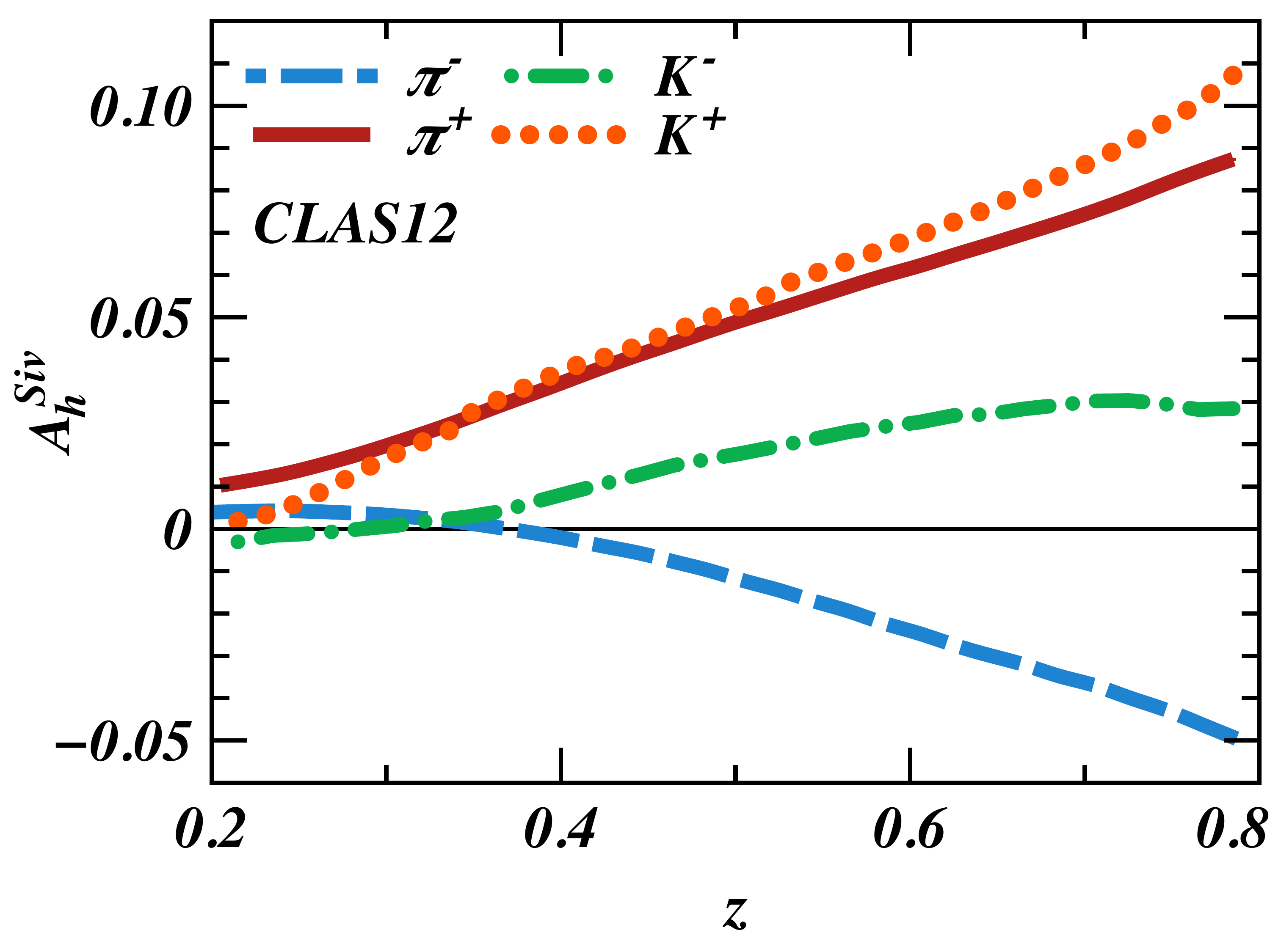}
}
\\\vspace{-0.2cm}
\subfigure[] {
\includegraphics[width=\ImM]{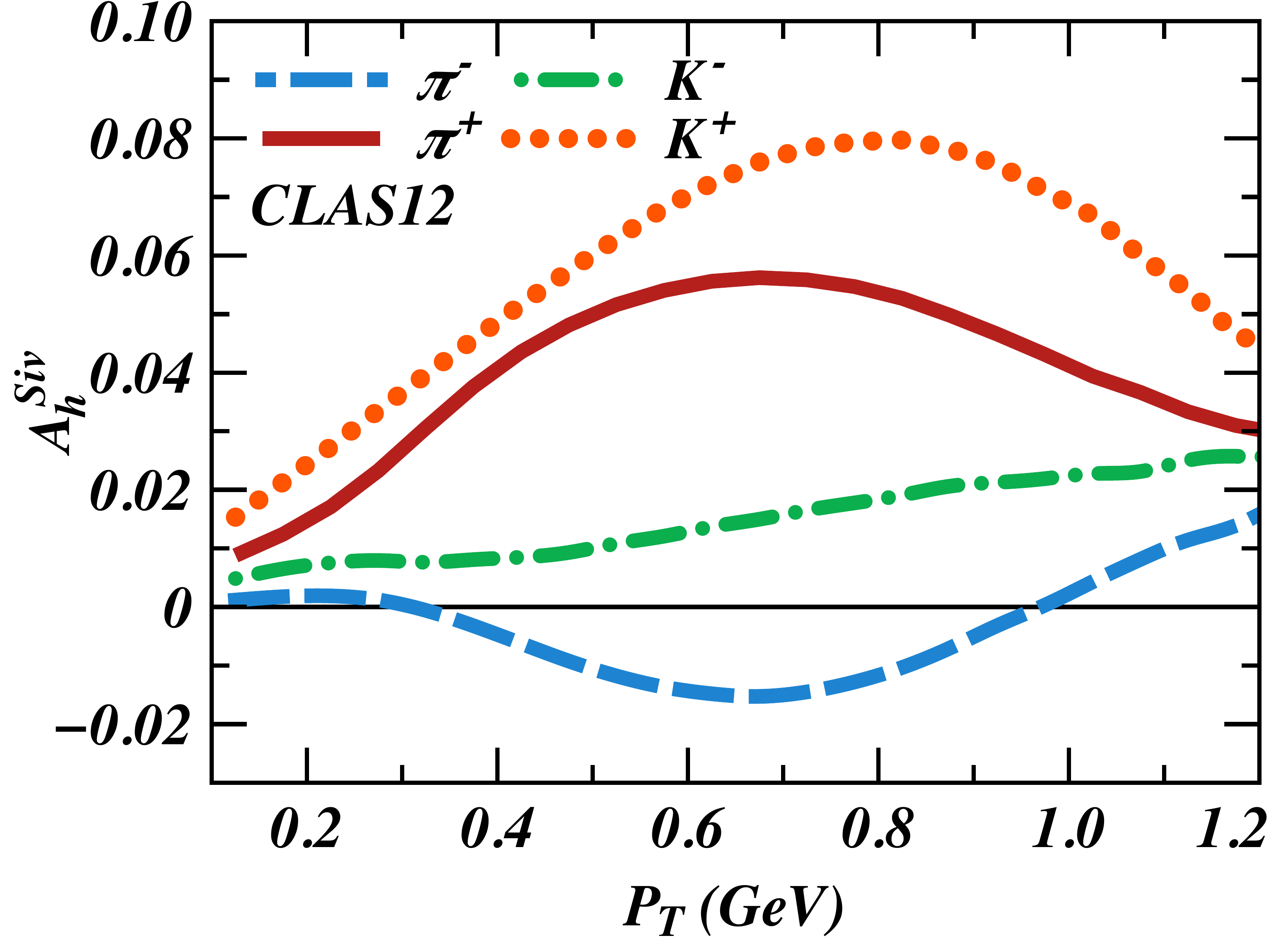}
}
%
\caption{Predictions for the Sivers asymmetry in charged hadron production at CLAS12 versus (a) $x$, (b) $z$ and (c) $P_T$.}
\label{PLOT_CLAS12_SIV_1H}
\end{figure}

  The plots in Fig.~\ref{PLOT_EIC_SSA_XF} depict the results of the SSAs for charged pions and kaons versus $x_F$, analogous to those depicted in Fig.~\ref{PLOT_EIC_TOY_SSA_XF} for the toy model. We see that the nonzero Sivers SSAs are also present in the TFR ($x_F<0$), with opposite sign to those in the CFR ($x_F>0$) except for $\pi^-$. This and the nontrivial shapes and unequal sizes of the SSAs for the other mesons are governed by the Sivers PDF parametrizations and the complicated hadronization dynamics in the CFR and TFR regions. Thus, measuring these SSAs over the entire range of $x_F$ (or rapidity) will provide additional information to improve our understanding of the complete hadronization of both the knocked out parton and the target remnant. 

One of the important challenges in phenomenological fits is to disentangle the flavor dependence of the Sivers PDF from the measurements for a small set of final particle types. The difficulty is that in the corresponding cross section the Sivers PDF is convoluted with the fragmentation function to the specific hadron and these are summed over the flavor of the parton. Thus having SSA measurements for as large a number of final hadron types as possible would allow us to remove some or all of the assumptions about the flavor dependence of the corresponding Sivers PDFs. In experiment though, the practical limitations of the detectors limit this ability to only a few hadrons, such as charged pions and kaons. Thus it is important to gain phenomenological understanding of the relative contributions of each of the parton flavors to the Sivers term in the cross section.  In Fig.~\ref{PLOT_EIC_QFLAV} we depict the results for the  rates of struck quark flavor for the selected events that contribute to the count rates of a given hadron for charge pion and kaon final states and the two different sets of energies. We see that the charm sea becomes significant at large CM energies, though we set the charm Sivers function to be zero. Thus this should be in part responsible for weakening of the SSA signals with increasing energies, as we saw earlier. We have seen in the toy model calculations that with the increased energy there was a slight reduction of the SSAs, even when all the light quarks were assigned a large Sivers PDF. We should keep in mind here that these integrated rates hide a much more complicated picture when considering the full $x$, $z$ and $P_T$ dependences.

\begin{figure}[tb]
\centering 
\includegraphics[width=\ImMM]{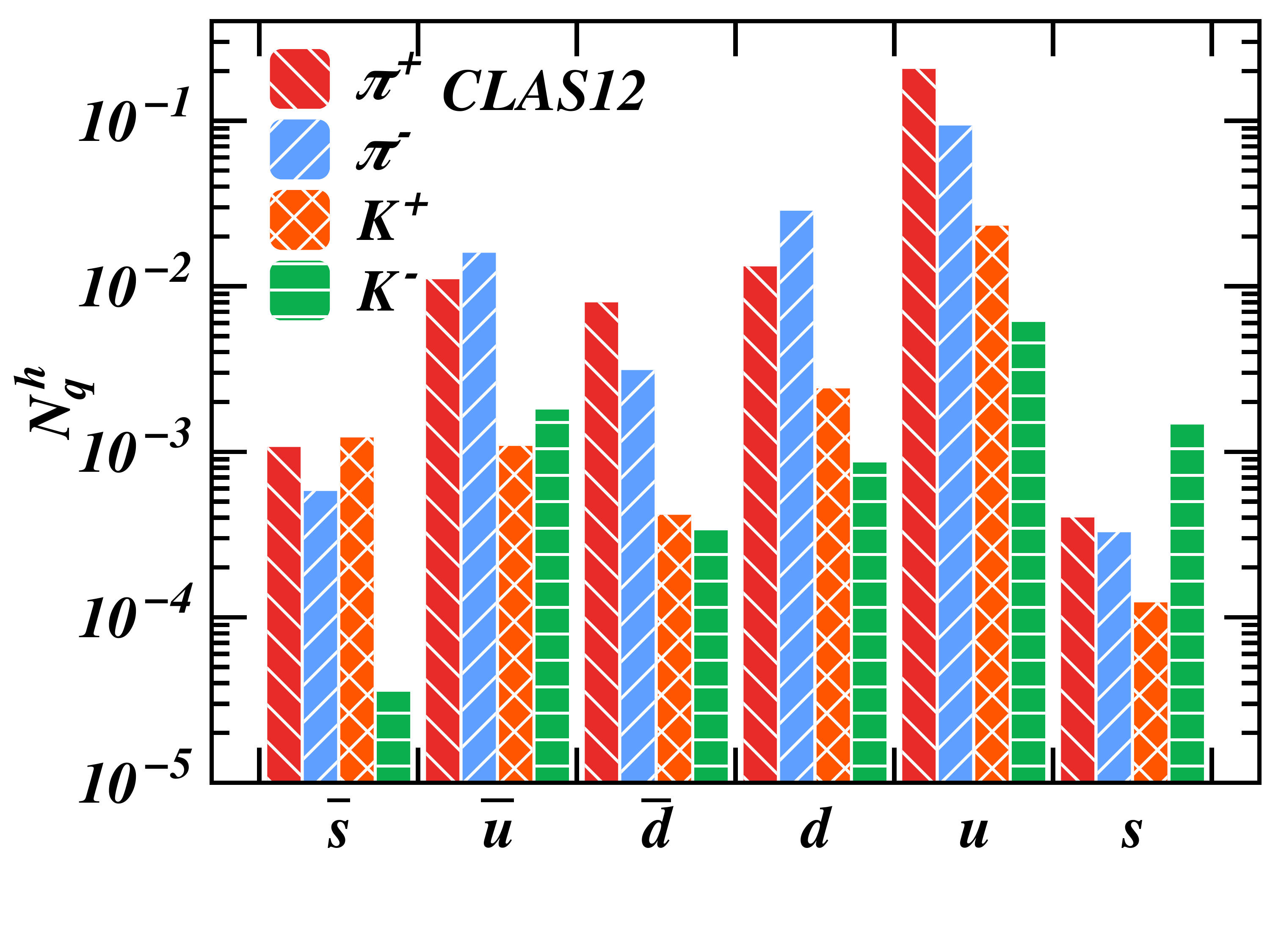}
\vspace{-0.2cm}
\caption{Predictions for the relative rates for the flavor of the struck quark that produces charged hadrons in MC events at CLAS12, with all the relevant kinematical cuts.}
\label{PLOT_CLAS12_SIV_QUARK_ID}
\end{figure}
%
\begin{figure}[htb]
\centering 
\vspace{0.2cm}
\includegraphics[width=\ImMM]{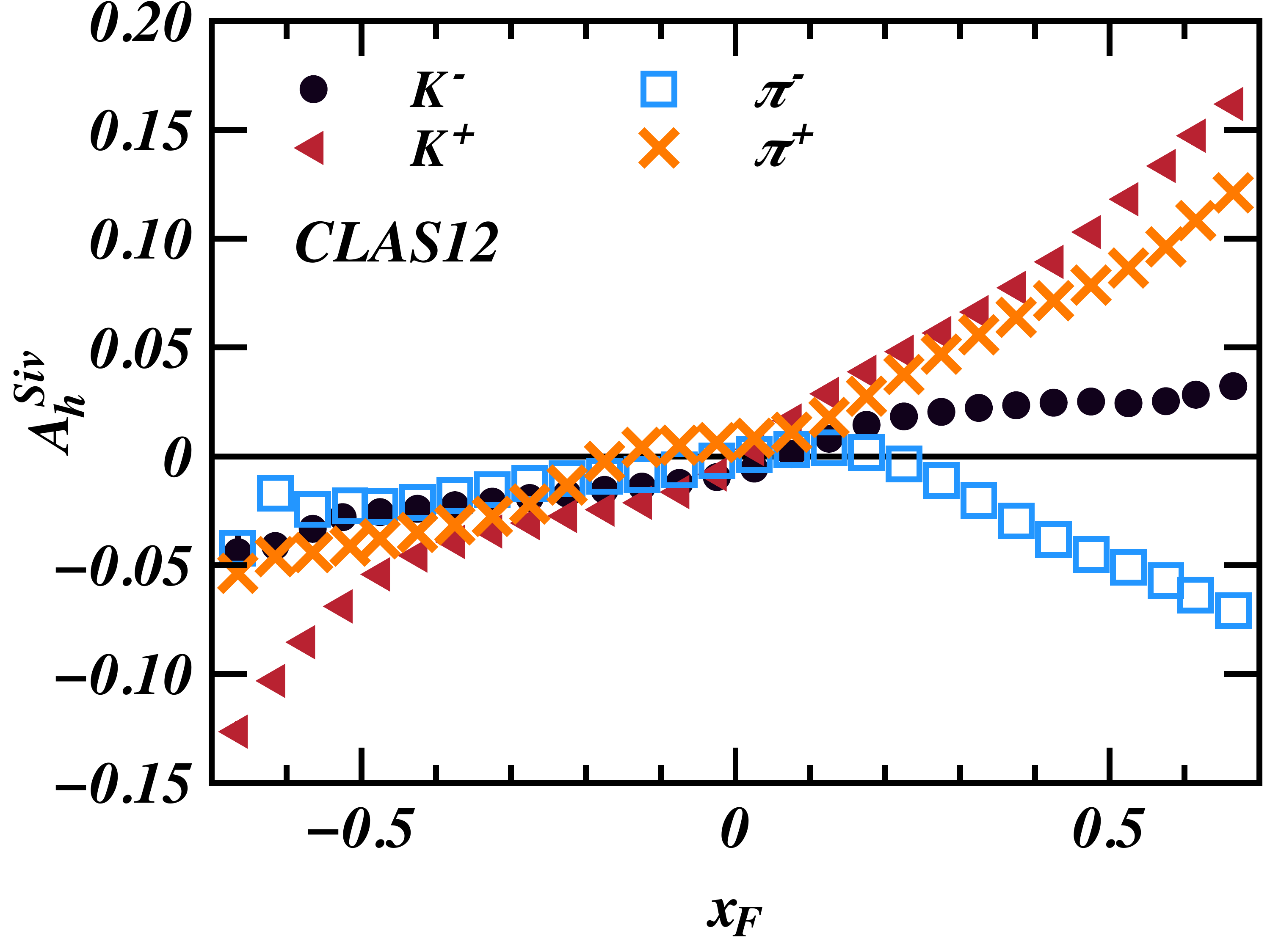}
\\\vspace{-0.2cm}
\caption{Predictions for SSAs for charged pions and kaons vs $x_F$ at CLAS12. The Sivers asymmetry is present both in the current and target fragmentation regions.}
\label{PLOT_CLAS12_SSA_XF}
\end{figure}

\subsection{Predictions for CLAS12}
\label{SUB_SEC_MLEPTO}

 In this subsection we study the single hadron SSAs for the upcoming CLAS12 experiment~\cite{JLAB12:CLAS12}, again using the \MPYTH MC event generator with the parameters slightly tuned to best suit the kinematics for the CLAS12 experiment at $11~\Ge$, that has been described in the beginning of this section. 

 The \MPYTH result for the one hadron Sivers SSAs for charged pions and kaons as functions of $x$, $z$ and $P_T$ are presented in Fig.~\ref{PLOT_CLAS12_SIV_1H} (a), (b) and (c), respectively. In Fig.~\ref{PLOT_CLAS12_SIV_QUARK_ID} (drawn in the $\log$ scale) we show the analogous plots to Fig.~\ref{PLOT_EIC_QFLAV} with the rates of the produced hadrons after all the relevant cuts have been applied. We note that the Sivers SSA  for $\pi^+$ is large and positive, but the SSA for $K^+$ is even bigger. On the other hand, the rates for $K^+$ production are about an order of magnitude smaller than for $\pi^+$. Thus much higher statistics would be required to measure $K^+$ SSAs versus those for $\pi^+$. The SSAs for $\pi^-$ and $K^-$ are relatively smaller, with the latter being produced at a rate that is much less than even $K^+$, making it very challenging to measure at CLAS12. 
  
  Finally, the plots in Fig.~\ref{PLOT_CLAS12_SSA_XF} depict the results of the SSAs for charged pions and kaons versus $x_F$. It is interesting to note that the SSAs here are larger in magnitude than those plotted for EIC in Fig.~\ref{PLOT_EIC_SSA_XF}, thus they should be easier to access in the experiment. Additionally, we notice here that the kinematical cuts employed for CLAS12 (on the missing mass, etc) limit the $x_F$ region of these mesons to about $\pm0.75$.

 Here we would underline the importance of the new, higher-statics measurements of the Sivers effect in one hadron SIDIS. This will allow us to improve the extractions of Sivers PDFs, especially in the valence region, as well as allowing us to shrink the error bars of the corresponding parametrizations coming from the current experimental uncertainties. Further, development and testing on real world data of a complete analysis tool chain for extracting the Sivers PDF, that accounts for all the non-DIS and parton showering effects for CLAS12, would be of high importance both for CLAS12 and future EIC experiments.

\section{Monte Carlo predictions for two hadron SSA}
\label{SEC_MC_TWO_HADRON}

  The one hadron, SIDIS production off a transversely polarized target has been long established as one of the principal processes for measuring the Sivers PDF. Nonetheless, the phenomenological extractions of the Sivers PDF still remain challenging. This is especially true for the flavor dependence, where we can only measure a limited number of different hadron types in the final state. Two hadron SIDIS production~\cite{Kotzinian:2014lsa,Kotzinian:2014gza,Kotzinian:2014hoa} provides a complimentary approach to extract the Sivers function with very different systematic errors both  from theory and experiment. Here again the relevant term of the cross section is a convolution of the Sivers PDF with an unpolarized dihadron fragmentation function, summed over the type of the parton. This approach is promising, because with two detected hadrons in the final state we have a much wider basis for extracting the flavor dependence of the Sivers PDF. For a given number of types of the detected hadrons $N_h$, the number of possible dihadron pairs is given by $N_h\cdot(N_h+1)/2$, as we can have pairs with the same types of hadrons. Thus it would be important to estimate the viability of such dihadron SSA measurements at both CLAS12 and EIC using our MC generators. Here we can measure SSAs corresponding to modulations with respect to $\vf_{Siv,T}= \vf_{T}-\vf_S$  and $\vf_{Siv,R}= \vf_{R}-\vf_S$ [see Eqs.~(\ref{EQ_2H_X_SEC_INT_R},\ref{EQ_2H_X_SEC_INT_T})], that we will simply refer to as "T" and "R" modes.
%
\begin{figure}[tb]
\centering 
\subfigure[] {
\hspace{-0.2cm}
\includegraphics[width=\ImMM]{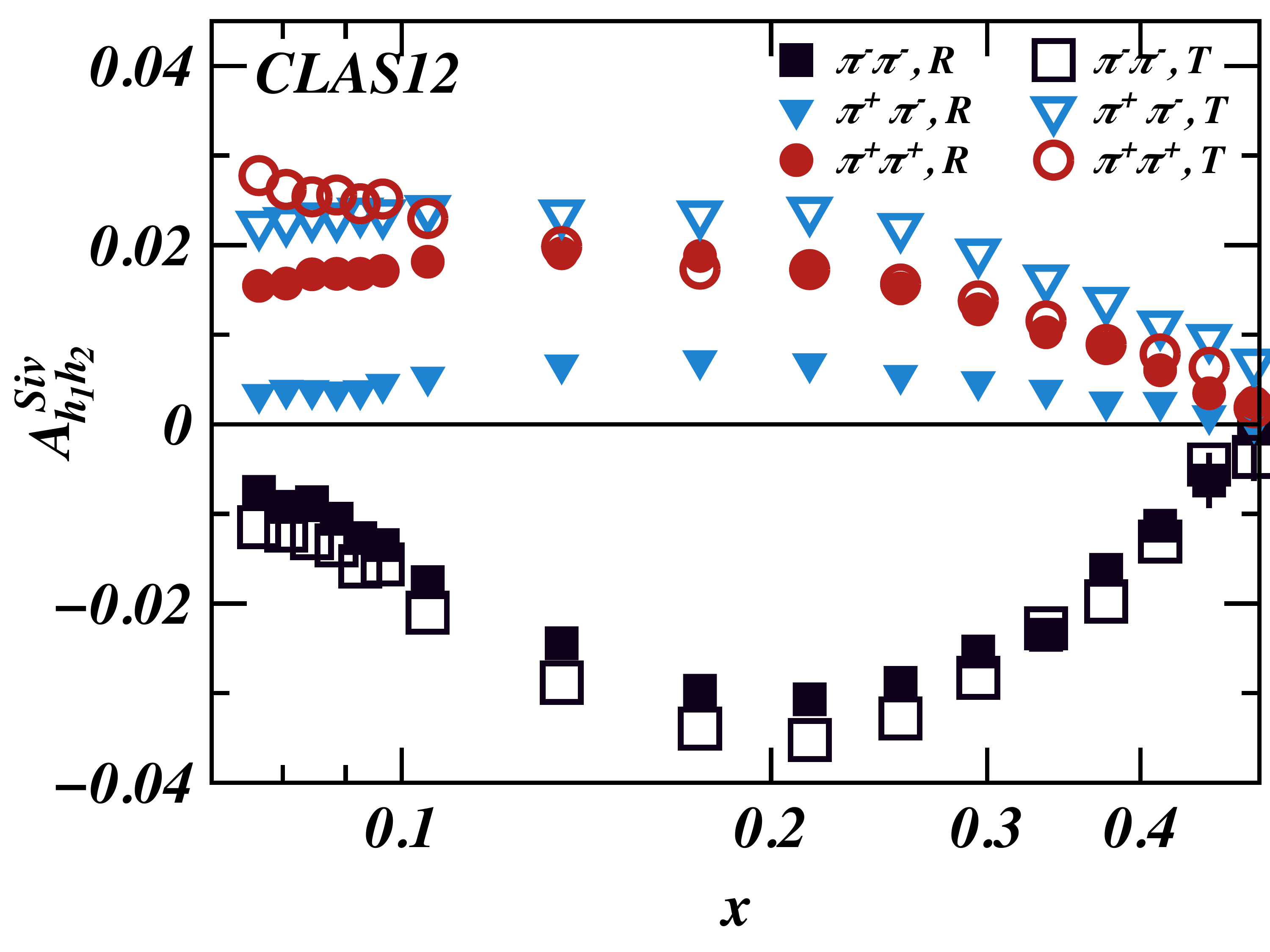}}
\hspace{-0.2cm}
\subfigure[] {
\includegraphics[width=\ImMM]{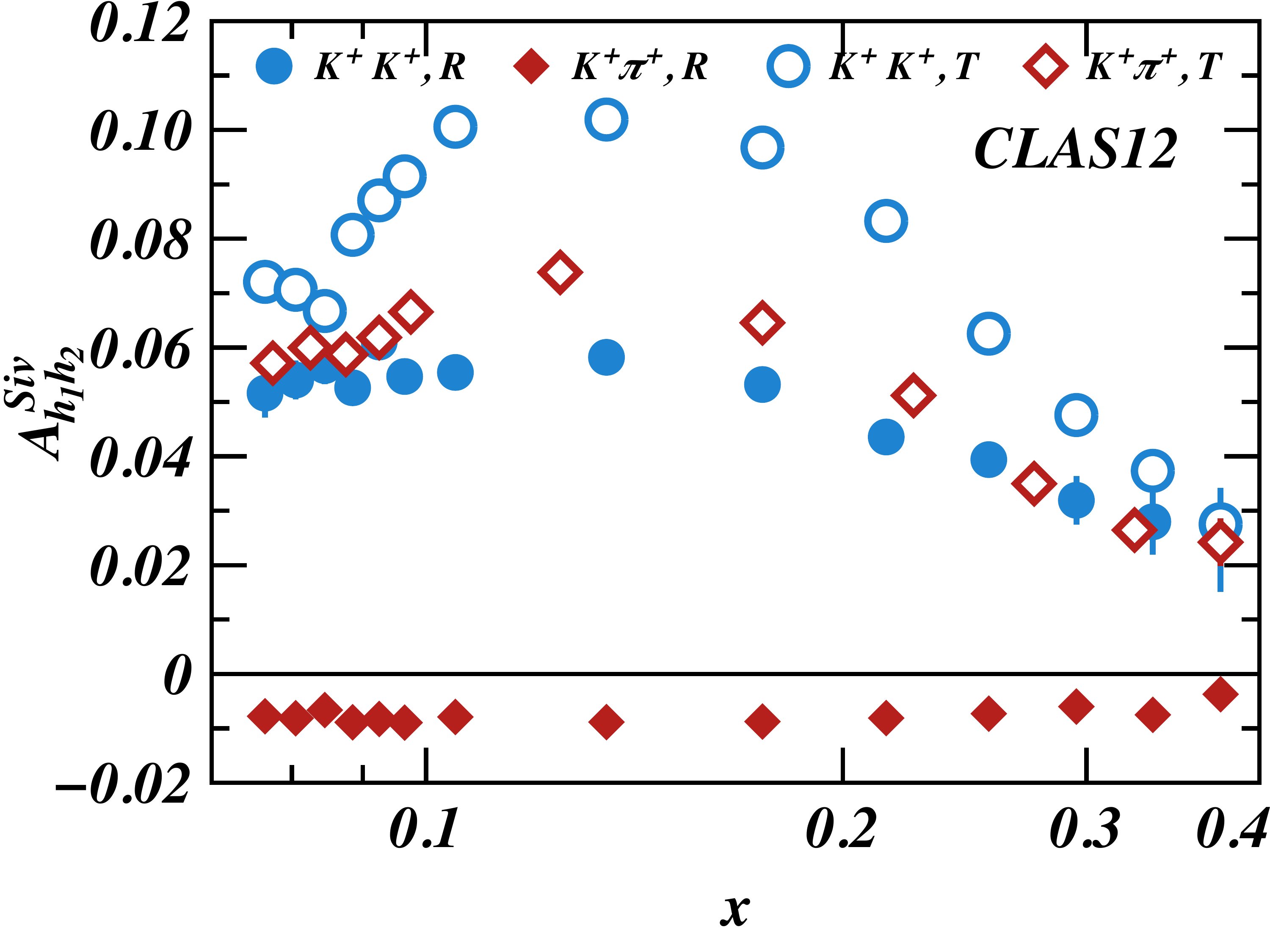}
}
\\\vspace{-0.2cm}
\caption{Predictions for Sivers SSA  for charged (a) pion and (b) kaon-inclusive pair production off proton target at CLAS12 versus $x$.}
\label{PLOT_SIV_2H_X}
\end{figure}
%
\begin{figure}[tbh]
\centering 
\subfigure[] {
\includegraphics[width=\ImM]{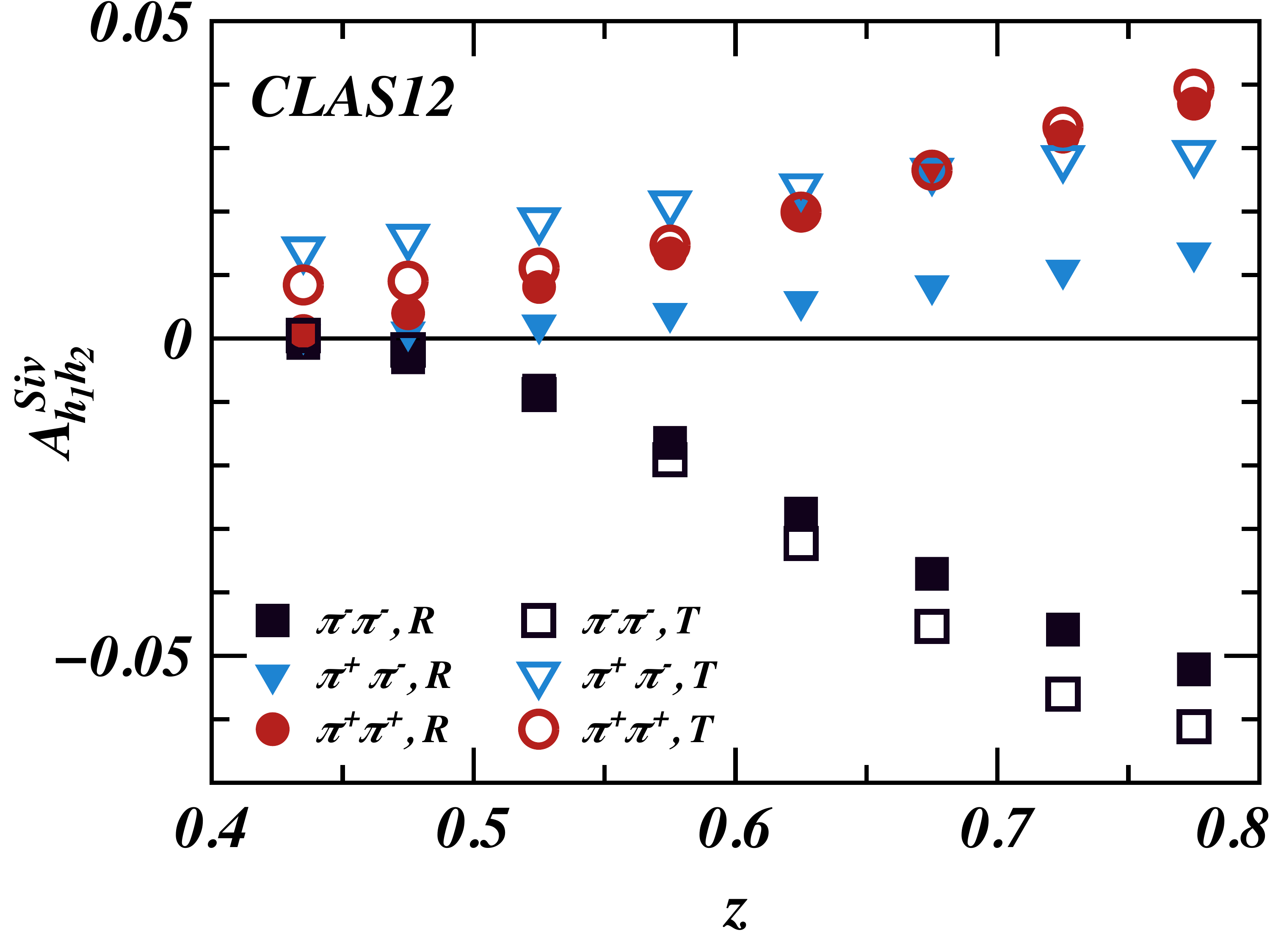}
}
\\\vspace{-0.2cm}
\subfigure[] {
\includegraphics[width=\ImM]{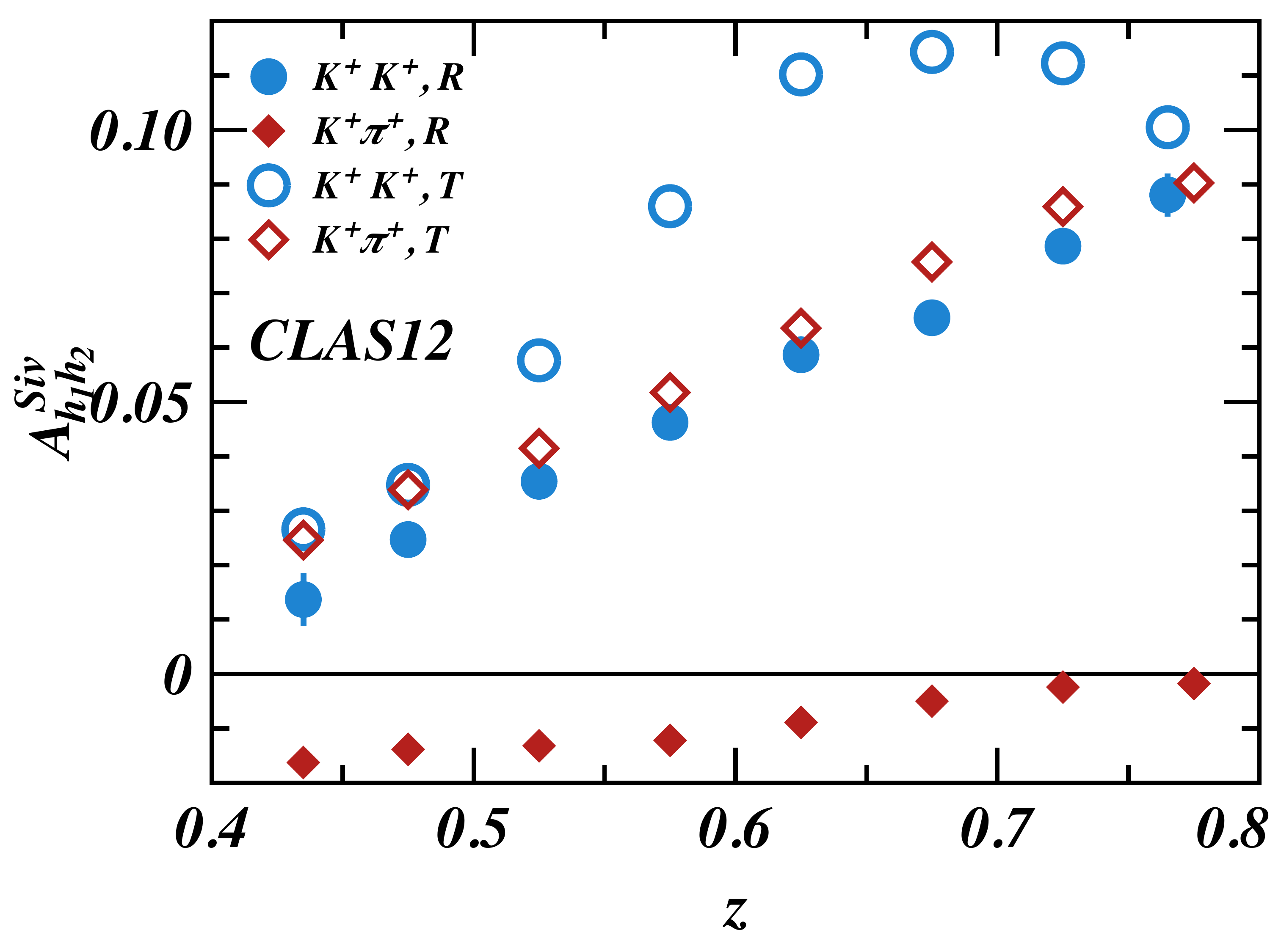}
}
\\\vspace{-0.2cm}
\caption{Predictions for Sivers SSA  for charged (a) pion and (b) kaon-inclusive pair production off proton target at CLAS12 versus $z$.}
\label{PLOT_SIV_2H_Z}
\end{figure}
%
\begin{figure}[tbh]
\centering 
\subfigure[] {
\includegraphics[width=\ImM]{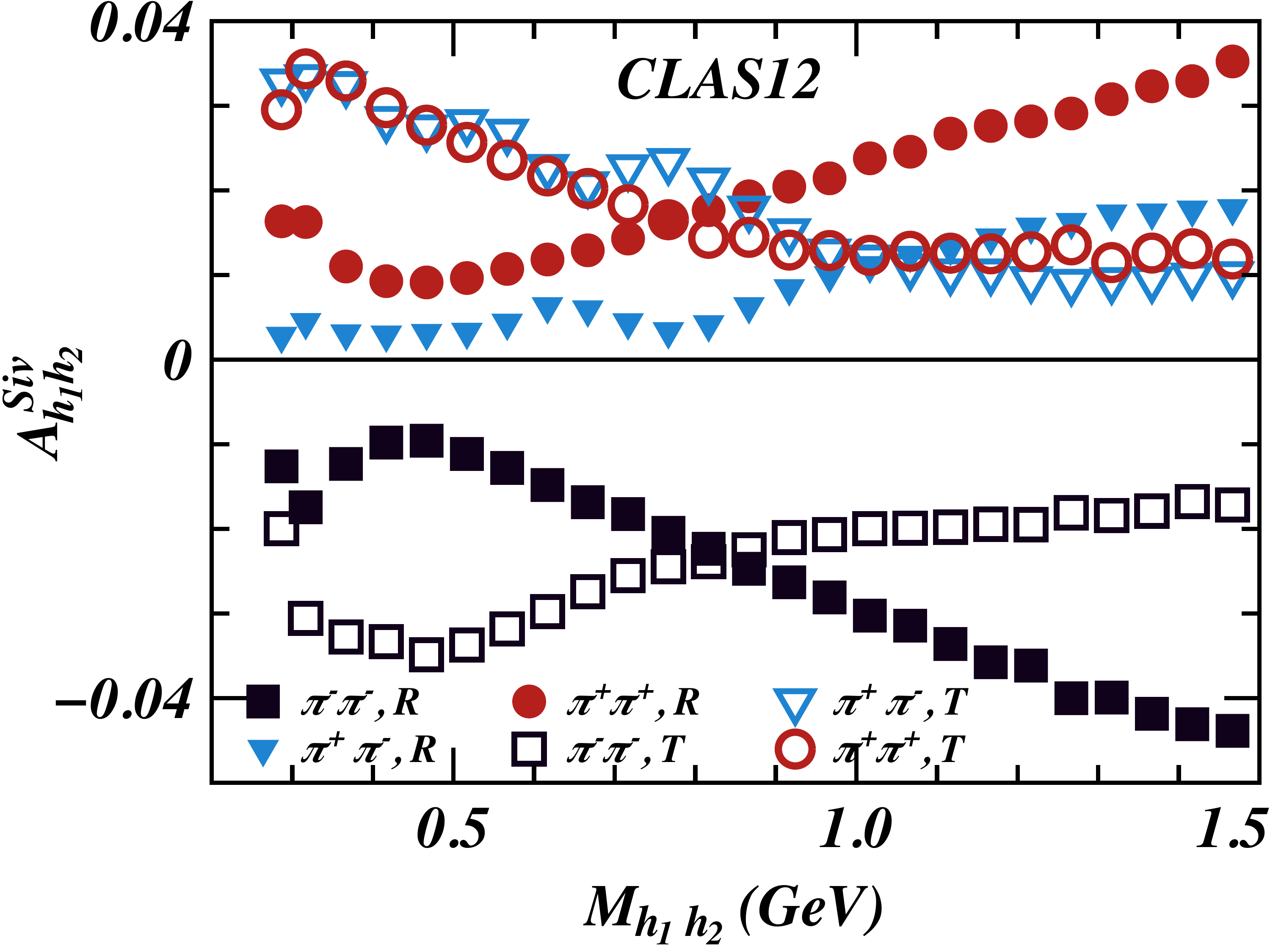}
}
\\\vspace{-0.2cm}
\subfigure[] {
\includegraphics[width=\ImM]{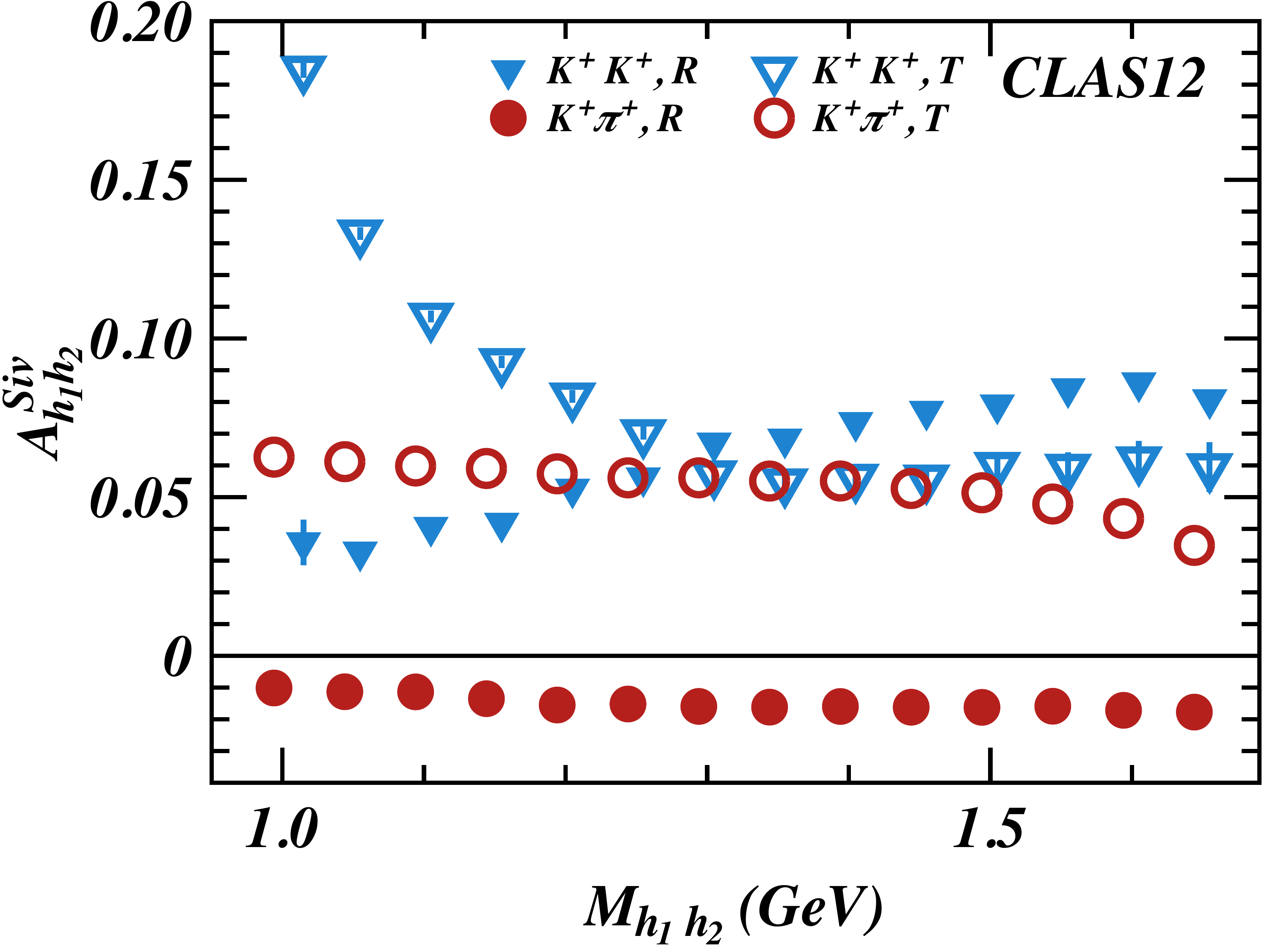}
}
\\\vspace{-0.2cm}
\caption{Predictions for Sivers SSA  for charged (a) pion and (b) kaon-inclusive pair production off proton target at CLAS12 versus $M_{h_1h_2}$.}
\label{PLOT_SIV_2H_MH}
\end{figure}
%
\begin{figure}[!hb]
\centering 
\subfigure[] {
\includegraphics[width=0.27\paperwidth]{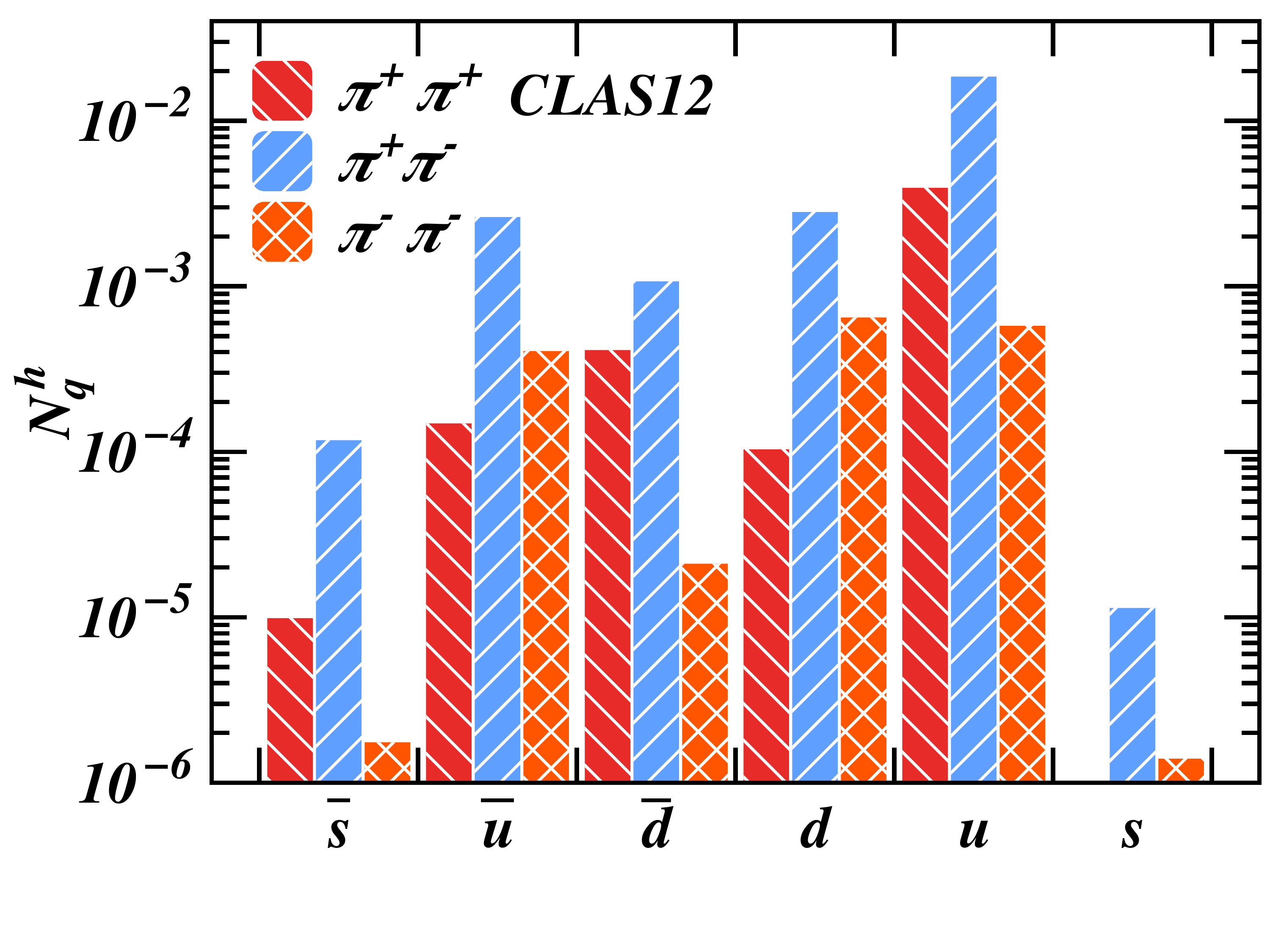}
}
\\\vspace{-0.2cm}
\subfigure[] {
\includegraphics[width=0.27\paperwidth]{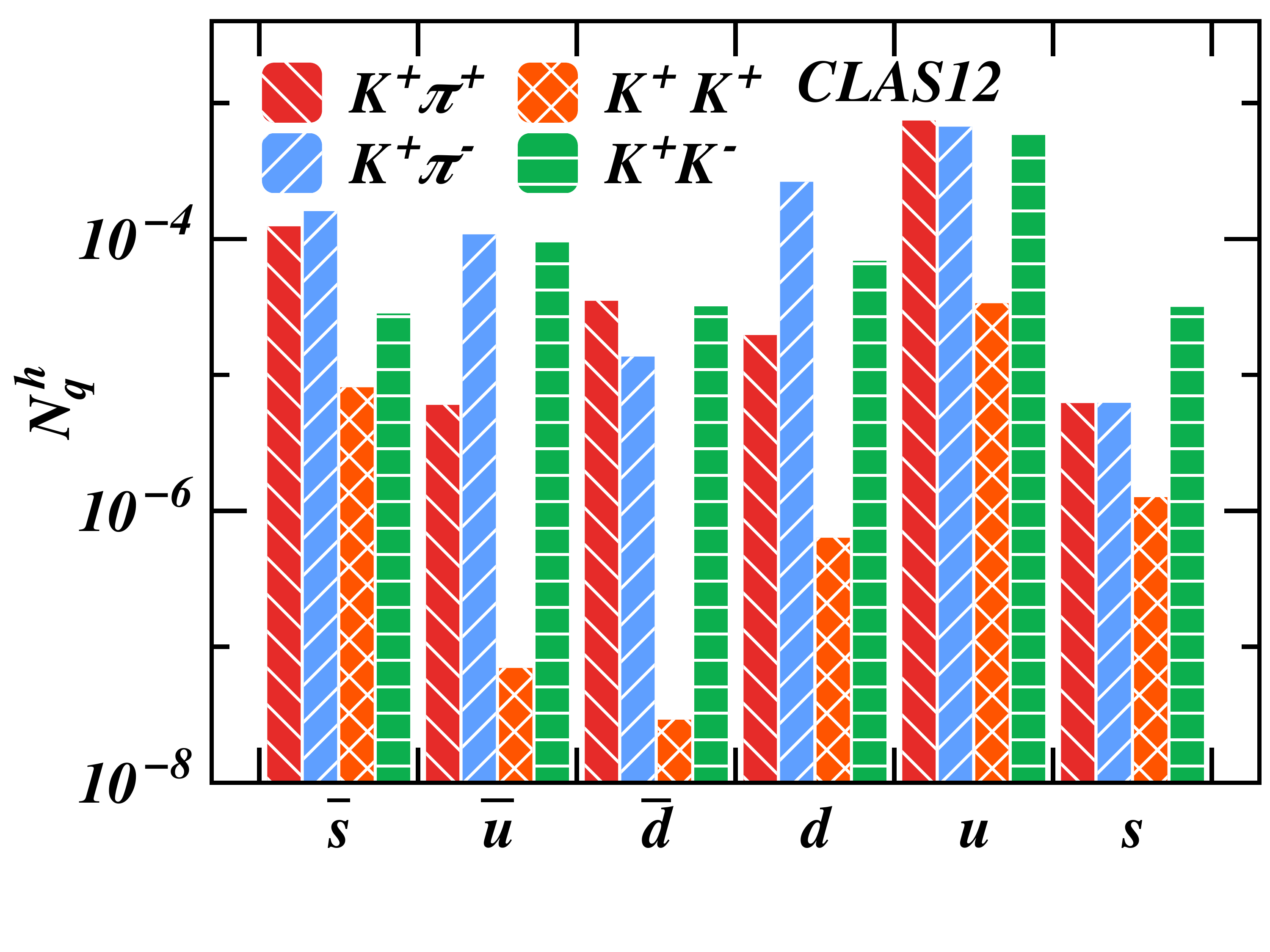}
}
\\\vspace{-0.2cm}
\caption{Predictions for the relative rates for the flavor of the struck quark that produces charged (a) pion and (b) kaon-inclusive pair in MC events at CLAS12, with all the relevant kinematical cuts.}
\label{PLOT_CLAS12_SIV_QUARK_DIHAD_ID}
\end{figure}

\subsection{Predictions for CLAS12}
\label{SEC_MLEPTO}

\subsubsection{Dihadron SSAs}

 Here we consider the dihadron pairs of pions and kaons. In identical pairs (e.g. $\pi^+\pi^+$), we choose $h_1$ as the hadron with the larger value of $z_i$ in the pair. Figures.~\ref{PLOT_SIV_2H_X} show the results for both "T" and "R" type SSAs for dihadron pairs as functions of $x$. We see that the pairs involving $K^+$ have the largest SSAs. Further, the modulations involving the $\vf_{Siv,T}$ are larger in magnitude than those for the same pair involving $\vf_{Siv,R}$. Both SSAs, but especially those for $\vf_{Siv,R}$, can be enhanced significantly be choosing asymmetric cuts on the momenta of the pair, as has been already shown in COMPASS kinematics in~\cite{Kotzinian:2014gza} and for EIC kinematics later in this sections. The dependences of these SSAs on the total energy fraction $z$ and invariant mass $M_{h_1h_2}$ of the pairs are presented in Figs.~\ref{PLOT_SIV_2H_Z} and~\ref{PLOT_SIV_2H_MH}. It is interesting to note that for the same charged pion pairs, the "T" and "R" SSAs trade strength as we increase  $M_{h_1h_2}$. 

 Again, we can learn about the relative rates of the different hadron pair production and their dependence on the struck quark's flavor from the plots in Fig.~\ref{PLOT_CLAS12_SIV_QUARK_DIHAD_ID}, that are the analogues of those for the single hadron production in Fig.~\ref{PLOT_CLAS12_SIV_QUARK_ID}. Here we note that $\pi^+\pi^-$ pairs are produced at the highest rate, while the rate for $K^+K^-$ pairs is about $100$ times smaller. Thus measuring kaon pairs would be more challenging than the pion pairs, but would be very important in understanding the flavor structure of the Sivers PDF.

\subsection{Predictions for EIC}
\label{SUBSEC_EIC_MPYTHIA}

\begin{figure}[tb]
\centering 
\subfigure[] {
\includegraphics[width=\ImM]{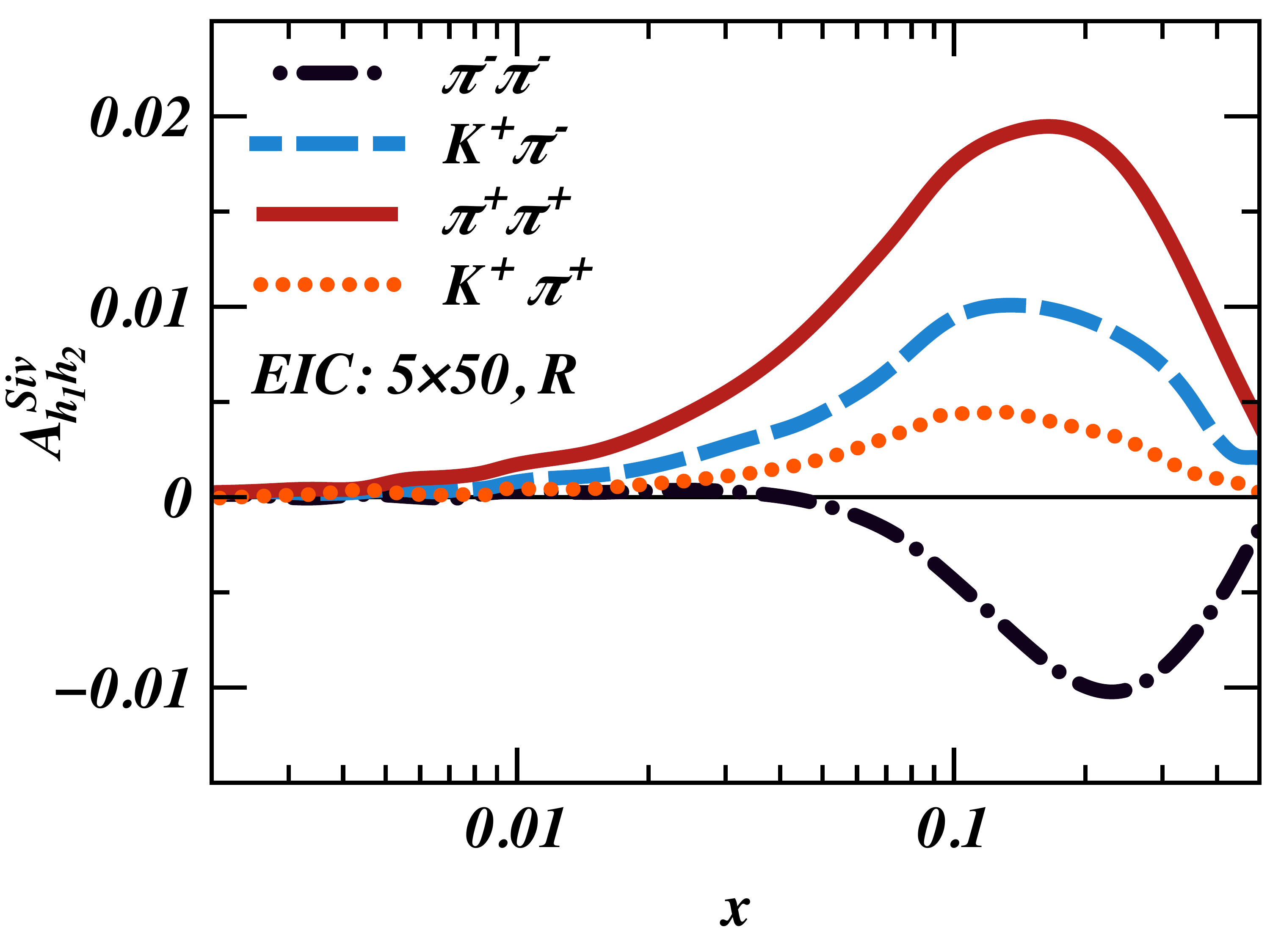}}
\\\vspace{-0.2cm}
\subfigure[] {
\includegraphics[width=\ImM]{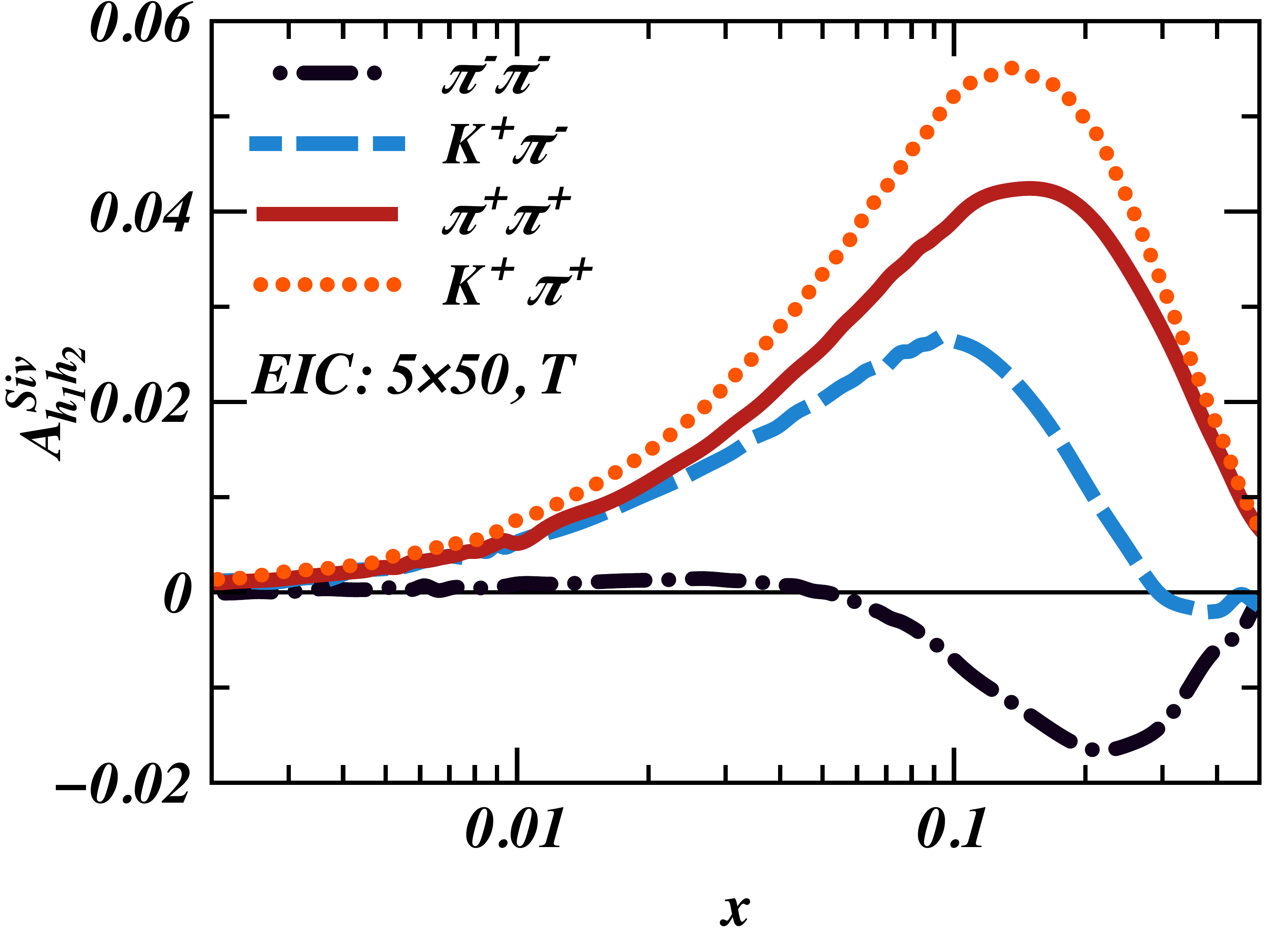}
}
\\\vspace{-0.2cm}
\caption{EIC $5\times50$ results vs $x$ for (a) "R" and (b) "T" type SSAs.}
\label{PLOT_EIC_SSA_2H_X_5x50}
\end{figure}

  Here we present the dihadron SSAs for EIC kinematics. First, we study the $x$  dependence of  "R" and "T" type SSAs for various charged hadron paris for $5\times 50$ kinematics, as depicted in Fig.~\ref{PLOT_EIC_SSA_2H_X_5x50} (a) and (b) respectively. Here again the SSAs involving the $K^+$ are the largest ones. Moreover, we see the different behaviors of these SSAs that would be very important in extracting the Sivers PDF from them.
 
\begin{figure}[tb]
\centering 
\subfigure[] {
\includegraphics[width=\ImM]{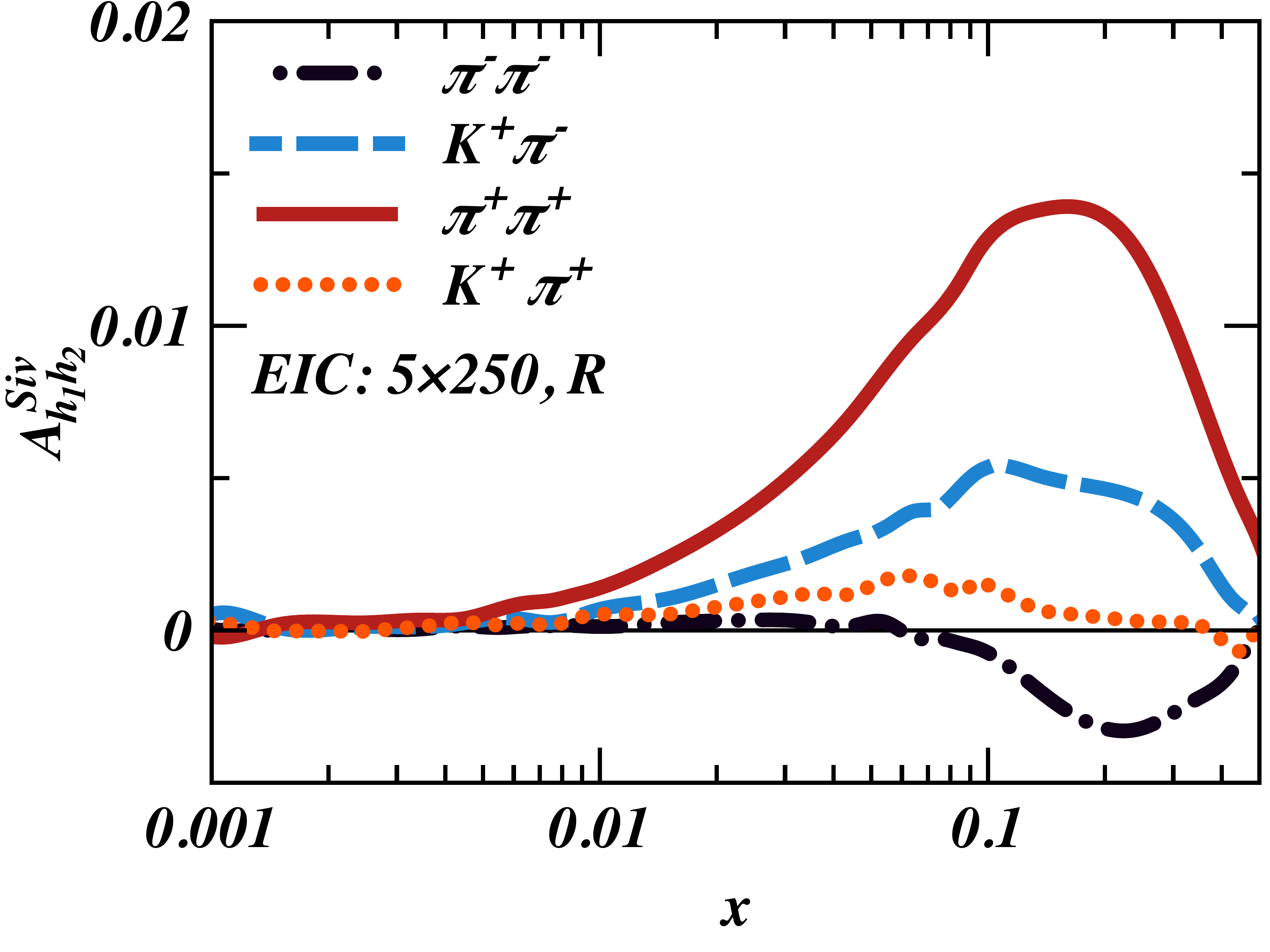}
}
\\\vspace{-0.2cm}
\subfigure[] {
\includegraphics[width=\ImM]{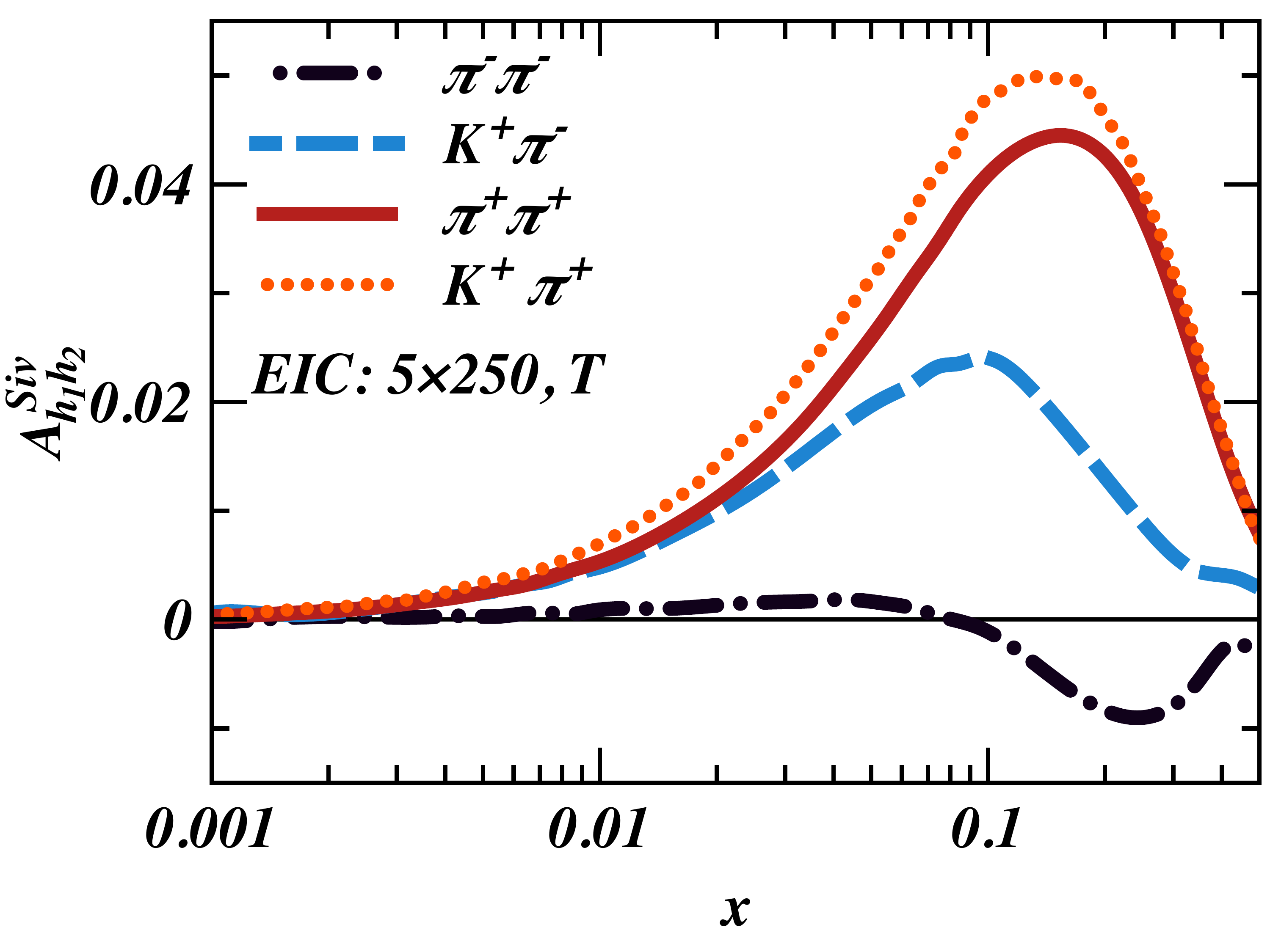}
}
\\\vspace{-0.2cm}
\caption{EIC $5\times250$ results vs $x$ for (a) "R" and (b) "T" type SSAs.}
\label{PLOT_EIC_SSA_2H_X_5x250}
\end{figure}

In Fig.~\ref{PLOT_EIC_SSA_2H_X_5x250} (a) and (b) we present the $5\times250$ results for both "R" and "T" type SSAs as functions of $x$ for the same pairs as in Fig.~\ref{PLOT_EIC_SSA_2H_X_5x50}. Again, similar to the one hadron case, we see decrease of the SSAs as we increase the CM energy. The asymmetric cuts on the momenta of hadron in the pairs should enhance these SSAs (especially in "R" mode), as demonstrated for $\pi^+\pi^-$ pairs in the next subsection. 

\begin{figure}[tbh]
\centering 
\subfigure[] {
\includegraphics[width=\ImM]{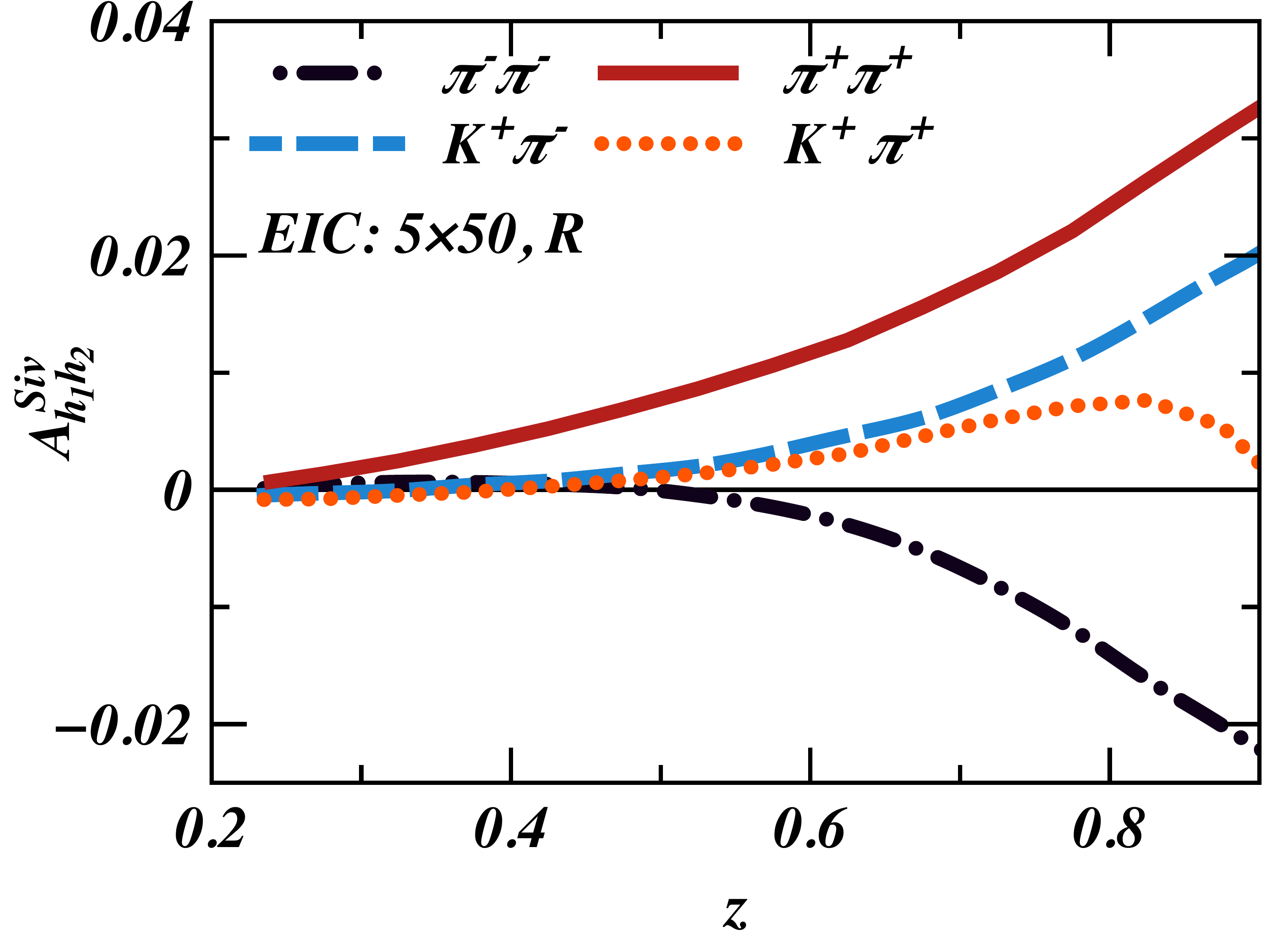}}
\\\vspace{-0.2cm}
\subfigure[] {
\includegraphics[width=\ImM]{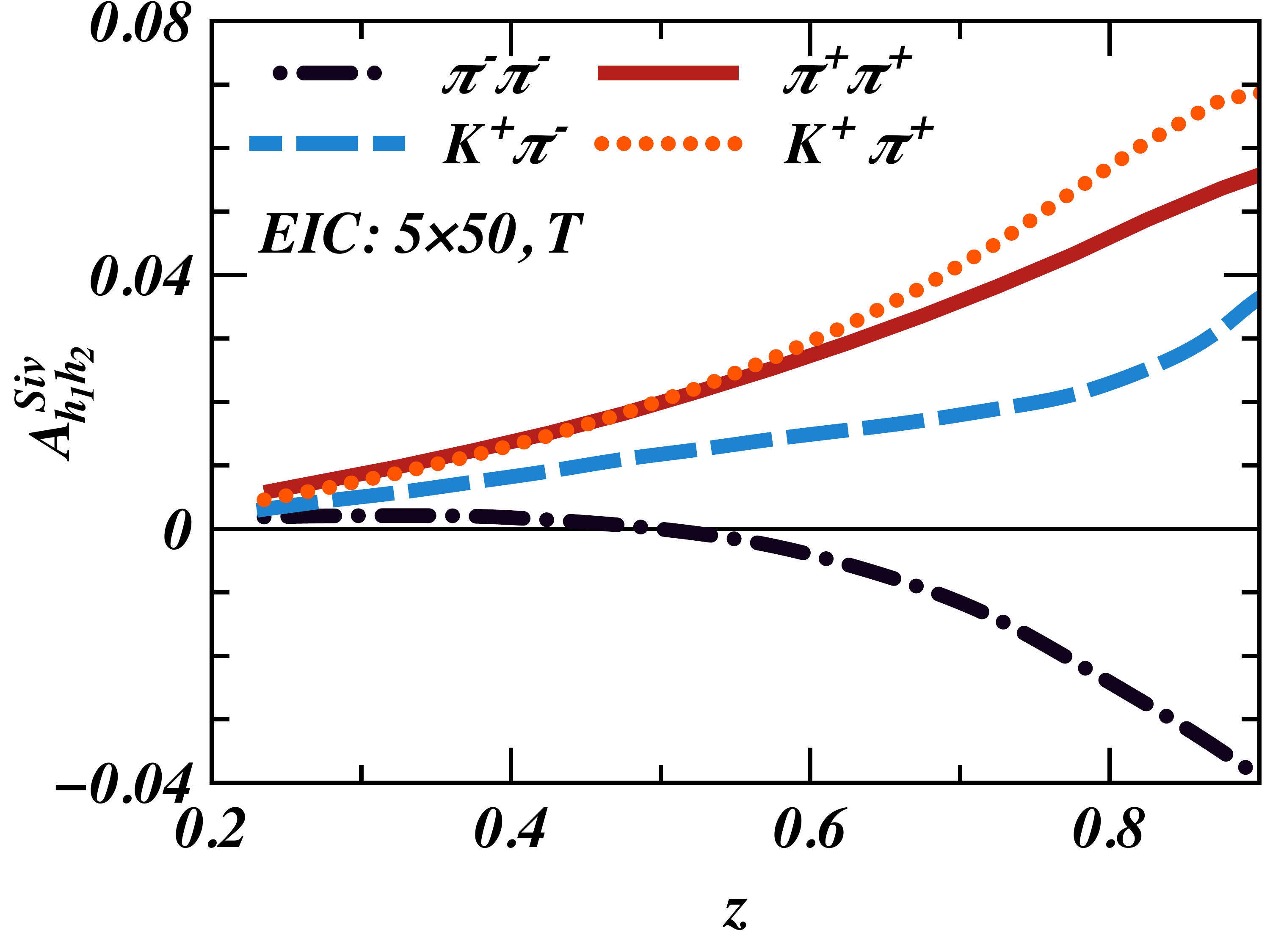}
}
\\\vspace{-0.2cm}
\caption{EIC $5\times50$ results vs $z$ for (a) "R" and (b) "T" type SSAs.}
\label{PLOT_EIC_SSA_2H_Z_5x50}
\end{figure}
%
\begin{figure}[htb]
\centering 
\subfigure[] {
\includegraphics[width=\ImM]{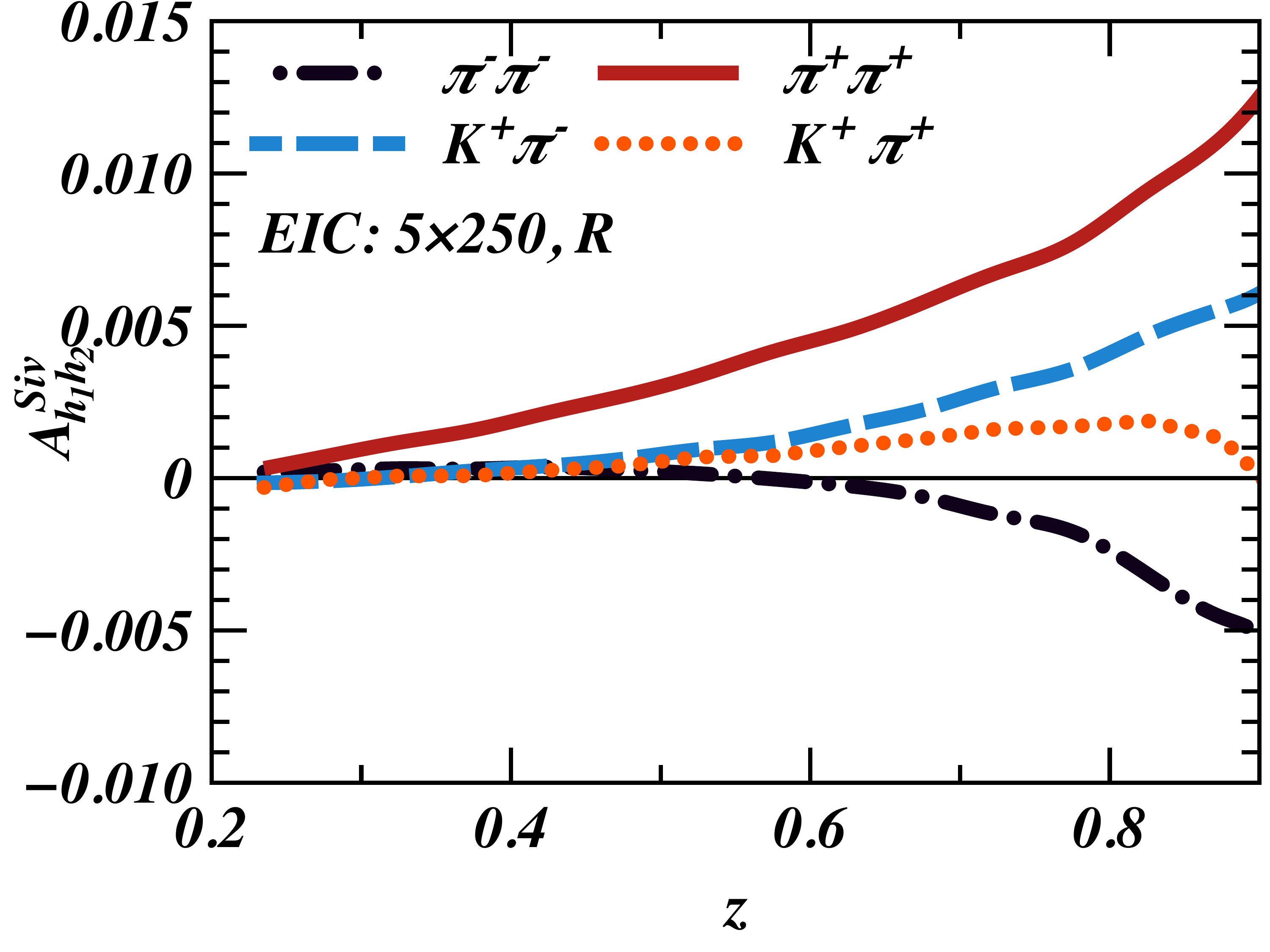}
}
\\\vspace{-0.2cm}
\subfigure[] {
\includegraphics[width=\ImM]{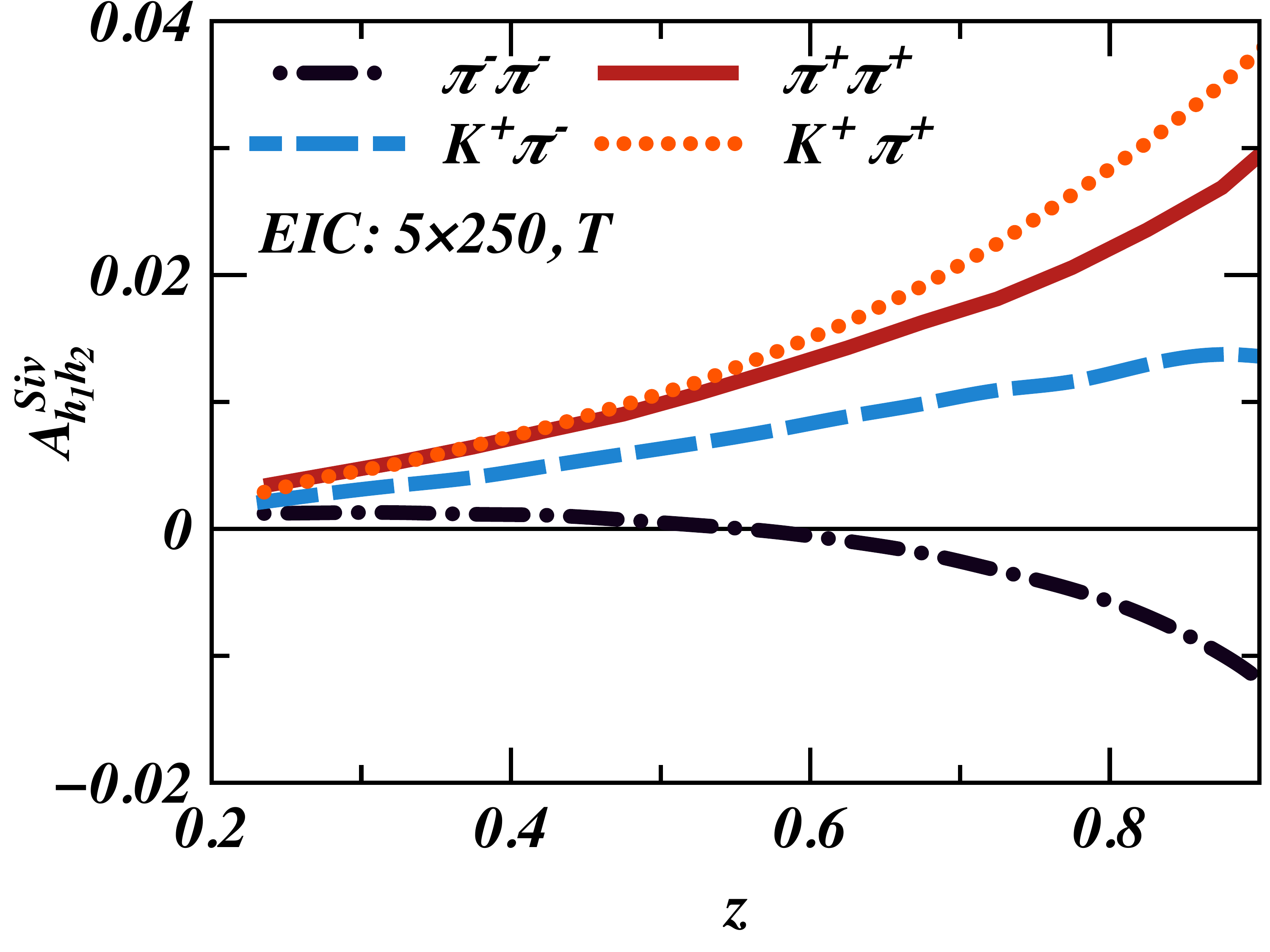}
}
\\\vspace{-0.2cm}
\caption{EIC $5\times250$ results vs $z$ for (a) "R" and (b) "T" type SSAs.}
\label{PLOT_EIC_SSA_2H_Z_5x250}
\end{figure}

The plots in Figs.~\ref{PLOT_EIC_SSA_2H_Z_5x50}, \ref{PLOT_EIC_SSA_2H_Z_5x250} depict the analogous results for both energies, but now as functions of $z$. Here the rise of most SSAs with $z$ can be interpreted by simplistic arguments that the hadrons (and pairs) with larger $z$ carry larger fractions of the struck quark's transverse momentum. 

\begin{figure}[htb]
\centering 
\subfigure[] {
\includegraphics[width=\ImM]{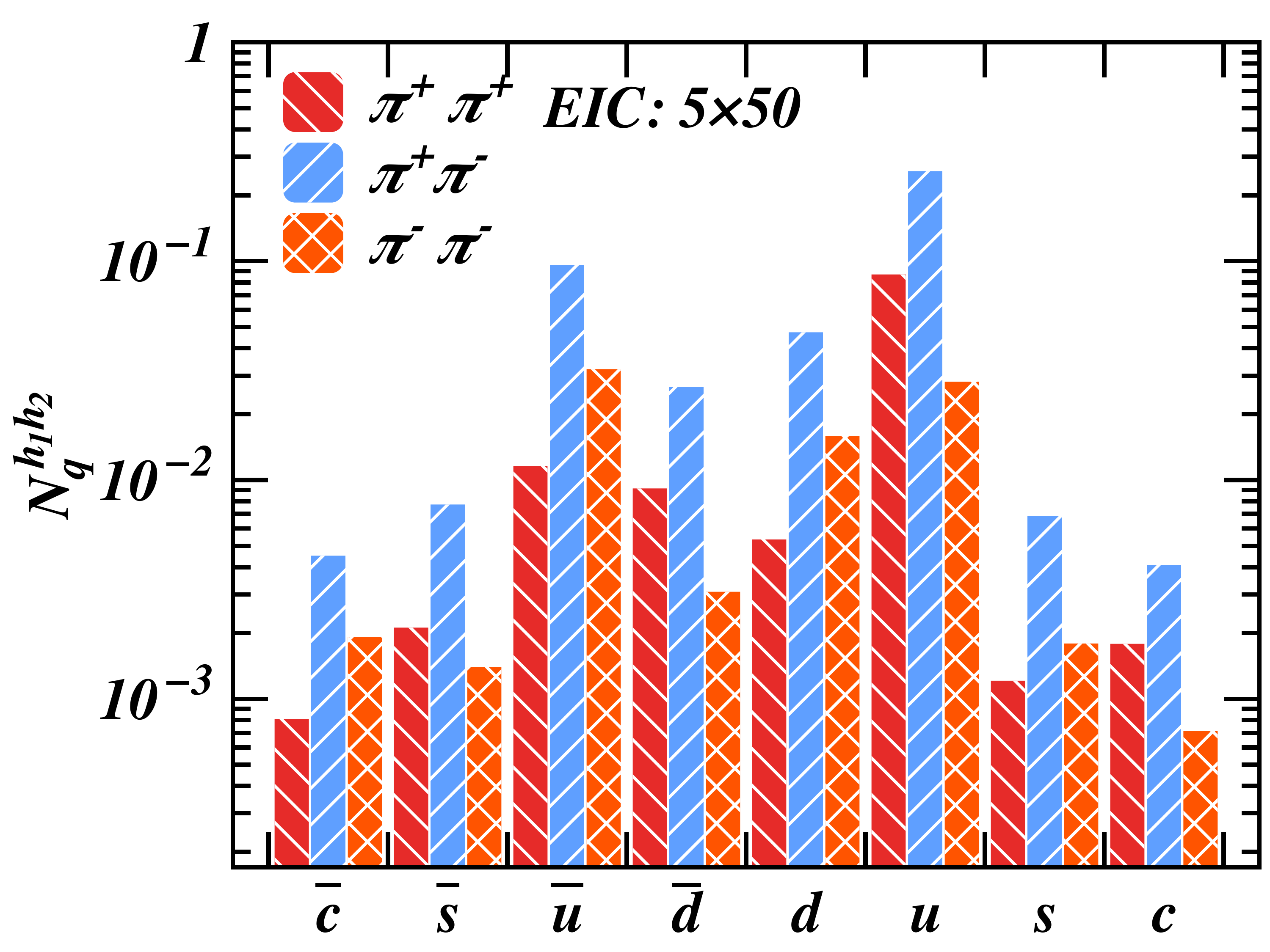}
}
\\\vspace{-0.2cm}
\subfigure[] {
\includegraphics[width=\ImM]{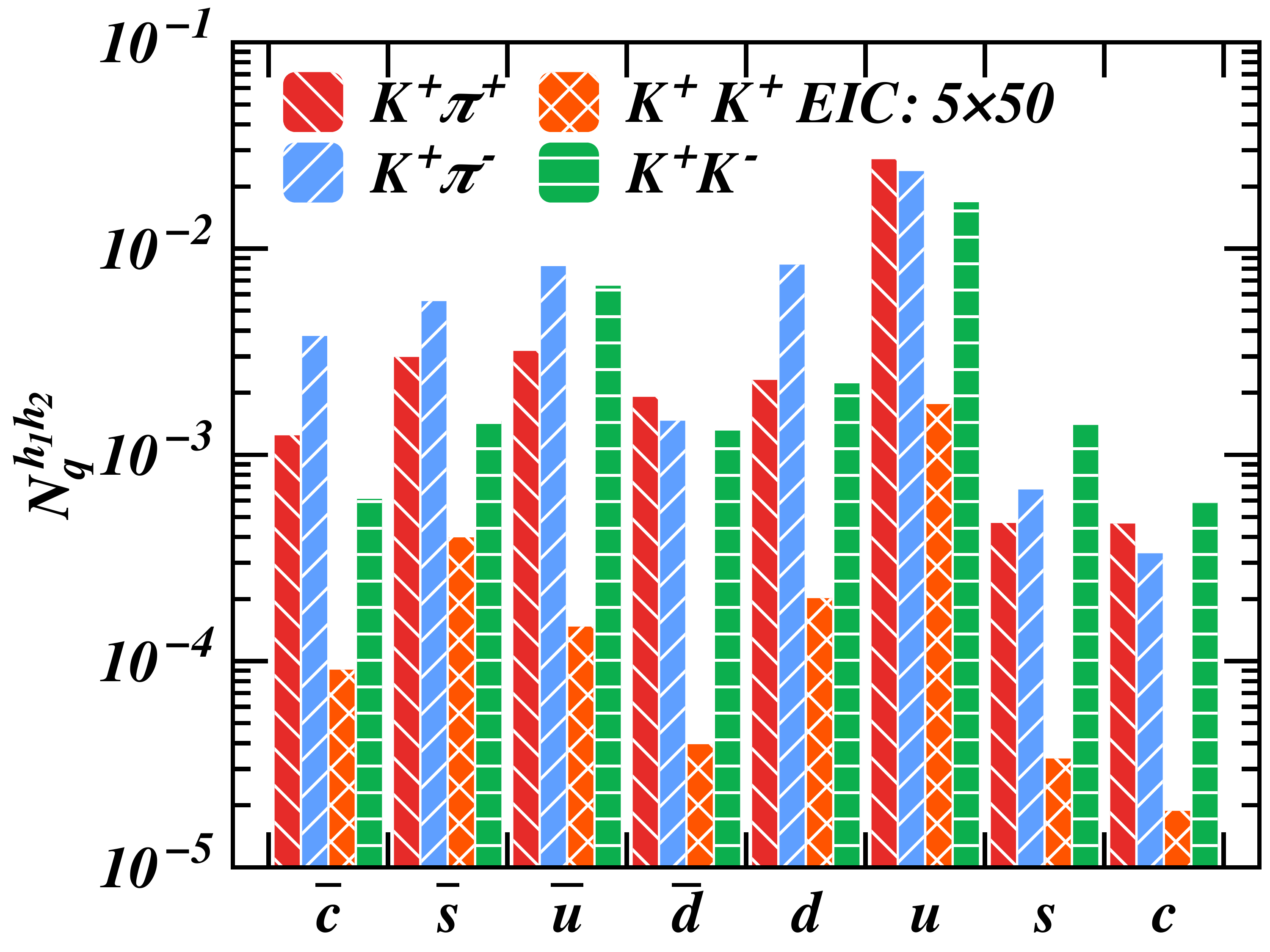}
}
\\\vspace{-0.2cm}
\caption{EIC $5\times50$ predictions for the relative rates for the flavor of the struck quark that produces charged (a) pion and (b) kaon-inclusive pair in MC events with all the relevant kinematical cuts.}
\label{PLOT_EIC_SIV_QUARK_DIHAD_ID}
\end{figure}

 Finally, it would be important to once again examine the relative rates of the different hadron pair production and their dependence on the struck quark's flavor, as depicted on the plots in Fig.~\ref{PLOT_EIC_SIV_QUARK_DIHAD_ID}.  Here we notice a significant increase in the rates of production from all the quarks compared to CLAS12 results shown in Fig.~\ref{PLOT_CLAS12_SIV_QUARK_DIHAD_ID}. Thus EIC should provide a higher precision measurements for the relevant SSAs than CLAS12 for the same number of DIS events. This is of course due to the higher average multiplicities and energies of the produced hadrons.

\subsubsection{Dihadron SSAs with asymmetric cuts}

 Following Refs.~\cite{Kotzinian:2014lsa,Kotzinian:2014gza}, here we also impose asymmetric cuts ($z_1>0.3$ and $P_{1T}>0.3~\Ge$) on the momenta of the pair. These cuts enhance the SSAs, especially the "R" type, but reduce the rate of detecting such pairs and increasing the statistical noise.  Thus the cuts might be optimized to gain the maximum signal to noise ratio for the particular experiment. Here we simply demonstrate the effects of the sample choice of the cuts on the dihadron SSAs.
  
\begin{figure}[htb]
\centering 
\hspace{-0.2cm}
\subfigure[] {
\includegraphics[width=\ImM]{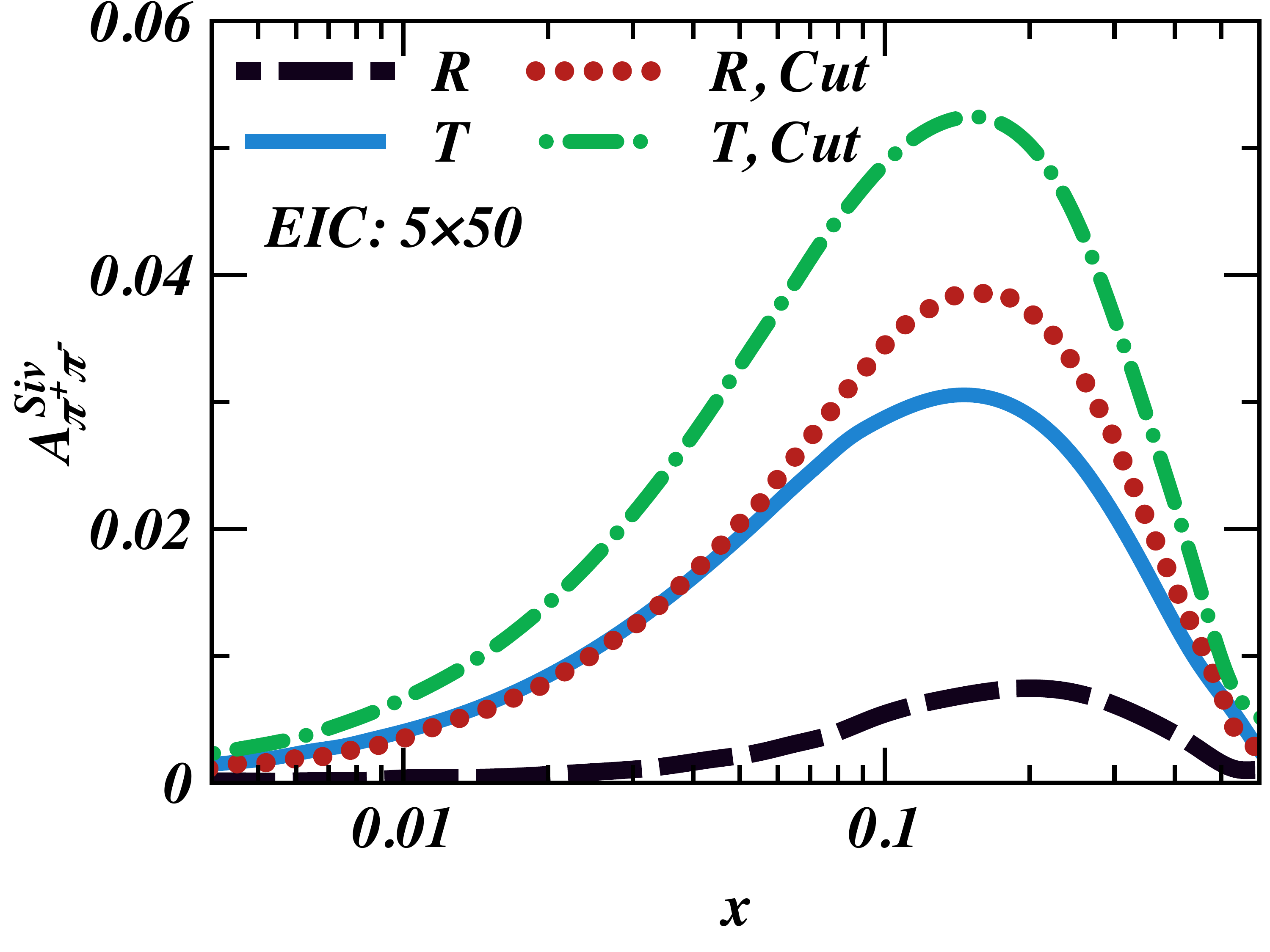}}
\hspace{-0.2cm}
\subfigure[] {
\includegraphics[width=\ImM]{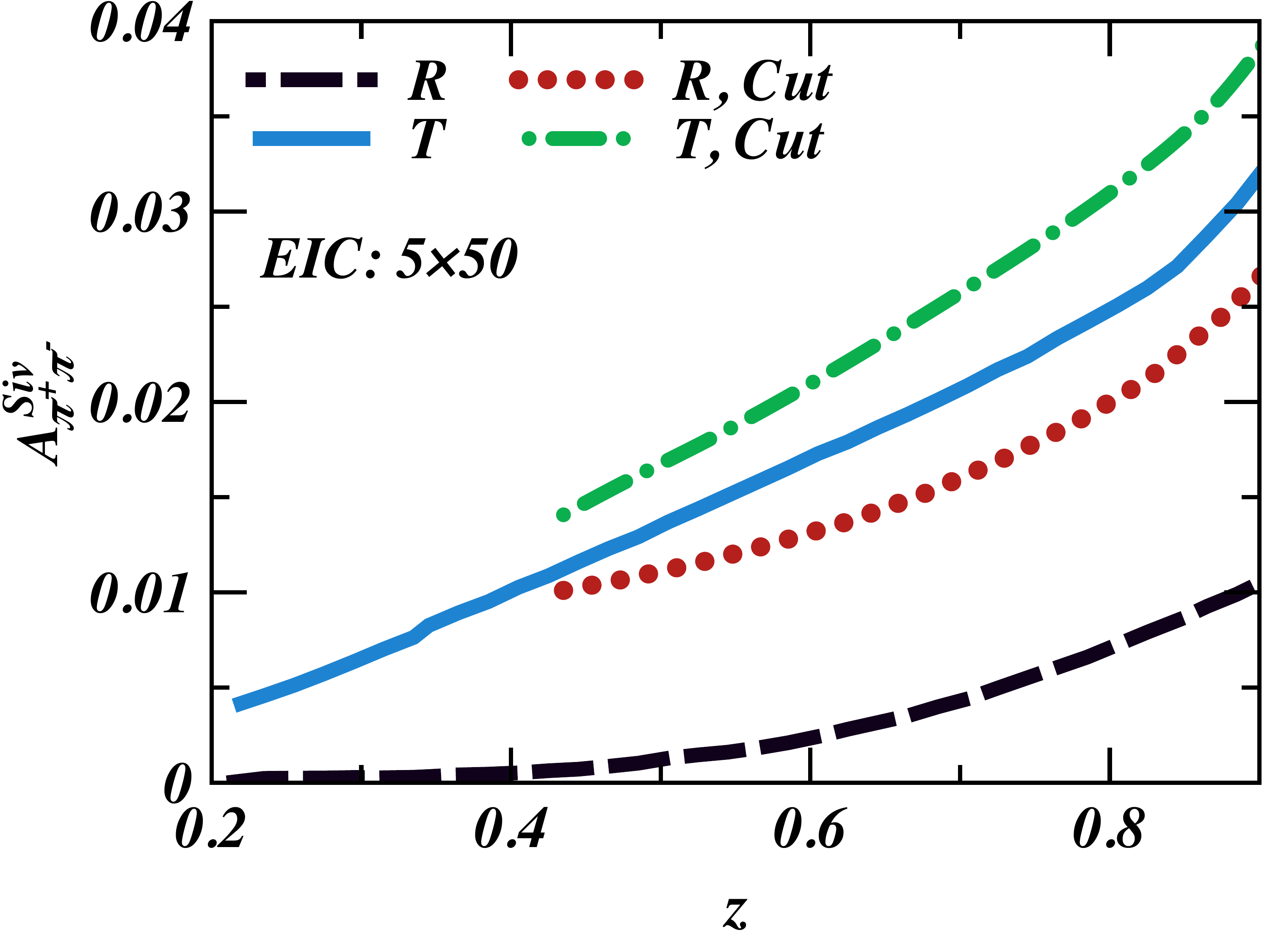}
}
\\\vspace{-0.2cm}
\caption{EIC results for Sivers asymmetry versus (a) $x$ and (b) $z$ with asymmetric cuts applied to the momenta of the hadrons in the pair.}
\label{PLOT_EIC_SIV_2H_X_Z_CUT}
\end{figure}

   Here we present plots analogous to those from the previous subsection, but now with the cuts applied. The  dependences of the SSAs for various charged pairs on $x$ and $z$ are depicted in Fig.~\ref{PLOT_EIC_SIV_2H_X_Z_CUT} (a) and (b), respectively. We note that in general, the SSAs are significantly enhanced, especially those for "R" type modulations. The tradeoff here is that the rates for producing the pairs are reduced by roughly three to four times. Thus a careful optimization for SSA enhancement/rate decrease can be made for particular experiment to achieve the best signal.
   
\section{Conclusions}
\label{SEC_CONC}

 The measurement of the Sivers PDF remains one of the top priorities of the current and future experiments in hadronic physics, with SIDIS on a transversely polarized nucleon target being one of the most viable channels. There are many challenges in achieving this goal, both in measuring the SSAs and extracting the PDF from them using phenomenological analysis of the data. Our goal was to use the state-of-the-art knowledge of the Sivers PDF to make predictions for the upcoming experiments at CLAS12 and the planned EIC. We used the well-established MC event generators \LEPT and \PYTH, modifying them to include the Sivers modulation of the struck quark's transverse momentum using the phenomenologically extracted Sivers PDFs from HERMES and COMPASS measurements.
 
  First we studied the single hadron SIDIS production in Sec.~\ref{SEC_MC_ONE_HADRON}. We used a toy model to study the relevant contributions of sea and valence quarks to the SSAs for EIC kinematics at various energies. At the lower CM energies, the contributions of both sea and valence quarks were comparable in different regions of $x$, while the sea dominates the SSAs for the larger CM energies as one would intuitively expect.  Next, we explored the contributions of non-DIS and parton showering effects on the SSAs, as in most of the phenomenological analyses these are omitted. We found a significant dilution of the Sivers SSAs when these processes were included in the MC simulations, strongly suggesting that the current extraction of the Sivers PDFs are significantly underestimated.  Nevertheless, a comparative analysis of these effects both for COMPASS and EIC showed that these PDFs can be used to estimate the SSAs if we only include the DIS events, since the modifications of both unpolarized and polarized parts of the cross sections are very similar for both kinematics in this model. Then we presented the SSA results for various hadron production at EIC, and also explored their dependence on the CM energy. In the current parametrizations, the sea quark Sivers PDFs are small and poorly constrained, leading to decreasing SSAs with increasing CM energy. Similar results were also presented for CLAS12.
  
  In Sec.~\ref{SEC_MC_TWO_HADRON} we presented the SSA predictions for various charged hadron pairs in both EIC and CLAS12 kinematics. We have noticed that they are of a similar size to the single hadron ones, thus making them accessible experimentally. Further, asymmetric  cuts on the momenta of the hadrons in the pair allow one to significantly enhance these asymmetries at the cost of reduced statistics. Moreover, the dihadron channel provides a large basis for disentangling the flavor dependence of the Sivers PDF. The measurement itself is complementary to the highly publicized two hadron studies of the transversity PDF via the interference fragmentation mechanism, with several current analyses of the experimental data and a number of proposals for future experiments using CLAS12 and SoLID at JLab. In fact, the dihadron Sivers analysis has been already included in the SoLID proposal~\cite{JLAB12:SOLID} for transversity measurements on a neutron target, that would provide very exciting test for the Sivers PDFs extracted from proton and deuteron data. On the phenological side, these dihadron measurements would complement the single hadron measurements in the global data fits. The unknown, fully unintegrated dihadron fragmentation functions needed for such analysis also need study, but a very similar challenge exists also for the dihadron transversity measurements (there we need two dihadron FFs). Then these dihadron FFs can be studied in a similar fashion, such as using $e^+e^-$ data on dihadron multiplicities, MC and model studies.
  
  The measurements of the SSAs for hadrons produced in the TFR will allow one to study the structure of the nucleon through the fracture functions. These objects, though more complicated than the ordinary PDFs and FFs, will provide important new information. The hadronization in \MPYTH and \MLEPT is governed by the Lund string model, that differs from the usual QCD factorized approach that describes the hadron production in the CFR with a convolution of PDFs and FFs and in the TFR using two additional independent fracture functions. In particular, in \MPYTH we take into account only the correlation between the nucleon's transverse polarization and the transverse momentum of the struck quark, which propagates to both CFR and TFR hadrons by the Lund fragmentation model. On the other hand, in the TFR factorized formalism there are two independent fracture functions describing the correlation of both the struck quark's and the produced hadron's transverse momenta with the transverse spin of the nucleon. Thus it would be crucial to test experimentally, whether the unified description of Sivers SSAs in both CFR and TFR can be achieved with \MPYTH. Our results of the Sivers SSAs for charged hadrons showed a sizable signal in the TFR, comparable in size to that in the CFR.
  
  In conclusion, we have developed a tool based on very successful MC generators for studying the Sivers effect in experiments. Our Sivers SSA predictions for both CLAS12 and EIC indicate that these should be accessible experimentally and will provide information on the Sivers PDF in a wide range of kinematical variables. We noted in particular the power of dihadron studies in pinning down the dependence of the Sivers PDF on the flavor of the struck quark. We did not include the TMD evolution of the Sivers PDF that would weaken the signal at EIC energies. Further, a detailed analysis of the real world experimental limitations, such as detector acceptance effects, statistics with the specific proposed time frame of data collection, etc, will of course create more challenges in their measurements but these considerations are out of scope of this article. We have also demonstrated using the \MPYTH MC generator, that the current extractions of the Sivers PDF might be significantly underestimated. Thus improvements in estimating the true relative size of the Sivers term in the cross section would be needed in such future analyses.  
  
\section*{Acknowledgements}

 We would like to thank Stefano Melis for providing us with the parameters of the fits for the Sivers function for the sea quarks. This work was supported by the Australian Research Council through Grants No. FL0992247 (AWT), No. CE110001004 (CoEPP) and by the University of Adelaide. A.K. was partially supported by INFN Torino unit, and thanks CSSM and CoEPP at the University of Adelaide for the hospitality during his visit when part of this work has been completed. E.C.A. acknowledges support by the U.S. Department of Energy under Contract No. {DE-SC0012704}. The Jefferson Science Associates (JSA) operates the Thomas Jefferson National Accelerator Facility for the United States Department of Energy under Contract No. DE-AC05-06OR23177. 


\bibliographystyle{apsrev}
\bibliography{fragment}

\begin{thebibliography}{37}
\expandafter\ifx\csname natexlab\endcsname\relax\def\natexlab#1{#1}\fi
\expandafter\ifx\csname bibnamefont\endcsname\relax
  \def\bibnamefont#1{#1}\fi
\expandafter\ifx\csname bibfnamefont\endcsname\relax
  \def\bibfnamefont#1{#1}\fi
\expandafter\ifx\csname citenamefont\endcsname\relax
  \def\citenamefont#1{#1}\fi
\expandafter\ifx\csname url\endcsname\relax
  \def\url#1{\texttt{#1}}\fi
\expandafter\ifx\csname urlprefix\endcsname\relax\def\urlprefix{URL }\fi
\providecommand{\bibinfo}[2]{#2}
\providecommand{\eprint}[2][]{\url{#2}}

\bibitem[{\citenamefont{Sivers}(1990)}]{Sivers:1989cc}
\bibinfo{author}{\bibfnamefont{D.~W.} \bibnamefont{Sivers}},
  \bibinfo{journal}{Phys.Rev.} \textbf{\bibinfo{volume}{D41}},
  \bibinfo{pages}{83} (\bibinfo{year}{1990}).

\bibitem[{\citenamefont{Anselmino
  et~al.}(2005{\natexlab{a}})\citenamefont{Anselmino, Boglione, D'Alesio,
  Kotzinian, Murgia et~al.}}]{Anselmino:2005nn}
\bibinfo{author}{\bibfnamefont{M.}~\bibnamefont{Anselmino}},
  \bibinfo{author}{\bibfnamefont{M.}~\bibnamefont{Boglione}},
  \bibinfo{author}{\bibfnamefont{U.}~\bibnamefont{D'Alesio}},
  \bibinfo{author}{\bibfnamefont{A.}~\bibnamefont{Kotzinian}},
  \bibinfo{author}{\bibfnamefont{F.}~\bibnamefont{Murgia}},
  \bibnamefont{et~al.}, \bibinfo{journal}{Phys.Rev.}
  \textbf{\bibinfo{volume}{D71}}, \bibinfo{pages}{074006}
  (\bibinfo{year}{2005}{\natexlab{a}}), \eprint{hep-ph/0501196}.

\bibitem[{\citenamefont{Airapetian et~al.}(2009)}]{Airapetian:2009ae}
\bibinfo{author}{\bibfnamefont{A.}~\bibnamefont{Airapetian}}
  \bibnamefont{et~al.} (\bibinfo{collaboration}{HERMES Collaboration}),
  \bibinfo{journal}{Phys.Rev.Lett.} \textbf{\bibinfo{volume}{103}},
  \bibinfo{pages}{152002} (\bibinfo{year}{2009}), \eprint{0906.3918}.

\bibitem[{\citenamefont{Airapetian et~al.}(2005)}]{Airapetian:2004tw}
\bibinfo{author}{\bibfnamefont{A.}~\bibnamefont{Airapetian}}
  \bibnamefont{et~al.} (\bibinfo{collaboration}{HERMES Collaboration}),
  \bibinfo{journal}{Phys.Rev.Lett.} \textbf{\bibinfo{volume}{94}},
  \bibinfo{pages}{012002} (\bibinfo{year}{2005}), \eprint{hep-ex/0408013}.

\bibitem[{\citenamefont{Adolph et~al.}(2012{\natexlab{a}})}]{Adolph:2012sp}
\bibinfo{author}{\bibfnamefont{C.}~\bibnamefont{Adolph}} \bibnamefont{et~al.}
  (\bibinfo{collaboration}{COMPASS Collaboration}),
  \bibinfo{journal}{Phys.Lett.} \textbf{\bibinfo{volume}{B717}},
  \bibinfo{pages}{383} (\bibinfo{year}{2012}{\natexlab{a}}),
  \eprint{1205.5122}.

\bibitem[{\citenamefont{Adolph et~al.}(2014)\citenamefont{Adolph, Akhunzyanov,
  Alexeev, Alexeev, Amoroso et~al.}}]{Adolph:2014zba}
\bibinfo{author}{\bibfnamefont{C.}~\bibnamefont{Adolph}},
  \bibinfo{author}{\bibfnamefont{R.}~\bibnamefont{Akhunzyanov}},
  \bibinfo{author}{\bibfnamefont{M.}~\bibnamefont{Alexeev}},
  \bibinfo{author}{\bibfnamefont{G.}~\bibnamefont{Alexeev}},
  \bibinfo{author}{\bibfnamefont{A.}~\bibnamefont{Amoroso}},
  \bibnamefont{et~al.} (\bibinfo{year}{2014}), \eprint{1408.4405}.

\bibitem[{\citenamefont{Qian et~al.}(2011)}]{Qian:2011py}
\bibinfo{author}{\bibfnamefont{X.}~\bibnamefont{Qian}} \bibnamefont{et~al.}
  (\bibinfo{collaboration}{Jefferson Lab Hall A Collaboration}),
  \bibinfo{journal}{Phys.Rev.Lett.} \textbf{\bibinfo{volume}{107}},
  \bibinfo{pages}{072003} (\bibinfo{year}{2011}), \eprint{1106.0363}.

\bibitem[{\citenamefont{Anselmino et~al.}(2009)}]{Anselmino:2008sga}
\bibinfo{author}{\bibfnamefont{M.}~\bibnamefont{Anselmino}}
  \bibnamefont{et~al.}, \bibinfo{journal}{Eur. Phys. J.}
  \textbf{\bibinfo{volume}{A39}}, \bibinfo{pages}{89} (\bibinfo{year}{2009}),
  \eprint{0805.2677}.

\bibitem[{\citenamefont{Collins and Rogers}(2015)}]{Collins:2014jpa}
\bibinfo{author}{\bibfnamefont{J.}~\bibnamefont{Collins}} \bibnamefont{and}
  \bibinfo{author}{\bibfnamefont{T.}~\bibnamefont{Rogers}},
  \bibinfo{journal}{Phys.Rev.} \textbf{\bibinfo{volume}{D91}},
  \bibinfo{pages}{074020} (\bibinfo{year}{2015}), \eprint{1412.3820}.

\bibitem[{\citenamefont{Anselmino et~al.}(2012)\citenamefont{Anselmino,
  Boglione, and Melis}}]{Anselmino:2012aa}
\bibinfo{author}{\bibfnamefont{M.}~\bibnamefont{Anselmino}},
  \bibinfo{author}{\bibfnamefont{M.}~\bibnamefont{Boglione}}, \bibnamefont{and}
  \bibinfo{author}{\bibfnamefont{S.}~\bibnamefont{Melis}},
  \bibinfo{journal}{Phys.Rev.} \textbf{\bibinfo{volume}{D86}},
  \bibinfo{pages}{014028} (\bibinfo{year}{2012}), \eprint{1204.1239}.

\bibitem[{\citenamefont{Echevarria et~al.}(2014)\citenamefont{Echevarria,
  Idilbi, Kang, and Vitev}}]{Echevarria:2014xaa}
\bibinfo{author}{\bibfnamefont{M.~G.} \bibnamefont{Echevarria}},
  \bibinfo{author}{\bibfnamefont{A.}~\bibnamefont{Idilbi}},
  \bibinfo{author}{\bibfnamefont{Z.-B.} \bibnamefont{Kang}}, \bibnamefont{and}
  \bibinfo{author}{\bibfnamefont{I.}~\bibnamefont{Vitev}},
  \bibinfo{journal}{Phys.Rev.} \textbf{\bibinfo{volume}{D89}},
  \bibinfo{pages}{074013} (\bibinfo{year}{2014}), \eprint{1401.5078}.

\bibitem[{\citenamefont{Matevosyan
  et~al.}(2011{\natexlab{a}})\citenamefont{Matevosyan, Thomas, and
  Bentz}}]{Matevosyan:2011zza}
\bibinfo{author}{\bibfnamefont{H.~H.} \bibnamefont{Matevosyan}},
  \bibinfo{author}{\bibfnamefont{A.~W.} \bibnamefont{Thomas}},
  \bibnamefont{and} \bibinfo{author}{\bibfnamefont{W.}~\bibnamefont{Bentz}},
  \bibinfo{journal}{AIP Conf.Proc.} \textbf{\bibinfo{volume}{1374}},
  \bibinfo{pages}{387} (\bibinfo{year}{2011}{\natexlab{a}}).

\bibitem[{\citenamefont{Matevosyan
  et~al.}(2011{\natexlab{b}})\citenamefont{Matevosyan, Thomas, and
  Bentz}}]{Matevosyan:2010hh}
\bibinfo{author}{\bibfnamefont{H.~H.} \bibnamefont{Matevosyan}},
  \bibinfo{author}{\bibfnamefont{A.~W.} \bibnamefont{Thomas}},
  \bibnamefont{and} \bibinfo{author}{\bibfnamefont{W.}~\bibnamefont{Bentz}},
  \bibinfo{journal}{Phys.Rev.} \textbf{\bibinfo{volume}{D83}},
  \bibinfo{pages}{074003} (\bibinfo{year}{2011}{\natexlab{b}}),
  \eprint{1011.1052}.

\bibitem[{\citenamefont{Kotzinian
  et~al.}(2014{\natexlab{a}})\citenamefont{Kotzinian, Matevosyan, and
  Thomas}}]{Kotzinian:2014lsa}
\bibinfo{author}{\bibfnamefont{A.}~\bibnamefont{Kotzinian}},
  \bibinfo{author}{\bibfnamefont{H.~H.} \bibnamefont{Matevosyan}},
  \bibnamefont{and} \bibinfo{author}{\bibfnamefont{A.~W.}
  \bibnamefont{Thomas}}, \bibinfo{journal}{Phys.Rev.Lett.}
  \textbf{\bibinfo{volume}{113}}, \bibinfo{pages}{062003}
  (\bibinfo{year}{2014}{\natexlab{a}}), \eprint{1403.5562}.

\bibitem[{\citenamefont{Kotzinian
  et~al.}(2014{\natexlab{b}})\citenamefont{Kotzinian, Matevosyan, and
  Thomas}}]{Kotzinian:2014gza}
\bibinfo{author}{\bibfnamefont{A.}~\bibnamefont{Kotzinian}},
  \bibinfo{author}{\bibfnamefont{H.~H.} \bibnamefont{Matevosyan}},
  \bibnamefont{and} \bibinfo{author}{\bibfnamefont{A.~W.}
  \bibnamefont{Thomas}}, \bibinfo{journal}{Phys.Rev.}
  \textbf{\bibinfo{volume}{D90}}, \bibinfo{pages}{074006}
  (\bibinfo{year}{2014}{\natexlab{b}}), \eprint{1405.5059}.

\bibitem[{\citenamefont{Kotzinian
  et~al.}(2014{\natexlab{c}})\citenamefont{Kotzinian, Matevosyan, and
  Thomas}}]{Kotzinian:2014hoa}
\bibinfo{author}{\bibfnamefont{A.}~\bibnamefont{Kotzinian}},
  \bibinfo{author}{\bibfnamefont{H.~H.} \bibnamefont{Matevosyan}},
  \bibnamefont{and} \bibinfo{author}{\bibfnamefont{A.~W.} \bibnamefont{Thomas}}
  (\bibinfo{year}{2014}{\natexlab{c}}), \eprint{1407.6572}.

\bibitem[{JLA({\natexlab{a}})}]{JLAB12:CLAS12}
\bibinfo{howpublished}{\url{https://www.jlab.org/exp_prog/proposals/12/PR12-12-009.pdf}}.

\bibitem[{JLA({\natexlab{b}})}]{JLAB12:SOLID}
\bibinfo{howpublished}{\url{https://www.jlab.org/exp_prog/proposals/14/E12-10-006A.pdf}}.

\bibitem[{\citenamefont{Kotzinian}(2014)}]{Kotzinian:2014uya}
\bibinfo{author}{\bibfnamefont{A.}~\bibnamefont{Kotzinian}}
  (\bibinfo{year}{2014}), \eprint{1408.6674}.

\bibitem[{\citenamefont{Accardi et~al.}(2012)\citenamefont{Accardi, Albacete,
  Anselmino, Armesto, Aschenauer et~al.}}]{Accardi:2012qut}
\bibinfo{author}{\bibfnamefont{A.}~\bibnamefont{Accardi}},
  \bibinfo{author}{\bibfnamefont{J.}~\bibnamefont{Albacete}},
  \bibinfo{author}{\bibfnamefont{M.}~\bibnamefont{Anselmino}},
  \bibinfo{author}{\bibfnamefont{N.}~\bibnamefont{Armesto}},
  \bibinfo{author}{\bibfnamefont{E.}~\bibnamefont{Aschenauer}},
  \bibnamefont{et~al.} (\bibinfo{year}{2012}), \eprint{1212.1701}.

\bibitem[{\citenamefont{Bacchetta et~al.}(2007)\citenamefont{Bacchetta, Diehl,
  Goeke, Metz, Mulders et~al.}}]{Bacchetta:2006tn}
\bibinfo{author}{\bibfnamefont{A.}~\bibnamefont{Bacchetta}},
  \bibinfo{author}{\bibfnamefont{M.}~\bibnamefont{Diehl}},
  \bibinfo{author}{\bibfnamefont{K.}~\bibnamefont{Goeke}},
  \bibinfo{author}{\bibfnamefont{A.}~\bibnamefont{Metz}},
  \bibinfo{author}{\bibfnamefont{P.~J.} \bibnamefont{Mulders}},
  \bibnamefont{et~al.}, \bibinfo{journal}{JHEP}
  \textbf{\bibinfo{volume}{0702}}, \bibinfo{pages}{093} (\bibinfo{year}{2007}),
  \eprint{hep-ph/0611265}.

\bibitem[{\citenamefont{Bianconi et~al.}(2000)\citenamefont{Bianconi, Boffi,
  Jakob, and Radici}}]{Bianconi:1999cd}
\bibinfo{author}{\bibfnamefont{A.}~\bibnamefont{Bianconi}},
  \bibinfo{author}{\bibfnamefont{S.}~\bibnamefont{Boffi}},
  \bibinfo{author}{\bibfnamefont{R.}~\bibnamefont{Jakob}}, \bibnamefont{and}
  \bibinfo{author}{\bibfnamefont{M.}~\bibnamefont{Radici}},
  \bibinfo{journal}{Phys.Rev.} \textbf{\bibinfo{volume}{D62}},
  \bibinfo{pages}{034008} (\bibinfo{year}{2000}), \eprint{hep-ph/9907475}.

\bibitem[{\citenamefont{Airapetian et~al.}(2008)}]{Airapetian:2008sk}
\bibinfo{author}{\bibfnamefont{A.}~\bibnamefont{Airapetian}}
  \bibnamefont{et~al.} (\bibinfo{collaboration}{HERMES Collaboration}),
  \bibinfo{journal}{JHEP} \textbf{\bibinfo{volume}{06}}, \bibinfo{pages}{017}
  (\bibinfo{year}{2008}), \eprint{0803.2367}.

\bibitem[{\citenamefont{Adolph et~al.}(2012{\natexlab{b}})}]{Adolph:2012nw}
\bibinfo{author}{\bibfnamefont{C.}~\bibnamefont{Adolph}} \bibnamefont{et~al.}
  (\bibinfo{collaboration}{COMPASS Collaboration}),
  \bibinfo{journal}{Phys.Lett.} \textbf{\bibinfo{volume}{B713}},
  \bibinfo{pages}{10} (\bibinfo{year}{2012}{\natexlab{b}}), \eprint{1202.6150}.

\bibitem[{\citenamefont{Trentadue and Veneziano}(1994)}]{Trentadue:1993ka}
\bibinfo{author}{\bibfnamefont{L.}~\bibnamefont{Trentadue}} \bibnamefont{and}
  \bibinfo{author}{\bibfnamefont{G.}~\bibnamefont{Veneziano}},
  \bibinfo{journal}{Phys.Lett.} \textbf{\bibinfo{volume}{B323}},
  \bibinfo{pages}{201} (\bibinfo{year}{1994}).

\bibitem[{\citenamefont{Anselmino et~al.}(2011)\citenamefont{Anselmino, Barone,
  and Kotzinian}}]{Anselmino:2011ss}
\bibinfo{author}{\bibfnamefont{M.}~\bibnamefont{Anselmino}},
  \bibinfo{author}{\bibfnamefont{V.}~\bibnamefont{Barone}}, \bibnamefont{and}
  \bibinfo{author}{\bibfnamefont{A.}~\bibnamefont{Kotzinian}},
  \bibinfo{journal}{Phys.Lett.} \textbf{\bibinfo{volume}{B699}},
  \bibinfo{pages}{108} (\bibinfo{year}{2011}), \eprint{1102.4214}.

\bibitem[{\citenamefont{Kotzinian et~al.}(2012)\citenamefont{Kotzinian,
  Anselmino, and Barone}}]{Kotzinian:2011av}
\bibinfo{author}{\bibfnamefont{A.}~\bibnamefont{Kotzinian}},
  \bibinfo{author}{\bibfnamefont{M.}~\bibnamefont{Anselmino}},
  \bibnamefont{and} \bibinfo{author}{\bibfnamefont{V.}~\bibnamefont{Barone}},
  \bibinfo{journal}{Nuovo Cim.} \textbf{\bibinfo{volume}{C035N2}},
  \bibinfo{pages}{85} (\bibinfo{year}{2012}), \eprint{1110.5256}.

\bibitem[{\citenamefont{Diefenthaler}(2005)}]{Diefenthaler:2005gx}
\bibinfo{author}{\bibfnamefont{M.}~\bibnamefont{Diefenthaler}}
  (\bibinfo{collaboration}{HERMES Collaboration}), \bibinfo{journal}{AIP
  Conf.Proc.} \textbf{\bibinfo{volume}{792}}, \bibinfo{pages}{933}
  (\bibinfo{year}{2005}), \eprint{hep-ex/0507013}.

\bibitem[{\citenamefont{Alexakhin et~al.}(2005)}]{Alexakhin:2005iw}
\bibinfo{author}{\bibfnamefont{V.}~\bibnamefont{Alexakhin}}
  \bibnamefont{et~al.} (\bibinfo{collaboration}{COMPASS Collaboration}),
  \bibinfo{journal}{Phys.Rev.Lett.} \textbf{\bibinfo{volume}{94}},
  \bibinfo{pages}{202002} (\bibinfo{year}{2005}), \eprint{hep-ex/0503002}.

\bibitem[{\citenamefont{Bradamante}(2012)}]{Bradamante:2011xu}
\bibinfo{author}{\bibfnamefont{F.}~\bibnamefont{Bradamante}}
  (\bibinfo{collaboration}{COMPASS Collaboration}), \bibinfo{journal}{Nuovo
  Cim.} \textbf{\bibinfo{volume}{C035N2}}, \bibinfo{pages}{107}
  (\bibinfo{year}{2012}), \eprint{1111.0869}.

\bibitem[{\citenamefont{Kotzinian}(2005{\natexlab{a}})}]{Kotzinian:2005zs}
\bibinfo{author}{\bibfnamefont{A.}~\bibnamefont{Kotzinian}},
  \bibinfo{journal}{Proceedings of Transversity 2005,} pp.
  \bibinfo{pages}{228--235} (\bibinfo{year}{2005}{\natexlab{a}}),
  \eprint{hep-ph/0510359}.

\bibitem[{\citenamefont{Kotzinian}(2005{\natexlab{b}})}]{Kotzinian:2005zg}
\bibinfo{author}{\bibfnamefont{A.}~\bibnamefont{Kotzinian}}
  (\bibinfo{year}{2005}{\natexlab{b}}), \eprint{hep-ph/0504081}.

\bibitem[{\citenamefont{Avakian et~al.}(2012)}]{CLAS12:2012DH}
\bibinfo{author}{\bibfnamefont{H.}~\bibnamefont{Avakian}} \bibnamefont{et~al.}
  (\bibinfo{year}{2012}), \eprint{PR12-12-009}.

\bibitem[{EIC()}]{EIC:BNL}
\bibinfo{howpublished}{\url{https://wiki.bnl.gov/eic/}}.

\bibitem[{\citenamefont{Aschenauer et~al.}(2014)\citenamefont{Aschenauer,
  Baker, Bazilevsky, Boyle, Belomestnykh et~al.}}]{Aschenauer:2014cki}
\bibinfo{author}{\bibfnamefont{E.}~\bibnamefont{Aschenauer}},
  \bibinfo{author}{\bibfnamefont{M.}~\bibnamefont{Baker}},
  \bibinfo{author}{\bibfnamefont{A.}~\bibnamefont{Bazilevsky}},
  \bibinfo{author}{\bibfnamefont{K.}~\bibnamefont{Boyle}},
  \bibinfo{author}{\bibfnamefont{S.}~\bibnamefont{Belomestnykh}},
  \bibnamefont{et~al.} (\bibinfo{year}{2014}), \eprint{1409.1633}.

\bibitem[{\citenamefont{Anselmino
  et~al.}(2005{\natexlab{b}})\citenamefont{Anselmino, Boglione, D'Alesio,
  Kotzinian, Murgia et~al.}}]{Anselmino:2005ea}
\bibinfo{author}{\bibfnamefont{M.}~\bibnamefont{Anselmino}},
  \bibinfo{author}{\bibfnamefont{M.}~\bibnamefont{Boglione}},
  \bibinfo{author}{\bibfnamefont{U.}~\bibnamefont{D'Alesio}},
  \bibinfo{author}{\bibfnamefont{A.}~\bibnamefont{Kotzinian}},
  \bibinfo{author}{\bibfnamefont{F.}~\bibnamefont{Murgia}},
  \bibnamefont{et~al.}, \bibinfo{journal}{Phys.Rev.}
  \textbf{\bibinfo{volume}{D72}}, \bibinfo{pages}{094007}
  (\bibinfo{year}{2005}{\natexlab{b}}), \eprint{hep-ph/0507181}.

\bibitem[{\citenamefont{Sjostrand et~al.}(2006)\citenamefont{Sjostrand, Mrenna,
  and Skands}}]{Sjostrand:2006za}
\bibinfo{author}{\bibfnamefont{T.}~\bibnamefont{Sjostrand}},
  \bibinfo{author}{\bibfnamefont{S.}~\bibnamefont{Mrenna}}, \bibnamefont{and}
  \bibinfo{author}{\bibfnamefont{P.~Z.} \bibnamefont{Skands}},
  \bibinfo{journal}{JHEP} \textbf{\bibinfo{volume}{05}}, \bibinfo{pages}{026}
  (\bibinfo{year}{2006}), \eprint{hep-ph/0603175}.

\end{thebibliography}

\end{document}